%% file: KKCSG.paper.tex
\documentclass[11pt]{article}
\usepackage{jheppub}

\usepackage{amsmath, amsfonts, amssymb}
\usepackage{hyperref}
\usepackage{empheq}
\usepackage{booktabs}
\usepackage{mathrsfs}
\usepackage{xcolor}
\usepackage{hyperref}
\hypersetup{
	unicode,	% use with \texorpdfstring
	colorlinks,
	citecolor=blue,
	linkcolor=[rgb]{0,.3,.5},
	urlcolor=[rgb]{0,.3,.5}, %cyan,% pdftex
	bookmarksopen=true,
	bookmarksopenlevel=\maxdimen,
        linktocpage,
}

\newcommand{\ua}{{\ul a}}

\newcommand{\um}{{\ul m}}
\newcommand{\un}{{\ul n}}
\newcommand{\up}{{\ul p}}
\newcommand{\uq}{{\ul q}}
\newcommand{\ur}{{\ul r}}

\newcommand{\GL}{{\rm GL}}
\newcommand{\SU}{{\rm SU}}
\newcommand{\SO}{{\rm SO}}
\newcommand{\gU}{{\rm U}}

\newcommand{\q}{\theta}

\DeclareMathOperator{\sdet}{sdet}

\include{sugracom_v3}
\newcommand{\rep}[1]{\mathbf{#1}}

\newcommand{\bnabla}{{\bar\nabla}}
\newcommand{\znabla}{{\mathring\nabla}}
\newcommand{\hnabla}{{\hat\nabla}}

\newcommand{\bbD}{{\mathbb D}}
\newcommand{\bbA}{{\mathbb A}}

\newcommand{\Lie}{\mathscr{L}}
\newcommand{\HC}{\text{h.c.}}

\newcommand{\sym}[1]{\stackrel{\scriptstyle#1}{\mbox{\tiny $\smile$}}}

\newcommand{\ZZ}{\mathscr Z}

\makeatletter
\g@addto@macro\bfseries{\boldmath}
\makeatother

\numberwithin{equation}{section}

\title{4D $\cN=1$ Kaluza-Klein superspace}

\author{Katrin Becker}
\author{and Daniel Butter}
\affiliation{
George P. and Cynthia Woods
Mitchell Institute for 
Fundamental Physics and Astronomy, \\
Texas A\&{}M University.\\
College Station, TX 77843, USA}

\emailAdd{kbecker@physics.tamu.edu}
\emailAdd{dbutter@tamu.edu}

\abstract{
Motivated by recent efforts to encode 11D supergravity in 4D $\cN=1$ superfields,
we introduce a general covariant framework relevant for describing any higher dimensional
supergravity theory in external 4D $\cN=1$ superspace with $n$ additional
internal coordinates.
The superspace geometry admits both external and internal diffeomorphisms and provides
the superfields necessary to encode the components of the higher dimensional
vielbein, except for the purely internal sector, in a universal way
that depends only on the internal dimension $n$.
In contrast, the $\cN=1$ superfield content of the internal sector of the metric
is expected to be 
highly case dependent and involve covariant matter superfields, with additional 
hidden higher dimensional Lorentz and supersymmetry transformations realized in
a non-linear manner. }

\begin{document}
\maketitle

%%%%%%%%%%%%%%%%%%%%%%%%%%%%%%%%%%%%%%%%%%%%%%%%%%%%%%%%%%%%%%%%%%%%%%%%%%%%%%%%%
\section{Introduction and motivation}
%%%%%%%%%%%%%%%%%%%%%%%%%%%%%%%%%%%%%%%%%%%%%%%%%%%%%%%%%%%%%%%%%%%%%%%%%%%%%%%%%
A major difficulty in studying higher-dimensional supergravity theories is
the absence of a (finite) off-shell formulation. This leads to a number of complications,
a major one being the difficulty in writing down generic higher-derivative
supersymmetric actions. This is in sharp contrast to the situations in lower dimensions
(and fewer supersymmetries) where off-shell superspaces are available.

Standard techniques to address this always involve trade-offs. One can introduce
infinite auxiliary fields, using harmonic superspace \cite{GIKOS, GIOS}
(which is related to projective superspace
\cite{Karlhede:1984vr, Lindstrom:1989ne,GonzalezRey:1997qh,Kuzenko:2007cj,Kuzenko:2007hu,Kuzenko:2008ep})
or pure spinor superspace 
\cite{Cederwall:2009ez, Cederwall:2010tn,Cederwall:2011vy, Cederwall:2013vba},
but the former does not seem applicable beyond six dimensions and the latter
leads to a very complicated Batalin-Vilkovisky form whose on-shell component
structure proves difficult to extract (see discussions in
\cite{Chang:2014nwa, Berkovits:2018gbq}). One could take the opposite extreme --
eliminating auxiliary fields altogether -- 
by working in light cone superspace \cite{Mandelstam:1982cb,Metsaev:2004wv,Ananth:2005vg,Kallosh:2009db},
but this breaks manifest Lorentz symmetry and leads to other complications --
for example, having to work only with gauge-fixed physical degrees of freedom.

A plausible middle ground is to keep manifest some number of auxiliary fields
and some amount of supersymmetry by working in some convenient low
dimensional, low $\cN$ superspace. 4D $\cN=1$ superspace is the obvious choice,
given its relative simplicity and presence of certain features (e.g. holomorphic
superpotentials) absent in even simpler superspaces.
Already in 1983, Marcus, Sagnotti, and Siegel took this approach
with the prototypical globally supersymmetric case by showing how to recast 10D
super Yang-Mills in 4D $\cN=1$ language \cite{Marcus:1983wb}. This breaks
the 10D Lorentz group to $\SO(3,1) \times \SU(3) \times \gU(1)$, but keeps off-shell
1/4 of the supersymmetry.\footnote{This was extended to lower dimensions
in \cite{ArkaniHamed:2001tb}, motivated in part by brane-world scenarios.}

The natural next step, discussed already in the conclusion of \cite{Marcus:1983wb}, would
be to repeat the exercise for 11D supergravity, but as Marcus et al.\ noted even $\cN=2$
supergravity had not yet been fully written in $\cN=1$ superfields at that time.
In the intervening 35 years, a number of papers have examined how to rewrite
higher dimensional supergravity theories in $\cN=1$ superspace.
These have included 4D $\cN=2$ supergravity
\cite{Gates:1983ie, Gates:1984wc, Labastida:1984qa,Labastida:1984us,Labastida:1986md,
Butter:2010sc},
5D $\cN=1$ supergravity \cite{Linch:2002wg,Paccetti:2004ri,Abe:2004ar,Sakamura:2012bj},
and 6D $\cN=(1,0)$ supergravity \cite{Abe:2015yya,Abe:2017pvw}, but the 11D case has
remained open.

In the last few years, that remaining case has been explored step-by-step.
The initial papers \cite{Becker:2016xgv, Becker:2016rku} identified the structure of the
$\cN=1$ tensor hierarchy that descends from the M-theory 3-form and constructed the
unique cubic $\cN=1$ Chern-Simons action.
The superfields in this tensor hierarchy turned out to encode all the spin $\leq 1$ fields.
In particular, one of them contained a gauge-invariant 3-form field
with which one can endow a Riemannian 7-manifold with a $G_2$ structure \cite{Becker:2016edk}.
Remarkably, the $\cN=1$ Chern-Simons action, combined with a natural choice of K\"ahler
potential, led to a 4D scalar potential that reproduces the internal sector
of the 11D action. Most of the kinetic terms were also correctly reproduced, except
for those terms involving fields in the $\rep{7}$ of $G_2$. The explanation offered in
\cite{Becker:2016edk} was that the gravitino superfield, which encodes the additional
seven gravitini, should include auxiliary vector and tensor fields that when integrated
out modify the kinetic terms in the $\rep{7}$.
This was demonstrated indeed to be the case in \cite{Becker:2017zwe} where the
entire linearized action was written down in $\cN=1$ superspace. Then it was shown
in \cite{Becker:2018phr} that the full action for fields of spin $\leq 1$ could be linearly
coupled to the gravitino and graviton supermultiplets consistently using its supercurrent.
What remained was to include the gravitino and graviton couplings to all orders.

The main stumbling block to this task turns out to be $\cN=1$ superspace itself. Unlike in
globally supersymmetric cases, the superspace covariant derivatives carry geometric data
in their connections and these must be made dependent on the
internal coordinates. Put another way, one must introduce new ``internal'' derivatives in
$\cN=1$ superspace, and these must have non-trivial commutators with the ``external''
superspace derivatives. This introduces anew the old problem of solving superspace
Bianchi identities, but with the added wrinkle of an additional set of coordinates
and a slew of new superfields describing the mixed curvatures.

It turns out this can be done in a rather universal way, which seems as applicable to
minimal 5D supergravity as to 11D supergravity, although the details of intermediate cases
have not yet been worked out.
In this paper, we provide such a generic reformulation of
4D $\cN=1$ superspace with $n$ additional internal coordinates. (Our interest is
$n=7$, of course, but the formulae are agnostic to the specific choice.)
Our construction will be motivated by the requirement that it consistently
covariantize the 11D supergravity results. It will also make contact with existing
5D \cite{Sakamura:2012bj} and 6D results \cite{Abe:2017pvw}, where the linearized
version of this supergeometry was built explicitly out of prepotential superfields.

This paper is organized as follows. In section 2, we review some details of how 11D
supergravity is recast in 4D $\cN=1$ language to motivate a number of choices
we will make for the Kaluza-Klein supergeometry. Section 3 is devoted to a general
discussion of bosonic Kaluza-Klein geometry that readily generalizes to
superspace. In sections 4 through 6, we discuss how to solve the superspace
Bianchi identities. Section 4 provides a general discussion in terms of abstract
curvature superfields and shows how, with a certain minimal set of constraints,
the Bianchi identities can be rewritten in terms of simpler abstract curvature
operators. This leads to a set of six abstract operator equations that must be
satisfied. In section 5, we discuss the linearized solution to these identities,
and then in section 6, we address the full solution. The existence of full superspace
and chiral superspace actions is established in section 7. In the conclusion, we
sketch the remaining steps needed to rewrite 11D supergravity in $\cN=1$ language,
which will be the subject of a subsequent publication.

%%%%%%%%%%%%%%%%%%%%%%%%%%%%%%%%%%%%%%%%%%%%%%%%%%%%%%%%%%%%%%%%%%%%%%%%%%%%%%%%%
\section{Elements of 11D supergravity and an $\cN=1$ wishlist}
%%%%%%%%%%%%%%%%%%%%%%%%%%%%%%%%%%%%%%%%%%%%%%%%%%%%%%%%%%%%%%%%%%%%%%%%%%%%%%%%%

In order to lay the groundwork for the Kaluza-Klein superspace we will construct,
it will be helpful to sketch what is currently known about the rewriting of
11D supergravity in $\cN=1$ language \cite{Becker:2016xgv, Becker:2016rku,
Becker:2016edk, Becker:2017zwe, Becker:2018phr}. (See the introduction
of \cite{Becker:2018phr} for more details.)

%%%%%%%%%%%%%%%%%%%%%%%%%%%%%%%%%%%%%%%%%%%%%%%%%%%%%%%%%%%%%%%%%%%%%%%%%%%%%%%%%
\subsection{A sketch of 11D supergravity}\label{S:11DSugra.Sketch}
%%%%%%%%%%%%%%%%%%%%%%%%%%%%%%%%%%%%%%%%%%%%%%%%%%%%%%%%%%%%%%%%%%%%%%%%%%%%%%%%%

Locally, we decompose 11D spacetime into four external coordinates $x^m$ and seven internal coordinates $y^\um$.
Four local supersymmetries are made manifest by introducing local Grassmann coordinates
$(\theta^\mu, \bar\theta_\dmu)$, which are combined with $x^m$ to give an external
4D $\cN=1$ superspace.
The 11D spectrum comprises a metric, 3-form, and a 32-component gravitino, each of which
must be decomposed into 4D $\cN=1$ multiplets. The structure of the 3-form is easiest
to understand as its abelian gauge structure has a unique encoding in $\cN=1$ superspace.
The component form decomposes directly into a tensor hierarchy of forms
\begin{align}
C_3 \quad \rightarrow \quad
    C_{\ul{mnp}}~, \quad
    C_{m \,\ul{np}}~, \quad
    C_{m n \,\up}~, \quad
    C_{m n p}
\end{align}
extending from a 0-form to a 3-form in external spacetime.
The $\cN=1$ superspace encoding of such a tensor hierarchy is known. 
In terms of $\cN=1$ superfields, it comprises a
chiral superfield $\Phi_{\ul{mnp}}$, a real vector superfield $V_{\ul{mn}}$, a chiral
spinor superfield $\Sigma_{\um \,\alpha}$, and a real superfield $X$.
They transform under abelian gauge transformations as (in form notation)
\begin{subequations}\label{E:TH.Trafo}
\begin{align}
\delta \Phi &= \pa \Lambda~, \\
\delta V &= \tfrac{1}{2i} (\Lambda - \bar \Lambda) - \pa U~, \\
\delta \Sigma_\alpha &= -\tfrac{1}{4} \bar\cD^2 \cD_\alpha U
    + \pa \Upsilon_\alpha
    + \cW_\alpha \lrcorner \Lambda~, \\
\delta X &= \tfrac{1}{2i} (\cD^\alpha \Upsilon_\alpha - \bar \cD_\dalpha \bar\Upsilon^\dalpha)
    - \omega_{\mathsf h} (\cW_\alpha, U)
\end{align}
\end{subequations}
with $\lrcorner$ and $\pa$ denoting the interior product and de Rham differential on
the internal space, and where we have used the shorthand
\begin{align}
\omega_{\mathsf h}(\chi_\alpha, v) &:= 
    {\chi^\alpha} \lrcorner {\cD}_\alpha v 
	+ \bar \chi_{\dalpha} \lrcorner \bar {\cD}^{\dalpha} v 
	+\frac12 \left( {\cD}^\alpha \chi_\alpha \lrcorner v 
		+ \bar {\cD}_{\dalpha} \bar \chi^{\dalpha} \lrcorner v 
	\right)
~.
\end{align}
The gauge parameters are a chiral superfield $\Lambda_{\ul{mn}}$, a real superfield $U_\um$,
and a chiral spinor superfield $\Upsilon_\alpha$. The 4D $\cN=1$ derivatives involve a Kaluza-Klein
connection, $\cD := \rd - \Lie_\cA$, which acts via the internal Lie derivative, i.e.
$\Lie_\cA U := \cA \lrcorner \pa U+ \pa (\cA \lrcorner U)$.
The prepotential $\cV^\um$ for the connection $\cA$ describes the $\cN=1$ vector multiplet that
includes the Kaluza-Klein vector component of the higher dimensional vielbein. In this
covariant formulation, the Kaluza-Klein prepotential does not appear explicitly, but rather
only via the covariant derivative and its chiral field strength $\cW_\alpha{}^\um$ obeying
the usual Bianchi identities,
\begin{align}
\bar \cD_\dalpha \cW_\alpha{}^\um = 0~, \qquad
\cD^\alpha \cW_\alpha{}^\um = \bar \cD_\dalpha \bar \cW^\dalpha{}^\um~.
\end{align}

The field strength superfields invariant under the gauge transformations \eqref{E:TH.Trafo}
are given in form notation as
\begin{subequations}\label{E:TH.Curvs}
\begin{align}
E &= \pa \Phi~, \\
F &= \tfrac{1}{2i} (\Phi - \bar \Phi) - \pa V~, \\
W_\alpha &= -\tfrac{1}{4} \bar\cD^2 \cD_\alpha V
    + \pa \Sigma_\alpha
    + \cW_\alpha \lrcorner \Phi~, \\
H &= \tfrac{1}{2i} (\cD^\alpha \Sigma_\alpha - \bar \cD_\dalpha \bar\Sigma^\dalpha)
    - \pa X - \omega_{\mathsf h} (\cW_\alpha, V)~, \\
G &= -\tfrac{1}{4} \bar \cD^2 X + \cW^\alpha \lrcorner \Sigma_\alpha
\end{align}
\end{subequations}
and satisfy Bianchi identities
\begin{subequations}\label{E:TH.BIs}
\begin{align}
0 &= \pa E~, \\
0 &= \tfrac{1}{2i} (E - \bar E) - \pa F~, \\
0 &= -\tfrac{1}{4} \bar\cD^2 \cD_\alpha F
    + \pa  W_\alpha
    + \cW_\alpha \lrcorner E~, \\
0 &= \tfrac{1}{2i} (\cD^\alpha  W_\alpha - \bar \cD_\dalpha \bar W^\dalpha)
    - \pa H - \omega_{\mathsf h} (\cW_\alpha, F)~, \\
0 &= -\tfrac{1}{4} \bar \cD^2 H + \pa G + \cW^\alpha \lrcorner  W_\alpha ~.
\end{align}
\end{subequations}
From these superfields, one can construct the $\cN=1$ supersymmetrization of the 11D
Chern-Simons term:
\begin{align}\label{E:CSaction}
-12 \kappa^2\, S_{CS} &= \int \rd^{11}x\, \rd^2\theta\, \Big\{
	i \Phi \wedge \Big(E G + \frac{1}{2} W^\alpha \wedge W_\alpha
		- \frac{i}{4}\bar \cD^2 (F \wedge H) \Big)
	\eol & \qquad \qquad \qquad \qquad
	+ i \Sigma^\alpha \wedge \Big(
		E \wedge W_\alpha
		- \frac{i}{4} \bar\cD^2 (F \wedge \cD_\alpha F)
		\Big)
	\Big\}
	\eol & \quad
	+ \int \rd^{11}x\, \rd^4\theta\, \Big\{
	V \wedge \Big(
		E \wedge H
		+ F \wedge \cD^\alpha W_\alpha
		+ 2 \cD^\alpha F \wedge (W_\alpha - i \cW_\alpha \lrcorner F)
	\Big)
	\eol & \qquad \qquad \qquad \qquad
	- X E \wedge F
	\Big\}
	\eol & \quad
	+ \HC
\end{align}

It turns out that the on-shell field content of the $\cN=1$ superfields above involves
more than just the 3-form fields. They also encode all of the spin-1/2 components of
the 11D gravitino and all components of the 11D metric except for the purely external
part. It also turns out that the above superspace Chern-Simons action encodes
the kinetic terms for the 4D vector fields. Together with just one other K\"ahler-type
term (whose precise form does not concern us here), nearly the entire action of 11D supergravity
for the spin $\leq 1$ fields can be encoded in $\cN=1$ superspace \cite{Becker:2016edk}.

The above description turns out to miss a few critical elements. The external graviton and
spin-3/2 part of the gravitino must belong to additional higher superspin multiplets.
Naturally, the $\cN=1$ gravitino combines with the external graviton into a single
supermultiplet, described by a prepotential superfield
$H_{\alpha \dalpha} = (\sigma^a)_{\alpha \dalpha} H_a$, subject to a linearized
gauge transformation
\begin{align}
\delta H_{\alpha \dalpha} = D_\alpha \bar L_\dalpha - \bar D_\dalpha L_\alpha~.
\end{align}
This is the linearized prepotential of $\cN=1$ conformal supergravity.
The remaining seven spin-3/2 components of the gravitino live in a superfield 
$\Psi_{\um}{}_\alpha$, subject to the linearized transformations
\begin{align}\label{E:deltaPsi0}
\delta \Psi_\um{}_\alpha = \Xi_\um{}_\alpha + D_\alpha \Omega_\um + 2i \,\pa_\um L_\alpha~,
\end{align}
where $\Xi_\um{}_\alpha$ is a chiral spinor and $\Omega_\um$ is an unconstrained real superfield.
This describes the so-called $\cN=1$ conformal graviton multiplet.

The parameter $L_\alpha$ encodes local $\cN=1$ superconformal transformations, while $\Xi$ and $\Omega$
are usually interpreted as encoding extended supersymmetry. The matter fields of the tensor hierarchy necessarily
also vary under these transformations, but the precise form will not concern us here,
except for the following observation.
As discussed in \cite{Becker:2018phr}, the $\Xi$ transformations of $\Psi$ and 
the other matter fields take a very simple form and do not strongly constrain the action.
Therefore, we are going to take the point of view that \emph{$\Xi$ is not really an
extended supersymmetry transformation}, but rather a symmetry naturally associated
with the prepotential structure of the $\cN=1$ superspace we want to construct.
The transformation involving $\Omega$ will then
be interpreted as an honest extended supersymmetry transformation. Since any such
transformation necessarily breaks manifest $\cN=1$ supersymmetry, it will not play any
further role in our discussion.

If we ignore the $\Omega$ parameter in the linearized transformation \eqref{E:deltaPsi0},
it is possible to combine the linearized $H_{\alpha \dalpha}$ and $\Psi_{\um \alpha}$
into an abelian tensor hierarchy, just like the 3-form fields, where the Kaluza-Klein
gauge field appears encoded in the covariant derivatives. It turns out that a further
chiral spinor superfield $\Phi_{\um \un}{}_\alpha$ is needed.
The linearized gauge transformations read
\begin{subequations}
\begin{align}
\delta H_{\alpha \dalpha}
    &= \cD_\alpha \bar L_\dalpha
    - \bar \cD_\dalpha L_\alpha~, \\
\delta \Psi_1{}_\alpha &= \Xi_1{}_\alpha + 2i \,\pa L_\alpha~, \label{E:HPsiTrafo.b} \\  
\delta \Phi_2{}_\alpha &= -\pa \,\Xi_1{}_\alpha
\end{align}
\end{subequations}
where we have written the internal degrees of the forms explicitly. The derivatives $\cD$
include the Kaluza-Klein connection as with the matter fields. The corresponding curvatures
are
\begin{subequations}\label{E:LinSugraCurvs}
\begin{align}
W_{\gamma \beta \alpha}
    &= \frac{1}{16} \bar \cD^2 \Big[
    i \cD_\gamma{}^\dalpha \cD_\beta H_{\alpha \dalpha}
    + \cD_\gamma \cW_\beta \lrcorner {\Psi}_1{}_\alpha
    - 2 \cW_\gamma \lrcorner \cD_\beta {\Psi}_1{}_\alpha
    \Big]_{(\gamma\beta\alpha)}
%     \eol & \quad
    + \frac{1}{2i} \cW_{(\gamma} \lrcorner \cW_\beta \lrcorner \Phi_2{}_{\alpha)}~,
    \label{E:LinSugraCurvs.a}\\
X_1{}_{\alpha \dalpha}
    &= \frac{1}{2i} (\bar\cD_\dalpha \Psi_1{}_\alpha + \cD_\alpha \bar\Psi_1{}_\dalpha)
    + \pa H_{\alpha \dalpha}~,
    \label{E:LinSugraCurvs.b}\\
\Psi_2{}_\alpha &= \Phi_2{}_\alpha + \pa \Psi_1{}_\alpha~, 
    \label{E:LinSugraCurvs.c }\\
\Phi_3{}_\alpha &= \pa \Phi_2{}_\alpha~,
    \label{E:LinSugraCurvs.d}
\end{align}
\end{subequations}
and they satisfy the Bianchi identities\footnote{There
is an additional Bianchi identity of the form
$\cD_{\dalpha}{}^\beta \cD^{\gamma} W_{\gamma \beta \alpha}
+ \cD_{\alpha}{}^\dbeta \bar\cD^{\dgamma} \bar W_{\dgamma \dbeta \dalpha} = \cdots$,
but this occurs at rather high dimension, so we won't have much use for it.
\label{F:WBianchi}}
\begin{subequations}\label{E:LinSugraBI}
\begin{align}
\pa W_{\gamma \beta \alpha} &= \bar \cD^2 \Big[
    \frac{i}{16} \cD_\gamma{}^\dalpha \cD_\beta X_1{}_{\alpha \dalpha}
    - \frac{1}{16} \cD_\gamma \cW_\beta \lrcorner \Psi_2{}_\alpha
    + \frac{1}{8} \cW_\gamma \lrcorner \cD_\beta \Psi_2{}_\alpha
    \Big]_{(\gamma\beta\alpha)}
    + \frac{1}{2i} \cW_{(\gamma} \lrcorner \cW_\beta \lrcorner \Phi_3{}_{\alpha)}
    ~, \label{E:LinSugraBI.a} \\
\pa X_1{}_{\alpha \dalpha}
    &= \frac{1}{2i} (\cD_\alpha \bar\Psi_2{}_\dalpha
    + \bar \cD_\dalpha \Psi_2{}_\alpha)~, \label{E:LinSugraBI.b}\\
\pa \Psi_{2 \alpha} &= \Phi_{3 \alpha}~, \label{E:LinSugraBI.c} \\
\pa \Phi_{3 \alpha} &= 0~. \label{E:LinSugraBI.d}
\end{align}
\end{subequations}
The chiral spinor $\Phi_{2\, \alpha}$ is necessary to ensure gauge invariance
of the curvatures under $\Xi$ transformations. The $\Xi$ transformations are important
because they preserve the ``gauge-for-gauge'' symmetry of $L_\alpha$, whereby a shift
of $L_\alpha$ by a chiral spinor superfield can always be balanced by some compensating
gauge transformation elsewhere. For the case of $\Psi_{\um  \alpha}$, this requires
that $\Xi_{\um \alpha}$ shift by the internal derivative of that chiral spinor.
This is another reason to consider $\Xi$ transformations as part of the purely $N=1$ sector
and not an honest extended supersymmetry transformation.

It is puzzling that $\Phi_2{}_\alpha$ was not encountered in our prior
linearized analysis \cite{Becker:2017zwe} or in the supercurrent analysis
\cite{Becker:2018phr}.
The only sensible explanation is that when the prepotentials above are coupled to the tensor
hierarchy fields of 11D supergravity, it becomes possible to absorb $\Phi_2{}_\alpha$ by a field
redefinition. We will argue in the conclusion that this is indeed so.

%%%%%%%%%%%%%%%%%%%%%%%%%%%%%%%%%%%%%%%%%%%%%%%%%%%%%%%%%%%%%%%%%%%%%%%%%%%%%%%%%
\subsection{In search of a covariant completion}\label{S:WishList}
%%%%%%%%%%%%%%%%%%%%%%%%%%%%%%%%%%%%%%%%%%%%%%%%%%%%%%%%%%%%%%%%%%%%%%%%%%%%%%%%%

The complete action involving the spin $\leq 1$ fields was presented in
\cite{Becker:2018phr}, but this was only to linear order in $H_{\alpha \dalpha}$ and
$\Psi_{\um \alpha}$. The main obstruction was that unlike the tensor
hierarchy fields, $H_{\alpha \dalpha}$ and $\Psi_\um{}_\alpha$ are expected to appear
intrinsically non-polynomially in the action, just like the Kaluza-Klein
prepotential $\cV^\um$.\footnote{One could adopt a Wess-Zumino gauge and work to a certain
order in these prepotentials, but that is cumbersome in practice.}
The solution to this should be, just as with $\cV^\um$, to introduce new covariant
derivatives in which these prepotentials are encoded, so that they appear only
via minimal substitution and their associated field strengths. When $H_{\alpha \dalpha}$
is $y$-independent and decouples from the supergravity hierarchy, this can be done
using $\cN=1$ conformal superspace \cite{Butter:2009cp},
where new covariant derivatives $\nabla_A$ are introduced so that the only field strength
is the $\cN=1$ super-Weyl tensor $W_{\alpha\beta\gamma}$. Then the $L_\alpha$ 
\emph{pregauge} transformations
of the prepotential $H_{\alpha \dalpha}$ are absorbed into superdiffeomorphisms and all terms are
manifestly supercovariant.
However, if the supergraviton multiplet depends on $y^\um$, the corresponding supergeometry
must be rebuilt from scratch to accommodate this.
This means that we must look for an $\cN=1$ superspace involving not only the usual
covariant external derivatives $\nabla_A = (\nabla_a, \nabla_\alpha, \bar\nabla^\dalpha)$
but also curved internal derivatives. We can motivate the constraints
and structure of this new superspace by requiring it to consistently covariantize
the matter and actions we already have.

For example, one might expect that the internal derivatives should be valued in the
internal tangent space group, e.g. $\nabla_\ua$ with $\ua$ a vector of $\SO(n)$, but
this would lead to an immediate complication: we would need to introduce an internal
vielbein $e_\um{}^\ua$ absent in \cite{Becker:2018phr}. Now, this is not an independent
field, but is equivalent (up to an $\SO(n)$ transformation) to the internal metric
$g_{\um\un}$; however, the internal metric is, in our approach, a composite field related to the tensor
hierarchy field strength $F_{\ul{mnp}}$ via the $G_2$ structure relation
\begin{align}
\sqrt{g}\, g_{\um \un} = -\frac{1}{144} \veps^{\ul{p_1 \cdots p_7}}\, 
    F_{\ul{m \,p_1 p_2}} F_{\ul{p_3 p_4 p_5}} F_{\ul{p_6 p_7\, n}} ~.
\end{align}
This equation is complicated enough without trying to take its square root to define
$e_\um{}^\ua$. So we should avoid introducing any $\SO(n)$ symmetry and take
our internal derivative to carry a $\GL(n)$ index,
\begin{align}\label{E:IntDSketch}
\nabla_\um = \pa_\um + \cdots~,
\end{align}
where new connection terms (which may include external derivatives)
must be added. Presumably such terms would be absent in flat space.

Let's impose a number of additional conditions for the superspace we seek:
\begin{itemize}
\item 
In the absence of an internal vielbein, all the matter fields should be
$p$-forms on the internal manifold. This includes chiral superfields $\Phi_{\ul{mnp}}$ 
and $\Sigma_{\um \, \alpha}$.  The $\cN=1$ superspace
geometry must support the existence of such chiral superfields. This implies certain
constraints on the torsion and curvatures of the $\cN=1$ derivatives $\nabla_A$.
In addition, the structure of the Chern-Simons action requires the existence of chiral 
superspace in addition to full superspace, leading to additional restrictions.

\item In flat space, the Chern-Simons action as well as the entire structure of the 3-form
tensor hierarchy is written in differential form notation, and the internal derivative
appears only via the de Rham differential. We assume that the same should be true
of its curved space version. This means that we should only expect to build covariant
quantities using suitable internal covariant de Rham differentials.

\item A related point follows: we should avoid introducing a
metric-compatible affine connection into $\nabla_\um$, since we're going
to use only covariant de Rham differentials. This means the composite $g_{\ul{mn}}$ should
play no role in the superspace derivatives. In fact, we're going to assume that
\emph{none} of the superfields of the tensor hierarchy play a role in the construction
of this superspace. They should appear only as consistent ``matter'', that is covariant
superfields consistent with but not required by the supergeometry.
This means that the underlying supergeometry should be built only out of four 
fundamental prepotentials,
$H_{\alpha \dalpha}$, $\Psi_{\um}{}_\alpha$, $\Phi_{\um \un}{}_\alpha$, and $\cV^\um$.
The curvatures of the superspace geometry should be built
only out of their five corresponding field strengths,
$W_{\alpha\beta\gamma}$, $X_{\um}{}_{\alpha \dalpha}$, $\Psi_{\um \un}{}_\alpha$,
$\Phi_{\ul{mnp}}{}_\alpha$, and $\cW_\alpha{}^\um$. 

\item The tensor hierarchy transformation of $\Phi_{\ul{mnp}}$ should be something like
$\delta \Phi_3 = \nabla_1 \Lambda_2$ where $\Lambda_{\ul{mn}}$ is also chiral.
This means that $\nabla_\um$ should preserve chirality. But if $\nabla_\um$ were to commute
with both $\bar\nabla_\dalpha$ and $\nabla_\alpha$, then it would commute with
$\nabla_a$ as well, and the supergeometry would trivialize to be $y$-independent.
The only way to make sense of this is to assume $\delta \Phi_3 = \nabla_1^+ \Lambda_2$
where $\nabla_\um^+$ is a modified \emph{complex} version of $\nabla_\um$ that preserves chirality.
Similarly, there will be a $\nabla_\um^-$ for anti-chiral superfields, with
$\nabla_\um^- = (\nabla_\um^+)^*$. These should
also look roughly like \eqref{E:IntDSketch}, meaning that they differ from each
other and from $\nabla_\um$ only by connection terms. In fact, such derivatives have already
been built at the linearized level to describe 5D \cite{Sakamura:2012bj} and 
6D supergravity \cite{Abe:2017pvw} in $N=1$ superspace.

\end{itemize}

%%%%%%%%%%%%%%%%%%%%%%%%%%%%%%%%%%%%%%%%%%%%%%%%%%%%%%%%%%%%%%%%%%%%%%%%%%%%%%%%%
\section{Kaluza-Klein (super)geometry}
\label{S:KKGeo}
%%%%%%%%%%%%%%%%%%%%%%%%%%%%%%%%%%%%%%%%%%%%%%%%%%%%%%%%%%%%%%%%%%%%%%%%%%%%%%%%%

The first step towards constructing a Kaluza-Klein supergeometry is to understand
its differential geometry. It will be useful to first review how Kaluza-Klein decompositions
work in more familiar bosonic spaces.
Extending these results to superspace amounts to just extending commuting world and tangent
space coordinates to include anticommuting ones, i.e. replacing $m \rightarrow M$ 
and $a \rightarrow A$ below. This makes no difference in formulae but complicates
notation, so we will restrict to a bosonic space here for clarity.

In addition, this will allow us to overload notation in this section and use
$M$ and $A$ as indices for the coordinates and tangent space of the higher-dimensional theory
that we are decomposing. Since these higher-dimensional coordinates will only appear
here, we hope there will be no confusion later on when we restore super-indices.

\subsection{Decomposition of the vielbein}\label{S:KKGeo.1}
Suppose we begin with some $D$-dimensional bosonic space with local coordinates $\hat x^M$.
We are interested in locally decomposing this space into a $d$-dimensional ``external space''
with local coordinates $x^m$ and an $n$-dimensional ``internal space'' with local 
coordinates $y^\um$. If the $D=(d+n)$-dimensional space is equipped with a vielbein $\hat e_M{}^A$,
a natural choice for its decomposition is
\begin{align}\label{E:KKVielbein}
\hat e_M{}^A &= 
\begin{pmatrix}
e_m{}^a + A_m{}^\um \chi_\um{}^a \quad& A_m{}^\um e_\um{}^\ua \\
\chi_\um{}^a & e_\um{}^\ua
\end{pmatrix}~,
\end{align}
where we have split the tangent space index as $A = (a, \ua)$, with $a$ and $\ua$
indices associated to the external and internal tangent spaces.
We assume that the external and internal vielbeins $e_m{}^a$ and $e_\um{}^\ua$ are invertible,
with inverses $e_a{}^m$ and $e_\ua{}^\um$.
Then the inverse of \eqref{E:KKVielbein} is
\begin{align}
\hat e_A{}^M &= 
\begin{pmatrix}
e_a{}^m & - e_a{}^m A_m{}^\um \\
- e_\ua{}^\un \chi_\un{}^b e_b{}^m    \quad &   e_\ua{}^\um + e_\ua{}^\un \chi_\un{}^b e_b{}^n A_n{}^\um
\end{pmatrix}~.
\end{align}

In conventional Kaluza-Klein scenarios, the higher dimensional tangent space is
$\SO(D-1,1)$ and allows one to choose an upper triangular gauge where $\chi=0$.
But this choice will not be available to us for two reasons. The first is that
we are actually interested in the situation where $x^m$ above is extended to include
the $\theta$ variables of 4D $\cN=1$ superspace -- that is, $x^m$ will be extended
to $(x^m, \theta^\mu, \bar\theta_\dmu)$, with $a$ extended similarly to
$(a, \alpha, \dalpha)$; in this case, the local symmetries are insufficient to fix
all of $\chi$ to zero. The second reason is that manifest $\cN=1$ supersymmetry
actually seems to \emph{require} even the bosonic part of $\chi$ to be non-zero,
at least prior to a Wess-Zumino gauge fixing. We will elaborate on this momentarily.

The precise choice of decomposition above is motivated by how the fields
transform under external and internal diffeomorphisms. We denote these 
transformations by $\xi^m$ and $\Lambda^\um$ and embed them into higher dimensional
diffeomorphisms $\hat \xi^M$ via
$\hat\xi^m = \xi^m$ and $\hat\xi^\um = \Lambda^\um - \xi^m A_m{}^\um$.
Under internal diffeomorphisms,
\begin{subequations}\label{E:TrafoIntE}
\begin{align}
\delta_\Lambda e_m{}^a &= \Lambda^\un \pa_\un e_m{}^a~,\\
\delta_\Lambda \chi_\um{}^a &= \pa_\um \Lambda^\un \chi_\un{}^a + \Lambda^\un \pa_\un \chi_\um{}^a~,\\
\delta_\Lambda A_m{}^\um &= \pa_m \Lambda^\um - A_m{}^\un \pa_\un \Lambda^\um
    + \Lambda^\un \pa_\un A_m{}^\um \equiv \hat D_m \Lambda^\um ~, \\
\delta_\Lambda e_\um{}^\ua &= \pa_\um \Lambda^\un e_\un{}^\ua + \Lambda^\un \pa_\un e_\um{}^\ua~.
\end{align}
\end{subequations}
That is $e_m{}^a$ transforms as an internal scalar, while $\chi_\um{}^a$ and $e_\um{}^\ua$
transform as internal 1-forms, and $A_m{}^\um$ transforms as a connection.
Under external diffeomorphisms,
\begin{subequations}\label{E:TrafoExtE}
\begin{align}
\delta_\xi e_m{}^a &= \hat D_m \xi^n e_n{}^a + \xi^n \hat D_n e_m{}^a~,\\
\delta_\xi \chi_\um{}^a &= \xi^n \hat D_n \chi_\um{}^a + \pa_\um \xi^n e_n{}^a~,\\
\delta_\xi A_m{}^\um &= \xi^n F_{n m}{}^\um~, \\
\delta_\xi e_\um{}^\ua &= \xi^n \hat D_n e_\um{}^\ua~,
\end{align}
\end{subequations}
where we have defined $\hat D_m := \pa_m - \delta_\Lambda(A_m)$ as the covariant
external derivative (with $\xi^m$ understood to be an internal scalar). The fields
$e_m{}^a$ and $e_\um{}^\ua$ transform as an external 1-form and a scalar respectively.
$\chi_\um{}^a$ transforms as a scalar with an anomalous piece involving $\pa_\um \xi^n$,
and $A_m{}^\um$ transforms as a connection with field strength $F_{n m}{}^{\up}$ given by
\begin{align}\label{E:DefFmn}
F_{n m}{}^\up := 2 \,\pa_{[n} A_{m]}{}^\up - 2 A_{[n}{}^\un \pa_\un A_{m]}{}^\up~.
\end{align}
The field strength automatically obeys the Bianchi identity $\hat D_{[p} F_{n m]}{}^{\uq} = 0$
and transforms as an internal vector and external 2-form under internal and external
diffeomorphisms.

It is conspicuous in the transformation laws above that not all of the components of the higher
dimensional vielbein transform into each other. In particular, the external vielbein $e_m{}^a$,
the Kaluza-Klein gauge field $A_m{}^\um$, and the additional field $\chi_\um{}^a$ can be
separated from the internal vielbein $e_\um{}^\ua$. This is fortuitous, as this is exactly
the sort of situation we require. The two prepotentials $H_{\alpha \dalpha}$ and $\cV^\um$
already encode $e_m{}^a$ and $A_m{}^\um$. It is natural to suppose $\Psi_{\um \alpha}$
encodes $\chi_\um{}^a$, and this is not hard to see. At the linearized level,
$\delta \chi_\um{}^a = \pa_\um \xi{}^a$ arises from
$\delta \Psi_\um{}_\alpha = 2 i \,\pa_\um L_\alpha$ provided we identify $\chi$ as
\begin{align}\label{E:DefLinChi}
\chi_\um{}^a = \frac{1}{4} (\bar\sigma^a)^{\dalpha \alpha} (
    \bar D_\dalpha \Psi_\um{}_\alpha - D_\alpha \bar\Psi_\um{}_\dalpha) \vert_{\theta=0}~.
\end{align}
Observe that the opposite combination
$\bar D_\dalpha \Psi_{\um \alpha} + D_\alpha \bar \Psi_{\um \dalpha}$ is what contributes to
the field strength $X_\um{}^a$ \eqref{E:LinSugraCurvs.b}.

In 11D supergravity, it is possible to set the component field $\chi_\um{}^a$ (as well as
the bottom component of $X_\um{}^a$) to zero by an $\Omega$ transformation. That is, the
usual Lorentz gauge-fixing of Kaluza-Klein theory corresponds here to a
choice of Wess-Zumino gauge, and precisely this choice was made in the
linearized analysis of \cite{Becker:2017zwe}.
But Wess-Zumino gauge fixing is awkward at the superfield level. While we can set
the bottom component of $\chi$ to zero, higher $\theta$ components will survive,
and so we cannot discard it completely. It is simpler to just keep the $\Omega$ gauge
unfixed.

As we have already mentioned, the internal vielbein $e_\um{}^\ua$ should not play
any role. This is because, at least in the $\cN=1$ case, the internal metric is
not its own independent superfield but is encoded in the bottom component of the 3-form
field strength $F_{\ul{mnp}}$. Moreover, in the $\cN=1$ spectrum we have already constructed,
there is no internal Lorentz group. All superfields are in 4D representations or representations
of the internal diffeomorphism group $\GL(7)$. This means that as we build covariant external
and internal derivatives, we must forbid the use of $e_\um{}^\ua$ at any point.
It is an important fact that this will be possible.

\subsection{Covariant internal $p$-forms and a covariant de Rham differential}

The transformation rules \eqref{E:TrafoIntE} and \eqref{E:TrafoExtE}
motivate a uniform notion for how external and
internal covariant forms transform under external and internal diffeomorphisms.
We remark first that a field $\phi$ is a \emph{covariant scalar field} if it transforms as
\begin{align}\label{E:TrafoPhi}
\delta \phi = \xi^n \hat D_n \phi + \Lambda^\un \pa_\un \phi
\end{align}
under external and internal diffeomorphisms. This definition naturally arises by
taking $\phi$ to be a scalar on the full space, i.e.
$\delta \phi = \hat \xi^N \hat\pa_N \phi$, and then decomposing $\hat\xi^N$.
The extension to external or internal 1-forms is obvious.
A field $\omega_\um$ is a \emph{covariant internal 1-form} if it transforms as
\begin{align}
\delta \omega_\um = \xi^n \hat D_n \omega_\um + \Lambda^\un \pa_\un \omega_\um + \pa_\um \Lambda^\un \omega_\un~.
\end{align}
Similarly, a field $\omega_m$ is a \emph{covariant external 1-form} if it transforms as
\begin{align}
\delta \omega_m = \xi^n \hat D_n \omega_m + \hat D_m \xi^n \, \omega_n + \Lambda^\un \pa_\un \omega_m~.
\end{align}
Both of these definitions arise by requiring $\omega_\ua$ and $\omega_a$
to transform as scalars under external and internal diffeomorphisms, and then defining
$\omega_m := e_m{}^a \omega_a$ and $\omega_\um := e_\um{}^\ua \omega_\ua$.
This is quite natural if we have a 1-form $\hat\omega_M$ on the full spacetime. Then
the usual way to define its external and internal components is to flatten the
indices with the higher-dimensional vielbein and then to unflatten with the external
or internal vielbein, i.e. $\omega_m := e_m{}^a \hat e_a{}^M \hat \omega_M$ and 
$\omega_\um := e_\um{}^\ua \hat e_\ua{}^M \hat\omega_M$. 
This can be generalized to higher degree forms or mixed internal/external forms.
However, for the remainder of this section, we will mainly be interested in internal
$p$-forms, since the $\cN=1$ superfields we encounter for 11D supergravity will be
in such representations.

We now want to introduce a notion of internal and external covariant derivatives -- that is,
generalizations of $\pa_\um$ and $\pa_m$ that preserve covariance.
Let's start with a covariant scalar field $\phi$ transforming as \eqref{E:TrafoPhi}.
It is obvious that $\hat e_a{}^M \pa_M \phi$ and $\hat e_\ua{}^M \pa_M \phi$
transform covariantly, as these are just the external and internal components of $D_A \phi$.
This suggests the definitions
\begin{subequations}
\begin{align}
D_m \equiv e_m{}^a D_a &:= e_m{}^a \hat e_a{}^M \pa_M = \pa_m - A_m{}^\um \pa_\um ~, \\
D_\um &:= e_\um{}^\ua  \hat e_\ua{}^M \pa_M = 
    \pa_\um - \chi_\um{}^a e_a{}^m (\pa_m - A_m{}^\un \pa_\un )~.
\end{align}
\end{subequations}
Then $D_m \phi$ and $D_\um \phi$ indeed transform covariantly.
Note that the former coincides with $\hat D_m$ for a scalar field, but 
$D_\um$ is \emph{not} identical to $\pa_\um$. 
In neither case does $e_\um{}^\ua$ appear in the fundamental
definition of the derivative. 

For an internal 1-form $\omega_\um$, the situation is a bit more subtle.
The derivative $\hat D_m$ acts as
\begin{align*}
\hat D_m \omega_\um 
    = D_m \omega_\um - \pa_\um A_m{}^\un \omega_\un
\end{align*}
This turns out to be covariant under internal diffeomorphisms, but it fails to be covariant
under external ones. One finds that
\begin{align}
\delta_\xi (\hat D_n \omega_\um) &= \hat D_n \xi^p \hat D_p \omega_\um 
	+ \xi^p \hat D_p \hat D_n \omega_\um
	- \pa_\um \xi^p F_{p n}{}^\up \omega_\up~,
\end{align}
The extra third term can be cancelled if we use instead the combination
$\hat D_n \omega_\um + \chi_\um{}^m F_{m n}{}^\up \omega_\up$.
This suggests that we introduce an external $\GL(n)$ connection acting on the internal indices,
$\mathring \Gamma_m{}_\un{}^\up$, so that
\begin{align}
\mathring \nabla_n \omega_\um &:= D_n \omega_\um
	- \mathring\Gamma_{n \um}{}^\up \omega_\up~, \qquad 
	\mathring\Gamma_{n \um}{}^\up := \pa_\um A_n{}^\up - \chi_\um{}^a F_{a n}{}^\up~.
\end{align}
It is convenient here to consider the term $\pa_\um A_n{}^\up$, originally part
of the internal Lie derivative, as part of the $\GL(n)$ connection. Or to put it
another way, we define $\mathring \nabla_m$ in terms of $D_m$ instead of $\hat D_m$.
We have denoted this specific choice of $\GL(n)$ connection with a circle accent
$\mathring{\phantom{a}}$
to emphasize that it is the simplest choice to make. Any other connection $\Gamma$
will differ from $\mathring \Gamma$ by some tensor field. A different choice might seem artificial,
but when we choose natural $\cN=1$ superspace constraints, they will turn out to
lead to such a modified $\GL(n)$ connection.

Now consider the internal derivative of $\omega_\um$. From flat space experience,
we expect that we should make do by covariantizing the de Rham differential, that is,
$\pa_{[\um} \omega_{\un]}$. From the scalar field case, we expect to use
$\hat D_\um := \pa_\um - \chi_\um{}^a \hat D_a$ plus some additional piece.
Under internal diffeomorphisms, we find that
\begin{align}
\delta_\Lambda (\hat D_\un \omega_\um)
	&= \pa_\un \Lambda^\up \hat D_\up \omega_\um
	+ \pa_\um \Lambda^\up \hat D_\un \omega_\up
	+ \Lambda^\up \pa_\up \hat D_\un \omega_\um
	+ \pa_\un \pa_\um \Lambda^\up \omega_\up~.
\end{align}
This indeed becomes covariant if we antisymmetrize $\un$ and $\um$.
However, under external diffeomorphisms,
\begin{align}
\delta_\xi (\hat D_\un \omega_\um)
	= \xi^p D_p \hat D_\un \omega_\um
	+ \xi^p \pa_\un \pa_\um A_p{}^\up \omega_\up
	- \chi_\un{}^a \pa_\um \xi^m F_{am}{}^\up \omega_\up~.
\end{align}
While the second term drops out upon antisymmetrizing, the third term remains.
To cancel it, we now introduce an internal leg to the  $\GL(n)$ connection
so that the antisymmetric part of $\mathring\nabla_\un \omega_\um$ is covariant.
We write this contribution as
\begin{align*}
\mathring\nabla_\un \omega_\um &:= 
	\pa_\un \omega_\um
        - \chi_\un{}^a \nabla_a \omega_\um
        - \mathring\Gamma_{\un \um}{}^\up \omega_\up~, \qquad
\mathring\Gamma_{\un \um}{}^\up
	:= \frac{1}{2} \chi_\un{}^c \chi_\um{}^b F_{c b}{}^\up~.
\end{align*}
Note that there is a contribution to the $\GL(n)$ connection coming from
the second term, so that $\chi_\un{}^a e_a{}^n \mathring \Gamma_{n \um}{}^\up$ is
being added to the explicit $\mathring \Gamma_{\un \um}{}^\up$.

The generalization to internal $p$-forms is obvious, but with the caveat that only
the totally antisymmetric part of the internal covariant derivative is actually covariant.
\emph{In other words, we covariantize only the internal de Rham differential, not the
internal derivative in general.} 
It is remarkable that $\chi$ and $A$ alone are needed to build an internal covariant
de Rham differential.

Naturally, the next objects one might consider are external $p$-forms, with an eye to
generalize to mixed external/internal forms, but the situation grows more complicated.
For example, if $\omega_m$ is an external 1-form, we find that
$2 \hat D_{[n} \omega_{m]} + F_{n m}{}^\up \chi_\up{}^c e_c{}^p \omega_p$
transforms as an external 2-form. This might suggest introducing a connection for
the external coordinate indices, but we should avoid doing this.
Eventually, we want to
reproduce as much as possible the structure of existing 4D $\cN=1$ superspace, and
no affine connection plays any role there. Instead, one deals solely with the Lorentz
and other tangent space connections. In addition, experience with 11D supergravity
suggests we will deal only with covariant $\cN=1$ superfields without any external
coordinate indices, but only internal $\GL(n)$ indices (and possibly Lorentz spinor
or vector indices), and so such objects won't be directly encountered.\footnote{The superfields of the $\cN=1$ tensor hierarchy
discussed in section \ref{S:11DSugra.Sketch} will turn out to be components of mixed superforms
in superspace.}

\subsection{Including tangent space connections}
\label{S:KK.Hconn}

We want to include additional connections for gauge symmetries that act on
$e_m{}^a$ and other tensor fields. The prototypical example is Lorentz symmetry
but we will be rather general since later on we will be considering the $\cN=1$ superconformal
group. Suppose we have a group $\cH$ that acts on $e_m{}^a$ and $\chi_\um{}^a$ as
\begin{align}
\delta_\cH e_m{}^a = e_m{}^b \lambda^{x} f_{x b}{}^a~, \qquad
\delta_\cH \chi_\um{}^a = \chi_\um{}^b \lambda^{x} f_{x b}{}^a~,
\end{align}
where $\lambda^{x}$ is a local gauge parameter and we 
use $x,y,\cdots$ to label the generators $g_x$ of $\cH$. 
We assume $A_m{}^\um$ is invariant. We suppose further that
we are furnished an $\cH$ connection with external and internal components,
$h_m{}^x$ and $h_\um{}^x$, transforming under $\cH$ transformations as
\begin{align}
\delta_\cH h_m{}^x &= \hat D_m g^x + e_m{}^a g^y f_{y a}{}^x + h_m{}^y g^z f_{z y}{}^x~, \eol
\delta_\cH h_\um{}^x &= \pa_\um g^x + \chi_\um{}^a g^y f_{y a}{}^x + h_\um{}^y g^z f_{z y}{}^x~.
\end{align}
The constants $f$ should obey the Jacobi identity associated with a Lie algebra that extends
$\cH$ by a generator $P_a$, with commutation relations\footnote{We treat $g_x$ as an operator
acting from the left that takes a covariant field to a covariant field,
so that $\delta_1 \delta_2 \Phi = \lambda_1^x \lambda_2^y g_x g_y \Phi$.}
\begin{align}\label{E:FlatHPAlgebra}
[g_x, g_y] = - f_{x y}{}^z g_z~, \qquad [g_x, P_a] = -f_{x a}{}^b P_b - f_{x a}{}^y g_y~, \qquad
[P_a, P_b] = 0~.
\end{align}
One can check that the commutator of $\delta_\cH$ transformations reproduces the $[g,g]$ algebra.

Now we augment the covariant derivatives defined in the previous section
with the $\cH$-connections. At the same time, we will allow the $\GL(n)$ connection
$\Gamma$ to differ from the simplest choice $\mathring \Gamma$. Explicitly, we have
\begin{align}
\nabla_a &= e_a{}^m (
    \pa_m - A_m{}^\un \pa_\un - \Gamma_{m \un}{}^\up g_\up{}^\un
    - h_m{}^x g_x)~, \eol
\nabla_\um &= \pa_\um - \chi_\um{}^a \nabla_a
	- \Gamma_{\um \un}{}^\up g_\up{}^\un
	- h_\um{}^x g_x~.
\end{align}
The operator $g_\um{}^\un$ generates $\GL(n)$ transformations, i.e.
$g_\um{}^\un \omega_\up = \delta_\up{}^\un \omega_\um$.
In order for the above covariant derivatives to remain covariant with respect
to external and internal diffeomorphisms, we must take the $\cH$ connections to
transform as\footnote{Note the anomalous term in the transformation of $h_\um{}^x$ that rotates it into
$h_m{}^x$: it is similar in structure to the anomalous term in the transformation of $\chi_\um{}^a$
that rotates it into $e_m{}^a$.}
\begin{align}
\delta h_m{}^x &= \hat D_m \xi^n h_n{}^x + \xi^n \hat D_n h_m{}^x + \Lambda^\un \pa_\un h_m{}^x~, \eol
\delta h_\um{}^x &=
    \pa_\um \Lambda^\un h_\un{}^x
    + \pa_\um \xi^n h_n{}^x
    + \xi^n \hat D_n h_\um{}^x
    + \Lambda^\un \pa_\un h_\um{}^x~.
\end{align}
They are also $\cH$-covariant in the sense that if $\Phi$ is some field transforming as
$\delta_\cH \Phi = \lambda^x g_x \Phi$, then 
\begin{align}
\delta_\cH  \nabla_a \Phi &\equiv \lambda^x g_x  \nabla_a \Phi
    = \lambda^x ( \nabla_a g_x \Phi  - f_{x a}{}^b  \nabla_b \Phi)~, \eol
\delta_\cH  \nabla_\um \Phi &\equiv \lambda^x g_x  \nabla_\um \Phi
    = \lambda^x  \nabla_\um g_x \Phi~,
\end{align}
which amounts to the formal operator algebra
\begin{align}
[g_x,  \nabla_a] = -f_{x a}{}^b  \nabla_b - f_{x a}{}^y g_y~, \qquad
[g_x,  \nabla_\um] = 0~.
\end{align}
Above, $\nabla_a$ is playing the role of $P_a$ in the flat
algebra \eqref{E:FlatHPAlgebra}. The vanishing commutators of $P_a$ and
$\pa_\um$ with each other are replaced with field-dependent curvature tensors
\begin{align}
[\nabla_a, \nabla_b] 
	&= - T_{a b}{}^c \nabla_c 
	- \cL_{F_{a b}} 
	- \cR_{a b}{}_\um{}^\un g_\un{}^\um
	- R_{a b}{}^x g_x~, \eol{}
[\nabla_a, \nabla_\um] 
	&= - T_{a \um}{}^b \nabla_b
	- \cL_{F_{a \un}} 
	- \cR_{a \um \un}{}^\up g_\up{}^\un
	- R_{a \um}{}^x g_x~, \eol{}
[\nabla_\um, \nabla_\un]
	&= -T_{\um\un}{}^a \nabla_a
	- \cL_{F_{\um \un}} 
	- \cR_{\ul{mnp}}{}^\uq g_\uq{}^\up
	- R_{\um \un}{}^x g_x~,
\label{E:NablaComms}
\end{align}
where $T^a$ is the external torsion tensor, $F^\um$ is the internal Kaluza-Klein curvature,
$\cR_\um{}^\un$ is the $\GL(n)$ curvature, and $R^x$ is the $\cH$-curvature.
$\cL$ denotes the internal covariant Lie derivative, defined so that any
lower internal form indices of $F^\um$ are spectators, e.g.
\begin{align}
\cL_{F_{\um \un}} \omega_\up &:=
    2 \,F_{\um \un}{}^\uq \nabla_{[\uq} \omega_{\up]}
    + \nabla_\up ( F_{\um \un}{}^{\uq} \omega_\uq )
    = 
    F_{\um \un}{}^\uq \nabla_{\uq} \omega_{\up}
    + \nabla_\up F_{\um \un}{}^{\uq} \,\omega_\uq ~.
\end{align}
We have chosen to package the internal curvature term in \eqref{E:NablaComms} 
as a covariant Lie derivative
(rather than a covariant derivative) because this ensures covariance of the
curvature terms separately when the commutator acts on an internal $p$-form.
This amounts to a redefinition of the $\GL(n)$ curvature $\cR{}_\um{}^\un$.

The external torsion tensors in \eqref{E:NablaComms} are given by
\begin{align}
T_{n m}{}^a &= 2 D_{[n} e_{m]}{}^a + 2 \,e_{[n}{}^b h_{m]}{}^x f_{x b}{}^a + F_{n m}{}^\up \chi_\up{}^a~, \eol
T_{n \um}{}^a + \chi_\um{}^b T_{n b}{}^a &=
    \hat D_n \chi_\um{}^a - \pa_\um e_n{}^a
    + e_n{}^b\, h_\um{}^x f_{x b}{}^a
    - \chi_\um{}^b h_n{}^x f_{x b}{}^a~, \eol
T_{\um \un}{}^a + 2 \chi_{[\um}{}^c T_{c \un]}{}^a + \chi_\um{}^c \chi_\un{}^b T_{c b}{}^a
    &=
    2 \pa_{[\un} \chi_{\um]}{}^a
    + 2\,\chi_{[\un}{}^b\, h_{\um]}{}^x f_{x b}{}^a~.
\end{align}
Here one must plug the first equation into the second and both into the third to
solve for $T_{n \um}{}^a$ and $T_{\um\un}{}^a$, and then flatten external world indices
with $e_a{}^m$. Similarly, the $\cH$ curvatures are given by
\begin{align}
R_{m n}{}^x &= 2 \hat D_{[m} h_{n]}{}^x
    + 2 h_{[m}{}^y e_{n]}{}^b f_{b y}{}^x
    + h_m{}^y h_n{}^z f_{z y}{}^x
    + F_{m n}{}^\up h_\up{}^x~, \eol
R_{m \un}{}^x + \chi_\un{}^b R_{m b}{}^x &=
    \hat D_m h_\un{}^x - \pa_\un h_m{}^x
    + h_m{}^y \chi_\un{}^b f_{b y}{}^x
    + e_m{}^b h_\un{}^y f_{y b}{}^x
    \eol & \qquad
    + h_m{}^y h_\un{}^z f_{z y}{}^x~, \eol
R_{\um \un}{}^x + 2 \chi_{[\um}{}^c R_{c \un]}{}^x + \chi_\um{}^c \chi_\un{}^b R_{c b}{}^x
    &= 2 \pa_{[\um} h_{\un]} 
    + 2 h_{[\um}{}^x \chi_{\un]}{}^a f_{a y}{}^x
    + h_\um{}^y h_\un{}^z f_{z y}{}^x~.
\end{align}
We do not give explicit expressions the $\GL(n)$ curvatures, although
they can be worked out straightforwardly. $F_{a b}{}^\um$ is given by flattening
the form indices of $F_{mn}{}^\um$ in \eqref{E:DefFmn} with the external vielbein. The
expressions for the mixed $F_{a \un}{}^\um$ and internal $F_{\ul{pn}}{}^\um$
tensors can be worked out explicitly. However, it is more helpful to observe that
when $\Gamma$ is chosen to be $\mathring\Gamma$, one finds that
$\mathring F_{a \um}{}^\un = \mathring F_{\um \un}{}^\up = 0$.
Then deforming the $\GL(n)$ connection by a purely covariant pieces $\Delta \Gamma$, defined so that
\begin{align}
\nabla_a := \mathring \nabla_a - \Delta\Gamma_a{}_\un{}^\up \,g_\up{}^\un~, \qquad
\nabla_\um := \mathring \nabla_\um - \Delta\Gamma_\um{}_\un{}^\up \,g_\up{}^\un~,
\end{align}
one can show that
\begin{align}\label{eq:defDeltaGamma}
F_{a \um}{}^\un = \Delta \Gamma_a{}_\um{}^\un~, \qquad
F_{\um \un}{}^\up = 2\, \Delta\Gamma_{[\um \un]}{}^\up ~.
\end{align}

By construction the curvatures above must obey Bianchi identities,
\begin{align*}
\sum_{[abc]} [\nabla_a, [\nabla_b, \nabla_c]] = 0~, \qquad
\sum_{[ab]} [\nabla_a, [\nabla_b, \nabla_\um]] = - [\nabla_\um, [\nabla_a, \nabla_b]]~, \qquad \text{etc.}
\end{align*}
Particularly useful are the Kaluza-Klein field strength Bianchi identities, which read
\begin{align}
0 &= \sum_{[abc]} \Big(\nabla_c F_{a b}{}^\um
	+ T_{a b}{}^e F_{e c}{}^\um
	+ F_{a b}{}^\un F_{\un c}{}^\um
	\Big)~, \eol
0 &= \sum_{[ab]} \Big(
	- \cR_{a b}{}_\um{}^\un
	+ 2 \nabla_a F_{b \um}{}^\un
	+ T_{a b}{}^e F_{e \um}{}^\un
	+ 2 T_{\um a}{}^e F_{e b}{}^\un
	+ F_{a b}{}^\up F_{\up \um}{}^\un
	+ 2 F_{\um a}{}^\up F_{\up b}{}^\un
	\Big)~,\eol
0 &= \sum_{[\um\un]} \Big(
	- 2 \cR_{a \um \un}{}^\up
	+ \nabla_a F_{\um \un}{}^\up
	+ T_{\um \un}{}^c F_{c a}{}^\up
	+ 2 T_{a \um}{}^e F_{e \un}{}^\up
	+ F_{\um\un}{}^\uq F_{\uq a}{}^\up
	+ 2 F_{a \um}{}^\uq F_{\ul{qn}}{}^\up
	\Big)~,\eol
0 &= \sum_{[\ul{mnp}]} \Big(
	- \cR_{\ul{mnp}}{}^\uq
	+ T_{\um \un}{}^e F_{e \up}{}^\uq
	+ F_{\um \un}{}^\ur F_{\ul{rp}}{}^\uq
	\Big)~.
\end{align}
The first equation ensures that $F_{ab}{}^\um$ is covariantly closed.
The other three determine the parts of the $\GL(n)$ curvature $\cR$ that are antisymmetric
in lower internal indices.
Because we will only be constructing internal covariant de Rham differentials,
only the (internal) antisymmetric parts of $\cR$ will ever appear, and these are
completely determined in terms of the other quantities.

Finally, for reference we give the covariantized external diffeomorphisms of
the various connections, which arise by combining an external diffeomorphism with
$\xi^m = \xi^a e_a{}^m$ and an $\cH$ transformation with
$\lambda^x = - \xi^a e_a{}^m h_m{}^x$:
\begin{align}
\delta^{\rm cov}_\xi e_m{}^a &= 
    D_m \xi^a 
    + h_m{}^x \xi^c f_{c x}{}^a
    + e_m{}^b \xi^c (T_{c b}{}^a - F_{c b}{}^\up \chi_\up{}^a)~, \eol
\delta^{\rm cov}_\xi \chi_\um{}^a &= 
    \pa_\um \xi^a 
    + h_\um{}^x \xi^c f_{c x}{}^a
    + \xi^c (T_{c \um}{}^a + \chi_\um{}^b T_{c b}{}^a)~, \eol
\delta^{\rm cov}_\xi h_m{}^x &= e_m{}^b \xi^c (R_{c b}{}^x - F_{c b}{}^\up h_\up{}^x)
    + h_m{}^y \xi^b f_{b y}{}^x ~, \eol
\delta^{\rm cov}_\xi h_\um{}^x &= 
    \xi^c (R_{c \um}{}^x + \chi_\um{}^b R_{c b}{}^x)
    + h_\um{}^y \xi^b f_{b y}{}^x~, \eol
\delta^{\rm cov}_\xi A_m{}^\um &= e_m{}^b \xi^c F_{c b}{}^\um~.
\end{align}
These transformations are relevant when the relations discussed above
are promoted to superspace; then the fermionic component of $\xi^A$ is identified
with the local supersymmetry parameter. Then it is crucial that the above
transformations involve covariant tensors; this ensures that the
SUSY transformations are sensibly defined.

%%%%%%%%%%%%%%%%%%%%%%%%%%%%%%%%%%%%%%%%%%%%%%%%%%%%%%%%%%%%%%%%%%%%%%%%%%%%%%%%%
\section{The supergeometry of 4D $\cN=1$ Kaluza-Klein superspace}
%%%%%%%%%%%%%%%%%%%%%%%%%%%%%%%%%%%%%%%%%%%%%%%%%%%%%%%%%%%%%%%%%%%%%%%%%%%%%%%%%

Now we are in a position to start building the general supergeometry of 4D
$\cN=1$ Kaluza-Klein superspace. The first step is to extend the discussion of
section \ref{S:KKGeo} by allowing the external space considered there to be a
superspace. This is just a cosmetic change, promoting the coordinates $x^m$
to supercoordinates $z^M = (x^m, \theta^\mu, \bar\theta_\dmu)$ and the tangent
indices $a$ to $A = (a, \alpha, \dalpha)$.
This requires promoting the external vielbein $e_m{}^a$, the Kaluza-Klein
gauge field $A_m{}^\um$, and the additional field $\chi_m{}^m$ to superfields
in the obvious manner, i.e.
\begin{gather}
e_m{}^a \rightarrow E_M{}^A~, \qquad A_m{}^\um \rightarrow A_M{}^\um~, \qquad
\chi_\um{}^a \rightarrow \chi_\um{}^A~.
\end{gather}
However, we do not modify the internal space -- it remains a bosonic manifold
with a $\GL(n)$ index $\um$. Let us not reproduce every formula,
but only give a few that are directly relevant. The superspace external covariant
derivatives $\nabla_A$ and internal $\nabla_\um$ are given by
\begin{align}
\nabla_A &= E_A{}^M (\pa_M - A_M{}^\un \pa_\un - \Gamma_{M \un}{}^\up g_\up{}^\un)
	- H_A{}^x g_x~, \eol
\nabla_\um &= \pa_\um - \chi_\um{}^A \nabla_A
	- \Gamma_{\um \un}{}^\up g_\up{}^\un
	- H_\um{}^x g_x~,
\end{align}
with the $\GL(n)$ connections defined as
\begin{align}
\Gamma_{A \um}{}^\up & := E_A{}^N \pa_\um A_N{}^\up - \chi_\um{}^B F_{B A}{}^\up
    + \Delta \Gamma_A{}_\um{}^\un~, \eol
\Gamma_{\un \um}{}^\up 
	&:= 
	- \frac{1}{2} \chi_\un{}^B \chi_\um{}^A F_{A B}{}^\up
        + \Delta \Gamma_{\un \um}{}^\up
        + \chi_\un{}^A \Delta \Gamma_A{}_\um{}^\un
        ~,
\end{align}
where the $\Delta \Gamma$ terms transform covariantly.
The covariant derivative algebra reads
\begin{subequations} \label{E:CovDAlg}
\begin{align}
[\nabla_A, \nabla_B] 
	&= - T_{A B}{}^C \nabla_C
	- \cL_{F_{A B}} 
	- \cR_{A B}{}_\um{}^\un g_\un{}^\um
	- R_{A B}{}^x g_x~, \\
[\nabla_A, \nabla_\um] 
	&= - T_{A \um}{}^B \nabla_B
	- \cL_{F_{A \un}} 
	- \cR_{A \um \un}{}^\up g_\up{}^\un
	- R_{A \um}{}^x g_x~, \\
[\nabla_\um, \nabla_\un]
	&= -T_{\um\un}{}^A \nabla_A
	- \cL_{F_{\um \un}} 
	- \cR_{\ul{mnp}}{}^\uq g_\uq{}^\up
	- R_{\um \un}{}^x g_x~.
\end{align}
\end{subequations}

In superspace, one is not generally interested in the precise expressions for the
various torsions and curvatures in terms of the potentials. Rather, one imposes
some constraints on the torsions/curvatures and solves the Bianchi identities
in terms of some fundamental curvature superfields
(which obey Bianchi identities themselves).\footnote{In principle, the fundamental
curvature superfields as well as all the potentials can in turn be solved in terms of
prepotential superfields explicitly. Usually this is highly non-polynomial and not immediately
useful. Typically only the linearized solution around a given background (e.g. flat space)
is necessary.} These quantities, e.g. $W_{\alpha\beta\gamma}$ in $\cN=1$ conformal
superspace, or $W_{\alpha\beta\gamma}$, $R$, and $G_{\alpha \dalpha}$ in the conventional
$\cN=1$ Wess-Zumino superspace (see e.g. \cite{Wess:1992cp, Buchbinder:1998qv,Gates:1983nr}),
are, along with the covariant derivatives, supermeasures,
and any covariant matter superfields, sufficient to construct covariant Lagrangians.
In our case, we expect these curvature superfields to be built out of the basic curvatures
$W_{\alpha\beta\gamma}$, $X_{\um}{}_{\alpha \dalpha}$, $\Psi_{\um \un}{}_\alpha$,
$\Phi_{\ul{mnp}}{}_\alpha$, and $\cW_\alpha{}^\um$.

\subsection{Abstract solution of the Bianchi identities}
A great deal of progress can be made working almost entirely abstractly if a very
strong set of constraints is imposed from the beginning:\footnote{The last constraint is mainly
a \emph{conventional constraint} -- that is, a definition of the connections in
$\nabla_a$.}
\begin{align}\label{E:BasicConstraint}
\{\nabla_\alpha, \nabla_\beta\} = \{\bnabla_\dalpha, \bnabla_\dbeta\} = 0~, \qquad
\{\nabla_\alpha, \bar\nabla_\dbeta\} = -2i \nabla_{\alpha\dbeta}~.
\end{align}
These coincide with the constraints of $\cN=1$ super Yang-Mills and were shown in
$\cN=1$ conformal superspace to be the appropriate constraints to describe
$\cN=1$ conformal supergravity \cite{Butter:2009cp}. These imply the existence 
of a coordinate system and a gauge where
$\bar\nabla^\dalpha = \pa / \pa \bar\theta_\dalpha$. In such a gauge, covariantly chiral
superfields are simply independent of $\bar\theta$. Such a set of constraints is not
actually necessary for supergravity (conventional Wess-Zumino superspace does not satisfy
these constraints, for example), but we will find them to be the right constraints in
our case. 

An immediate consequence of these constraints is the simplification of the external
spinor/vector commutator to only a spin-1/2 part:
\begin{align}
[\nabla_\alpha, \nabla_{\beta\dbeta}] = 2 \,\eps_{\alpha\beta} \bar\cW_\dbeta~, \qquad
[\bnabla_\dalpha, \nabla_{\beta\dbeta}] = 2 \,\eps_{\dalpha\dbeta} \cW_\beta~, 
\end{align}
where $\cW_\alpha$ is a fermionic operator, that is, it has an expansion
$\cW_\alpha = \cW_\alpha{}^B \nabla_B + \cL_{\cW_\alpha{}^\um}
    + \cW_\alpha{}_\um{}^\un g_\un{}^\um
    + \cW_\alpha{}^x g_x$.
It must satisfy
\begin{align*}
\{\bar \nabla_\ad, \cW_\a\} =0~, \qquad
\{\nabla^\a, \cW_\a \} = \{\bar\nabla_\ad, \bar \cW^\ad\}~.
\end{align*}
The first relation implies that $\cW_\alpha$ is a chiral operator -- it takes
chiral superfields to chiral superfields. The second relation implies a
reality condition reducing by half the number of independent pieces in the
$\theta$ expansion of $\cW$. These two identities together guarantee that the
$[\nabla_{[A}, [\nabla_B, \nabla_{C]}]] = 0$ Bianchi identity holds.
The final external commutator is vector/vector and is determined by the Bianchi identities to be
\begin{align}
[\nabla_a, \nabla_b] &= 
    - \frac{i}{2} (\sigma_{a b})^{\alpha \beta} \{\nabla_\alpha, \cW_\beta\}
    + \frac{i}{2} (\bsigma_{a b})^{\dalpha \dbeta} \{\bnabla_\dalpha, \bar\cW_\dbeta\}~.
\end{align}
The upshot is that the external curvatures are completely determined
by $\cW_\alpha$.
Because the constraints \eqref{E:BasicConstraint} are the same as imposed in
$\cN=1$ super Yang-Mills, the solution looks formally identical to that case.

Now we turn to the mixed curvature. Identifying the mixed curvature operator $R_{\um A}$,
\begin{align}
[\nabla_\um, \nabla_A] = - R_{\um A}~,
\end{align}
we can abstractly solve the $[\nabla_A, [\nabla_B, \nabla_\um]] + \cdots = 0$
Bianchi identity. The lowest dimension identities involving spinor derivatives imply that
\begin{align}
R_{\um \,\alpha\dalpha} := (\sigma^a)_{\alpha \dalpha} R_{\um a} 
    = \frac{i}{2} \{\nabla_\alpha, R_{\um \dalpha} \}
    + \frac{i}{2} \{\bar\nabla_\dalpha, R_{\um\alpha} \}~, \qquad
\nabla_{(\alpha} R_{\um \beta)} = 0~.
\end{align}
Because of the constraints \eqref{E:BasicConstraint}, the second identity suggests to identify
$R_{\um \alpha}$ as the spinor derivative of some other operator. By redefining
$\nabla_\um$, one can always choose that operator to be imaginary, so that
\begin{align}\label{E:Rmalpha}
R_{\um\alpha} = i [\nabla_\alpha, X_\um]
\end{align}
for some real operator $X_\um$. This operator is thus responsible for generating all
of the mixed curvatures.

The existence of such a real operator lets us satisfy another of the entries
on our wish list in section \ref{S:WishList} -- the existence of
a modified internal derivative $\nabla_\um^+$ that preserves chirality:
\begin{align}
\nabla^\pm_\um := \nabla_\um \pm i X_\um~, \qquad
[\nabla_\alpha, \nabla_\um^-] = [\bar\nabla_\dalpha, \nabla_\um^+] = 0~.
\end{align}
Provided $X_\um$ preserves covariance (and we will ensure it does), $\nabla_\um^+$
provides a chirality-preserving internal de Rham differential.
The remainder of the $[\nabla_A, [\nabla_B, \nabla_\um]] + \cdots = 0$ Bianchi is then
solved provided
\begin{align*}
[\nabla_\um^+, \cW_\alpha] = -\frac{1}{4} [\bar \nabla_\dbeta, \{\bar\nabla^\dbeta, [\nabla_\alpha, X_\um]\}]
\end{align*}
This intertwines the external curvatures with the mixed curvatures, implying they
cannot be fixed separately.

For later use, we define $R^+_{\um \alpha}$ and
$R^-_{\um \dalpha}$ as the mixed spinor curvatures arising from $\nabla^+_\um$
and $\nabla^-_\um$, respectively. They turn out to be twice the original
curvatures $R_{\um \alpha}$ and $R_{\um \dalpha}$,
\begin{align}
[\nabla^+_\um, \nabla_\alpha] \equiv - R^+_{\um \alpha} = -2 R_{\um \alpha}~, \qquad \qquad
[\nabla^-_\um, \bar\nabla_\dalpha] \equiv - R^+_{\um \dalpha} = -2 R_{\um \dalpha}~.
\end{align}
It is helpful to give $R^+_{\um \alpha}$ a name distinct from $R_{\um \alpha}$ because
when we expand them out in terms of derivatives and generators, we will write
$R^+_{\um \alpha}$ in terms of $\nabla_\um^+$ while $R_{\um \alpha}$ will be written
in terms of $\nabla_\um$. For example,
\begin{align}
R^+_{\um A} = 
    T^+_{\um A}{}^C \nabla_C + \cL^+_{\cF^+_{\um A}} + \cR^+_{\um A}{}_\un{}^\up g_\up{}^\un
    + R^+_{\um A}{}^x g_x~,
\end{align}
where $\cL^+$ denotes the covariant Lie derivative built from $\nabla_\um^+$.

Now let us address the internal curvature. The existence of chiral 
internal derivative $\nabla_\um^+$ suggests we should examine their curvatures, defined as
\begin{align}
[\nabla^+_\um, \nabla^+_\un] = - R_{\um\un}^+
\end{align}
and similarly for $R_{\um\un}^-$.
The content of the $[\nabla_A, [\nabla_\um, \nabla_\un]]$ Bianchi identity is now succinctly
encoded in two conditions. The first is that $R_{\um\un}^+$ is a chiral operator,
$[\bar\nabla^{\dalpha}, R_{\um\un}^+] = 0$.
The second condition is that $R_{\um\un}^+$ is related to the real $R_{\um\un}$ via
\begin{align}
R_{\um \un}^+ = R_{\um \un} - 2 i \nabla_{[\um} X_{\un]} + [X_\um, X_{\un}]~.
\end{align}
The real part of this expression defines $R_{\um \un}$, while its imaginary part
links the mixed curvatures to the internal curvatures.

The final Bianchi identity, $[[\nabla_{[\um}, \nabla_\un], \nabla_{\up]}] = 0$,
can equivalently be formulated in terms of $\nabla_\um^+$. It leads immediately to
$[\nabla_{[\up}^+, R^+_{\um \un]}] = 0$.

In summation, we have uncovered three basic operators: a complex spinor $\cW_\alpha$,
a real 1-form $X_{\um}$, and a complex 2-form $R^+_{\um \un}$ that must obey
six abstract Bianchi identities:
\begin{align}
\tag{BI.1} \label{BI.1} \{\bar \nabla_\ad, \cW_\a\} &= 0\\
\tag{BI.2} \label{BI.2}  \{\nabla^\a, \cW_\a \}&= \{\bar\nabla_\ad, \bar \cW^\ad\} \\
\tag{BI.3} \label{BI.3} [\nabla_\um^+, \cW_\alpha] &= -\frac{1}{4} [\bar \nabla_\dbeta, \{\bar\nabla^\dbeta, [\nabla_\alpha, X_\um]\}] \\
\tag{BI.4} \label{BI.4} [\bar\nabla^{\dalpha}, R_{\um\un}^+] &= 0 \\
\tag{BI.5} \label{BI.5} \frac{i}{4} R_{\um\un}^+ - \frac{i}{4} R_{\um\un}^- &= [\nabla_{[\um}, X_{\un]}]   \\
\tag{BI.6} \label{BI.6} [\nabla_{[\up}^+, R^+_{\um \un]}] &= 0 
\end{align}
In terms of these, the external curvatures $R_{A B} = - [\nabla_A, \nabla_B]$ are given by
\begin{gather}
R_{\alpha \beta} = 0~, \qquad
R_{\dalpha \dbeta} = 0~, \qquad
R_{\alpha \dbeta} = 2 i \, \nabla_{\alpha \dbeta}~, \eol
R_{\alpha b} = - (\sigma_{b})_{\alpha \dalpha} \bar \cW^{\dalpha}~, \qquad
R^{\dalpha}{}_{b} = (\bsigma_{b})^{\dalpha \alpha} \cW_{\alpha}~, \eol
R_{a b} = 
    \frac{i}{2} (\sigma_{a b})^{\alpha \beta} \{\nabla_\alpha, \cW_\beta\}
    -\frac{i}{2} (\bsigma_{a b})^{\dalpha \dbeta} \{\bnabla_\dalpha, \bar\cW_\dbeta\}~,
\end{gather}
the internal curvatures $R_{\um \un} = -[\nabla_\um, \nabla_\un]$ are given by
\begin{align}
R_{\um \un} = \frac{1}{2} (R^+_{\um \un} + R^-_{\um \un}) - [X_\um, X_\un]~,
\end{align}
and the mixed curvatures $R_{\um A} = -[\nabla_\um, \nabla_A]$ are given by
\begin{gather}
R_{\um \alpha} = i [\nabla_\alpha, X_\um]~, \qquad
R_{\um \dalpha} = -i [\bar\nabla_\dalpha, X_\um]~, \eol
R_{\um a} = 
    -\frac{1}{4} (\bsigma_a)^{\dalpha \alpha} \{\nabla_\alpha,[\bar\nabla_\dalpha, X_\um]\}
    +\frac{1}{4} (\bsigma_a)^{\dalpha \alpha} \{\bar\nabla_\dalpha, [\nabla_\alpha, X_\um]\}
\end{gather}
For reference, it is also useful to give the mixed curvatures when written in terms
of $\nabla_\um^+$:
\begin{gather}
R^+_{\um \alpha} = 2 i [\nabla_\alpha, X_\um]~, \qquad
R^+_{\um \dalpha} = 0~, \eol
R^+_{\um a} = \frac{1}{2} (\bsigma_a)^{\dalpha \alpha} \{\bar\nabla_\dalpha, [\nabla_\alpha, X_\um]\}~.
\end{gather}

Our goal in subsequent sections will be to impose further constraints on the operators
appearing above and to identify the fundamental curvature superfields that comprise them.
Before doing that, we need to elaborate a bit more on the structure group $\cH$ we will be
using.

\subsection{The superconformal structure group}
The conformal superspace approach to $\cN=1$ conformal supergravity introduced in
\cite{Butter:2009cp} involves choosing the generators $g_x$ to be
the set of Lorentz transformations $(M_{ab})$, dilatations and $\gU(1)_R$ transformations
($\bbD$ and $\bbA$), $S$-supersymmetry transformations ($S_\alpha$ and $\bar S^\dalpha$),
and finally special conformal boosts ($K_a$). Together with the covariant derivatives
$\nabla_A = (\nabla_a, \nabla_\alpha, \bnabla^\dalpha)$, they furnish a representation
of the $\cN=1$ superconformal algebra with (anti)commutators
\begingroup
\allowdisplaybreaks
\begin{alignat}{2}
[M_{ab} , \nabla_c ] &= 2 \,\eta_{c [a} \nabla_{b]} \ , &\quad
[M_{ab} , K_c] &= 2 \,\eta_{c[a} K_{b]} \ , \eol{}
[\mathbb D, \nabla_a] &= \nabla_a \ , &\quad  
[\mathbb D, K_a] &= - K_a \ , \eol{}
[K_a , \nabla_b] &= 2 \,\eta_{ab} \mathbb D + 2 M_{ab} ~, &\quad
[M_{ab} , M_{cd}] &= 2 \,\eta_{c[a} M_{b] d} - 2 \,\eta_{d [a} M_{b] c} \ , \eol[2ex]
[M_{ab}, \nabla_\gamma] &= -{(\sigma_{ab})_\gamma}^{\beta} \nabla_\beta~, &\quad
[M_{ab}, S_\gamma] &= -{(\sigma_{ab})_\gamma}^{\beta} S_\beta~, \eol{}
[M_{ab}, \bnabla^\dgamma] &= -{(\sigma_{ab})^\dgamma}_{\dbeta} \bnabla^\dbeta~, &\quad
[M_{ab}, \bar S^\dgamma] &= -{(\sigma_{ab})^\dgamma}_{\dbeta} \bar S^\dbeta~, \eol{}
[\mathbb D, \nabla_\alpha] &= \tfrac{1}{2} \nabla_\alpha, &\quad
[\mathbb D, S_\alpha] &= -\tfrac{1}{2} S_\alpha, \eol{}
[\mathbb D, \bar \nabla^\dalpha] &= \tfrac{1}{2} \bar \nabla^\dalpha~, &\quad
[\mathbb D, \bar S^\dalpha] &= -\tfrac{1}{2} \bar S^\dalpha~, \eol{}
[\mathbb A, \nabla_\alpha] &= -i \nabla_\alpha,&\quad 
[\mathbb A, S_\alpha] &= +i S_\alpha, \eol{}
[\mathbb A, \bar \nabla^\dalpha] &= +i \bar \nabla^\dalpha~, &\quad
[\mathbb A, \bar S^\dalpha] &= -i \bar S^\dalpha~, \eol[2ex]
[K_a, \nabla_\alpha] &= i (\sigma_{a})_{\alpha \dbeta} \bar S^{\dbeta}, &\quad
[K_a, \bar \nabla^\dalpha] &= i (\bsigma_a)^{\dalpha \beta} S_{\beta}~, \eol{}
[S_\alpha, \nabla_a] &= i (\sigma_{a})_{\alpha \dbeta} \bar \nabla^{\dbeta}, &\quad
[\bar S^\dalpha, \nabla_a] &= i (\bsigma_a)^{\dalpha \beta} \nabla_{\beta}~,\eol{}
\{S_\alpha, \nabla_\beta\} &= \epsilon_{\alpha \beta} ( 2 \,\bbD - 3 i \bbA) - 4 M_{\alpha \beta}~, &\quad
\{\bar S^\dalpha, \bar \nabla^\dbeta\} &= \epsilon^{\dalpha \dbeta} ( 2 \,\bbD + 3i \bbA) - 4 M^{\dalpha \dbeta}~, \eol{}
\{S_\alpha, \bar S_\dalpha \} &= 2i (\sigma^a)_{\alpha \dalpha} K_a~.
\end{alignat}%
\endgroup
The operators $g_x = \{M_{ab}, \bbD, \bbA, S_\alpha, \bar S^\dalpha, K_a\}$ are taken to
commute with $\nabla_\um$. Here we use $M_{\alpha \beta} = -\frac{1}{2} (\sigma^{ab})_{\alpha\beta} M_{ab}$ for the anti-self-dual part of $M_{ab}$ and similarly for $M^{\dalpha \dbeta}$.

If the $\nabla_A$ obeyed the flat $\cN=1$ superspace algebra, the algebra of the
operators $g_x$ and $\nabla_A$ would just be the $\cN=1$ superconformal algebra.
Because the $\nabla_A$ curvatures instead involve the curvature operator $\cW_\alpha$,
the flat superconformal algebra becomes deformed. This is the sense in which the
$\cN=1$ superconformal algebra has been gauged.

Consistency of the above relations with the algebra of covariant derivatives
implies that the basic curvature operators $\cW_\alpha$, $X_\um$, and $R^+_{\um \un}$ 
are conformal primaries. That is, their
(anti)commutators with $S_\alpha$, $\bar S^\dalpha$, and $K_a$ all vanish. These imply
a number of conditions on the various pieces of these operators, which were useful in
our analysis as checks, but we will not comment on them explicitly.
$\cW_\alpha$ additionally carries dilatation and $\gU(1)_R$ weights $3/2$ and $+1$, whereas
the other operators are inert.

\section{The linearized solution to the Bianchi identities}

In this section, we are going to sketch a solution to the Bianchi identities
\eqref{BI.1} -- \eqref{BI.6} at the linearized level, where it is possible to
be very explicit about how the prepotentials appear. This will allow us
also to make more transparent contact with the 5D \cite{Sakamura:2012bj} and
6D cases \cite{Abe:2017pvw}, which worked to linear order in the gravitino
superfield $\Psi_\um{}_\alpha$.

We treat the supergeometry as linearized around a nearly flat background, whose
only non-vanishing curvature is the Kaluza-Klein curvature. The background
covariant derivatives are $\bm\nabla_M = \cD_M = \pa_M - \cL_{\small A_M}$ and
$\bm\nabla_\um = \pa_\um$ with curvature operators
\begin{align}
\bm \cW_\alpha &= \cW_\alpha{}^\um \pa_\um + \pa_\um \cW_\alpha{}^\un g_\un{}^\um~, \qquad
\bm X_\um = 0~, \qquad
\bm R^+_{\um \un} = 0~.
\end{align}
The linearized fluctuations around this background are denoted
\begin{align}
\nabla_M = \bm\nabla_M + \delta \nabla_M~, \qquad
\nabla_\um = \pa_\um + \delta \nabla_\um
\end{align}
with linearized curvatures
\begin{align}
\cW_\alpha = \bm\cW_\alpha + \delta \cW_\alpha~, \qquad
X_\um = \delta X_\um~, \qquad
R^+_{\um\un} = \delta R^+_{\um\un}~.
\end{align}

The basic constraints \eqref{E:BasicConstraint} are solved (up to a gauge transformation)
by choosing
\begin{align}
\delta \nabla_\alpha = -i [\nabla_\alpha, \cV]~, \qquad
\delta \bar\nabla_\dalpha = +i [\bar\nabla_\dalpha, \cV]~,
\end{align}
with the linearized external curvature $\delta \cW_\alpha$ being given by
\begin{align}\label{E:VaryW}
\delta \cW_\alpha = -\frac{1}{4} [\bar\cD_\dbeta ,\{\bar \cD^\dbeta, [\cD_\alpha,\cV]\}]
    + i [\bm\cW_\alpha, \cV]~.
\end{align}
If we were discussing an abelian gauge theory, $\cV$ would be the vector
multiplet prepotential and $\delta\cW_\alpha$ would be its linearized field strength.
Here both become operators, whose form we will discuss shortly.
Preserving the chirality constraint $[\nabla_\um^+, \bar\nabla_\dalpha]=0$ then
tells us that
\begin{align}
\delta \nabla_\um^+ = i [\pa_\um, \cV] - \Lambda_\um~,
\end{align}
where $\Lambda_\um$ is a chiral operator, $[\bar \cD_\dalpha, \Lambda_\um]=0$.
It follows that $\delta \nabla_\um = -\frac{1}{2} (\Lambda_\um + \bar \Lambda_\um)$ and
\begin{align}
X_\um &= [\pa_\um, \cV] - \frac{1}{2i} (\Lambda_\um - \bar \Lambda_\um)~, \qquad
R^+_{\um \un} = 2 \,[\pa_{[\um}, \Lambda_{\un]}]~,
\end{align}

Specifying the linearized geometry amounts to specifying the operators $\cV$
and $\Lambda_\um$. There is some redundancy to this choice, as they can be taken
to transform under pregauge transformations
\begin{align}\label{E:LambdaPreGauge}
\delta \cV = \frac{1}{2i} (\Lambda - \bar \Lambda)~, \qquad
\delta \Lambda_\um = [\pa_\um, \Lambda]~,
\end{align}
where $\Lambda$ is a chiral operator.

\subsection{Structure of the prepotentials}
The operator $\cV$ is real but as yet unconstrained, with an expansion
\begin{align}
\cV &= H^A \cD_A 
    + \cV^\um \pa_\um
    + ( \pa_\um \cV^\un + \cV_\um{}^\un)  g_\un{}^\um
    + \frac{1}{2} \cV(M)^{ab} M_{ab}
    \eol & \quad
    + \cV(D) \,\mathbb D
    + \cV(A) \,\mathbb A
%     \eol & \quad
    + \cV(S)^\alpha S_\alpha + \cV(S)_\dalpha \bar S^\dalpha
    + \cV(K)^a K_a~.
\end{align}
We have denoted $\cV^A$ by $H^A$, which is common in superspace literature.
The superfield $H^a$ is the $\cN=1$ gravitational prepotential. The superfield 
$\cV^\um$ describes
fluctuations of the Kaluza-Klein prepotential about the background. All the
other prepotentials must be constrained in some way or turn out to be gauge
artifacts as a consequence of the pregauge freedom $\Lambda$. The proper
way to uncover the constraints is to take certain curvature tensors to vanish
and to derive conditions on the prepotentials from these.
We assume that $\cV$ is a primary operator, but, aside from 
$H^a$, $\cV^\um$, and $\cV_\um{}^\un$, the individual prepotentials in its
expansion are not primary.

The chiral operator $\Lambda_\um$ has a similar expansion
\begin{align}\label{E:LinearizedLambda}
\Lambda_\um &= 
    \Lambda_\um{}^A \cD_A
    + \Lambda_{\um,}{}^\un \pa_\un
    + (\pa_\un \Lambda_{\um,}{}^\up + \Lambda_{\um,\un}{}^\up) g_\up{}^\un
    + \frac{1}{2} \Lambda_\um(M)^{ab} M_{ab}
    \eol & \quad
    + \Lambda_\um(D) \,\mathbb D
    + \Lambda_\um(A) \,\mathbb A
    + \Lambda_\um(S)^\alpha S_\alpha
    + \Lambda_\um(S)_\dalpha \bar S^\dalpha
    + \Lambda_\um(K)^b K_b~.
\end{align}
The chirality constraint implies that the components of $\Lambda_\um{}^A$ are given by
\begin{gather}
\Lambda_\um{}^{\dalpha \alpha} = - \bar\cD^\dalpha \Psi_\um{}^\alpha~, \quad
\Lambda_\um{}^\alpha = \frac{i}{8} \bar\cD^2 \Psi_\um{}^\alpha~, \quad
\Lambda_\um{}_\dalpha  \phantom{=} \text{unconstrained}
\end{gather}
where $\Psi_\um{}^\alpha$ will play the role of the gravitino superfield.
The internal diffeomorphism and $\GL(n)$ parameters are given as
\begin{align}
\Lambda_{\um,}{}^\un = \varphi_{\um,}{}^\un - \cW^\beta{}^\un \Psi_\um{}_\beta ~, \qquad
\Lambda_{\um, \un}{}^\up = 
    \varphi_{\um, \un}{}^\up
    + \pa_\un \Psi_\um{}^\beta \cW_\beta{}^\up
\end{align}
where $\varphi_{\um,}{}^\un$ and $\varphi_{\um, \un}{}^\up$ are chiral superfields
and $\cW_\alpha{}^\um$ is the background Kaluza-Klein field strength.
The other parameters are found to be
\begin{gather}
\Lambda_\um(D) = \varphi_\um
    - \frac{1}{2} \bar\cD^\dgamma \Lambda_{\um \dgamma}~, \qquad
\Lambda_\um(A) = \frac{i}{2} \varphi_\um
    - \frac{3i}{4} \bar\cD^\dgamma \Lambda_\um{}_\dgamma~, \eol
\Lambda_\um(M)_{\dalpha \dbeta} = \bar\cD_{(\dalpha}  \Lambda_{\um \dbeta)}~,  \qquad
\Lambda_\um(M)_{\alpha\beta} = \varphi_\um{}_{\alpha\beta} ~, \eol
\Lambda_\um(S)_\alpha = \sigma_\um{}_\alpha ~, \qquad
\Lambda_\um(S)^{\dalpha} = \frac{1}{8} \bar\cD^2  \Lambda_\um{}^\dalpha~, \eol
\Lambda_\um(K)^{\dalpha \alpha} = 
    - i \bar\cD^\dalpha \sigma_\um{}^\alpha ~,
\end{gather}
where $\varphi_\um$ and $\varphi_{\um \alpha \beta}$ are chiral and
$\sigma_\um{}_\alpha$ is complex linear.
If $\Lambda_\um$ is required to be primary, then the extra superfields 
$\Lambda_\um{}_\dalpha$, $\varphi_\um$, $\varphi_\um{}_{\alpha\beta}$
and $\sigma_\um{}_\alpha$ are not primary and should be written in terms
of other superfields that are.

The operator $\Lambda$ describing pregauge transformations is identical to
$\Lambda_\um$, but with the $\um$ index deleted. We relabel
some of its components as
\begin{gather}
\Psi_\um{}^\alpha \rightarrow 2i\, L^\alpha~, \qquad
\varphi_{\um,}{}^\un \rightarrow \ell^\un~, \qquad
\varphi_{\um, \un}{}^\up \rightarrow \ell_\un{}^\up~, \eol
\varphi_\um \rightarrow \ell~, \qquad
\varphi_\um{}_{\alpha\beta} \rightarrow \ell_{\alpha\beta}~, \qquad
\sigma_\um{}_\alpha \rightarrow \sigma_\alpha~.
\end{gather}
We emphasize that the $\ell$'s above are chiral while $\sigma_\alpha$ is complex linear.
A few prepotentials can already be eliminated by a gauge choice using
the pregauge $\Lambda$ transformations.
$\Lambda_\dalpha$ and $\bar \Lambda^\alpha$ are unconstrained superfields
and can be used to fix $H_\dalpha$ and $H^\alpha$. In order to
keep $\cV$ as a primary operator, one actually should choose
\begin{align}
H^\alpha \stackrel{*}{=} -\frac{i}{8} \bar \cD_\dalpha H^{\dalpha \alpha}~, \qquad
H_\dalpha \stackrel{*}{=} -\frac{i}{8} \cD^\alpha H_{\alpha \dalpha}~.
\end{align}
We denote this equality with a $*$ to emphasize that this is a choice.
Similarly, the chiral superfield $\ell_\um{}^\un$ can be used to eliminate
$\varphi_{\um,}{}^\un$,
\begin{align}
\varphi_{\um,}{}^\un \stackrel{*}{=} 0~.
\end{align}
The other extra parameters in $\cV$ and $\Lambda_\um$
must be eliminated by imposing curvature constraints so that only
$H_{\alpha \dalpha}$, $\cV^\um$, and $\Psi_{\um \alpha}$ (and possibly
some chiral superfield $\Phi_{\um \un \alpha}$) remain.

\subsection{Choosing curvature constraints on $\cW_\alpha$ and $X_{\protect\um}$}
From the definition of $X_\um$, one can show that
\begin{subequations}
\begin{align}
X_\um{}^{\dalpha \alpha} &= \pa_\um H^{\dalpha \alpha} 
    - \frac{i}{2} (\bar\cD^\dalpha \Psi_\um{}^\alpha + \cD^\alpha \bar\Psi_\um{}^\dalpha)~, \\
X_\um{}^\alpha &= \pa_\um H^\alpha - \frac{1}{16} \bar \cD^2 \Psi_\um{}^\alpha 
    - \frac{i}{2} \bar \Lambda_\um{}^\alpha~, \\
X_\um{}_\dalpha &= \pa_\um H_\dalpha - \frac{1}{16} \cD^2 \bar\Psi_\um{}_\dalpha 
    + \frac{i}{2} \Lambda_\um{}_\dalpha~.
\end{align}
\end{subequations}
The presence of the unconstrained $\bar\Lambda_\um{}^\alpha$
and $\Lambda_\um{}_\dalpha$ mean $X_\um{}^\alpha$ and $X_\um{}_\dalpha$ can
be set however we wish, in analogy to $H^\alpha$ and $H_\dalpha$.
$X_\um{}^{\dalpha \alpha}$ matches the linearized curvature \eqref{E:LinSugraCurvs.b}.
The equations for $X_\um{}^\alpha$ and $X_\um{}_\dalpha$ can be interpreted as 
\emph{definitions} of
$\bar\Lambda_\um{}^\alpha$ and $\Lambda_\um{}_\dalpha$ in terms of these arbitrary
curvatures. If we want $X_\um$ to be a primary operator, the natural choice is
\begin{align}
X_\um{}^\alpha \stackrel{*}{=} -\frac{i}{8} \bar \cD_\dalpha X_\um{}^{\dalpha \alpha}~, \qquad
X_\um{}_\dalpha \stackrel{*}{=} -\frac{i}{8} \cD^\alpha X_\um{}_{\alpha \dalpha}~.
\end{align}
It follows that
\begin{align}
\Lambda_{\um\dalpha} \stackrel{*}{=} 
    - \frac{i}{8} \cD_\beta \bar\cD_\dalpha \Psi_\um{}^\beta~, \qquad
\bar \Lambda_{\um}{}^{\alpha} \stackrel{*}{=}
    \frac{i}{8} \bar\cD^\dbeta \cD^\alpha \bar\Psi_\um{}_\dbeta~.
\end{align}

Next, let's impose a constraint on $\delta\cW_\alpha$. The simplest constraint we
can impose is that $\cW_\alpha{}^b = \delta \cW_\alpha{}^b = 0$. Using \eqref{E:VaryW} and
being careful to account for the variation of the covariant derivatives in the
operator $\delta \cW_\alpha$, one finds
\begin{align}
\cV(M)_\alpha{}^\beta + \frac{1}{2} \delta_\alpha{}^\beta (\cV(D) - 2 i \cV(A))
    &= \cD_\alpha H^\beta
    - \delta_\alpha{}^\beta \bar \cD^\dalpha H_\dalpha
    - \frac{i}{4} \bar \cD_\dgamma \cD_\alpha H^{\dgamma \beta}
    \eol & \quad
    - \frac{i}{2} \cW_\alpha{}^\um \Psi_\um{}^\beta
    + \text{chiral superfield}~.
\end{align}
The entire expression appears under $\bar \cD^\dbeta$ and so there is an undetermined
chiral superfield on the right-hand side. Assuming $\cV$ is primary, this
chiral superfield is also primary. In fact, it can be eliminated using the 
chiral superfields $\ell$ and $\ell_{\alpha\beta}$ in the $\Lambda$
pregauge freedom:
\begin{align}
\cV(M)_\alpha{}^\beta + \frac{1}{2} \delta_\alpha{}^\beta (\cV(D) - 2 i \cV(A))
    &\stackrel{*}{=} \cD_\alpha H^\beta
    - \delta_\alpha{}^\beta \bar \cD^\dalpha H_\dalpha
    - \frac{i}{4} \bar \cD_\dgamma \cD_\alpha H^{\dgamma \beta}
    - \frac{i}{2} \cW_\alpha{}^\um \Psi_\um{}^\beta~.
\end{align}
From this and its complex conjugate, one can determine $\cV(M)_{ab}$, $\cV(D)$,
and $\cV(A)$.

Taking the same combination of $X_\um$'s, one can show that
\begin{align}\label{E:XDAMeq1}
X_\um(M)_\alpha{}^\beta + \frac{1}{2} \delta_\alpha{}^\beta (X_\um(D) - 2 i X_\um(A))
    &= \cD_\alpha X_\um{}^\beta
    - \delta_\alpha{}^\beta \bar \cD^\dalpha X_\um{}_\dalpha
    - \frac{i}{4} \bar \cD_\dgamma \cD_\alpha X_\um{}^{\dgamma \gamma}
    \eol & \quad
    + \frac{i}{2} \cW_\alpha{}^\un \Psi_{\un\um}{}^\beta
    - \frac{i}{2} \cW_\alpha{}^\un \Phi_{\un\um}{}^\beta
    \eol & \quad
    - \frac{1}{16} \bar \cD^2 \cD_\alpha \Psi_\um{}^\beta
    + \frac{i}{2} \varphi_\um \delta_\alpha{}^\beta + \frac{i}{2} \varphi_\um{}_\alpha{}^\beta~.
\end{align}
In computing the above, we have introduced a new field $\Phi_{\um\un}{}^\alpha$ and
required
\begin{align}
\Psi_{\un \um}{}^\beta = 2 \,\pa_{[\un} \Psi_{\um]}{}^\beta + \Phi_{\un \um}{}^\beta
\end{align}
to be invariant under $\Xi$ transformations. At this point, the introduction
of $\Phi_{\um\un}{}^\alpha$ was \emph{ad hoc} in order to ensure a $\Xi$ invariance
we are imposing by hand.
The last four terms of \eqref{E:XDAMeq1} are chiral potentials, while the
rest are curvatures. Making the choice
\begin{align}
    \frac{i}{2} \varphi_\um \delta_\alpha{}^\beta + \frac{i}{2} \varphi_\um{}_\alpha{}^\beta
    \stackrel{*}{=} \frac{1}{16} \bar \cD^2 \cD_\alpha \Psi_\um{}^\beta
    + \frac{i}{2} \cW_\alpha{}^\un \Phi_{\un\um}{}^\beta~,
\end{align}
which also determines the chiral prepotentials $\varphi_\um$ and $\varphi_\um{}_{\alpha\beta}$ 
separately, one finds
\begin{align}\label{E:XDAM.Lin}
X_\um(M)_\alpha{}^\beta + \frac{1}{2} \delta_\alpha{}^\beta (X_\um(D) - 2 i X_\um(A))
    &\stackrel{*}{=} \cD_\alpha X_\um{}^\beta
    - \delta_\alpha{}^\beta \bar \cD^\dalpha X_\um{}_\dalpha
    \eol & \quad
    - \frac{i}{4} \bar \cD_\dgamma \cD_\alpha X_\um{}^{\dgamma \beta}
    + \frac{i}{2} \cW_\alpha{}^\un \Psi_{\un\um}{}^\beta~.
\end{align}
From this expression, one can determine $X_\um(M)_{ab}$, $X_\um(D)$, and $X_\um(A)$.

Now let's compute another curvature in $\cW_\alpha$. It turns out that 
$\cW_\alpha{}^\beta$ vanishes as a consequence of
$\cW_\alpha{}^b$ vanishing. The next curvature is $\cW_{\alpha \dbeta}$.
Without going into great detail, one can show that
\begin{align}
\cW_{\alpha \dalpha} 
    = -2i \cW_\alpha{}^\um X_{\um \dalpha}
    - \frac{1}{2} \cW^{\beta \um} \cD_\beta X_{\um \alpha \dalpha}
    - \frac{1}{4} \cD^\beta\cW_\beta{}^\um X_{\um \alpha \dalpha}
    + i Y_{a} (\sigma^a)_{\alpha \dalpha}
\end{align}
where $Y_a$ is a real quantity given by
\begin{align}
Y_{\alpha \dalpha} &= 
    2 \cV(K)_{\alpha \dalpha} + 2 i \cD_\alpha \cV(S)_\dalpha + 2 i \bar\cD_\dalpha \cV(S)_\alpha
    - \frac{i}{4} \cD_\alpha \bar \cD^2 H_\dalpha
    - \frac{i}{4} \bar \cD_\dalpha \cD^2 H_\alpha
    \eol & \quad
    - \frac{1}{32} (\cD^\beta \bar \cD^2 \bar \cD_\dbeta + \bar \cD_\dbeta \cD^2 \bar\cD^\dbeta) H_{\alpha \dalpha}
    + \frac{1}{8} \cD^\beta \cW_\beta{}^\um (\cD_\alpha \bar \Psi_{\um \dalpha} - \bar \cD_\dalpha \Psi_{\um \alpha})
    \eol & \quad
    - \frac{i}{4} \cW^{\beta \um} \cD_\beta (\pa_\um H_{\alpha \dalpha} - i \bar \cD_\dalpha \Psi_{\um \alpha})
    + \frac{i}{4} \bar \cW_\dbeta{}^\um \bar\cD^\dbeta (\pa_\um H_{\alpha \dalpha} - i \cD_\alpha \bar\Psi_{\um \dalpha})~.
\end{align}
The choice of $\cV(K)_a$ amounts to a choice of $Y_a$. Two natural choices are
to fix $Y_a \stackrel{*}{=} 0$ or to choose $Y_a$ so that
$\cW_{\alpha \dalpha} \stackrel{*}{=} -\bar\cW_{\dalpha \alpha}$, but the specific
choice does not affect the following analysis.

Identifying $W_{\alpha\beta\gamma}$ as 
(proportional to) the totally symmetric part of $\cW_\alpha(M){}_{\beta \gamma}$, we find
\begin{align}
\cW_{(\alpha}(M)_{\beta \gamma)} = 2 i\, W_{\alpha \beta\gamma}
    = \cW_\alpha{}^\un \varphi_\un{}_{\beta \alpha}
    - \frac{1}{4} \bar \cD^2 \cD_{(\alpha} \cV(M)_{\beta \gamma)}~.
\end{align}
From the above expressions, one can show that
\begin{align}\label{E:Linearized.Wabc}
W_{\alpha\beta\gamma}
    \stackrel{*}{=}
    \frac{1}{16} \bar \cD^2 \Big[
    i \cD_\alpha{}^\dgamma \cD_\beta H_{\gamma \dgamma}
    + \cD_\alpha (\cW_\beta \lrcorner \Psi_1{}_\gamma)
    - \cW_\alpha \lrcorner \cD_\beta \Psi_1{}_\gamma
    \Big]_{(\alpha\beta\gamma)}
    - \frac{i}{2} \cW_{(\alpha} \lrcorner \cW_\beta \lrcorner \Phi_2{}_{\gamma)}~.
\end{align}
This is exactly the expression for $W_{\alpha\beta\gamma}$ that we have been seeking.
The remaining trace part is also a chiral superfield $Z_\alpha$. Writing
\begin{align}
\cW_{\alpha}(M)_{\beta\gamma}
	&= - \eps_{\alpha (\beta} Z_{\gamma)}
	+ 2 i \,W_{\alpha\beta\gamma}
\end{align}
we find that
\begin{align}\label{E:DefZalpha}
Z_\alpha &= 
    - \frac{1}{4} \bar \cD^2 \Big[
    4 \,\cV(S)_\alpha + \cD_\alpha \cV(D) + \frac{2i}{3} \cD_\alpha \cV(A)
    + \frac{i}{6} \cW_\alpha \lrcorner \cD_\beta \Psi_1{}^\beta
    - \frac{2}{3} \bar\cW^\dalpha \lrcorner X_{1 \alpha \dalpha}
    \Big]
    \eol & \quad
    + \frac{1}{3} \cW_\alpha \lrcorner \cW_{\beta} \lrcorner  \Phi_{2}{}^\beta~.
\end{align}
The remaining undetermined prepotential $\cV(S)_\alpha$ lets one choose 
$Z_\alpha$ however one wishes, at least in principle. A natural choice is
\begin{align}
Z_\alpha \stackrel{*}{=} 0 \quad \implies \quad \cV(S)_\alpha \,\, \text{determined}~.
\end{align}
This determines $\cV(S)_\alpha$ up to a complex linear superfield, which
corresponds to the pregauge freedom $\ell(S)_\alpha$ within the
chiral $\Lambda$ operator. A curious feature of this choice 
is that it seems to require a non-covariant expression for $\cV(S)_\alpha$, as one
must introduce a prepotential for the background $\cW_\alpha$ or for the
field $\Phi_2{}_\alpha$ in order to extract a $\bar \cD^2$ from the last term
in \eqref{E:DefZalpha}.

We have nearly exhausted all of the freedom to choose the components of the
operator $\cV$. The last element is $\cV_{\um}{}^\un$. This can be fixed by
observing that
\begin{align}
X_{\um}{}^\un
    = -\cV_\um{}^\un + \pa_\um \cV^\un + \frac{i}{2} (\Lambda_{\um,}{}^\un - \bar\Lambda_{\um,}{}^\un)~.
\end{align}
We will then make the simplifying choice
\begin{align}
X_\um{}^\un \stackrel{*}{=} 0 \quad \implies \quad \cV_\um{}^\un \stackrel{*}{=} \pa_\um \cV^\un + \frac{i}{2} (\Lambda_{\um,}{}^\un - \bar\Lambda_{\um,}{}^\un)~.
\end{align}

Now that all components of the operator $\cV$ have been fixed,
all components of $\cW_\alpha$ must now be determined, up to terms coming from
undetermined pieces in the chiral $\Lambda_\um$ operator. Indeed, we find for the
other dimension-1 components of $\cW_\alpha$ that
\begin{align}
\cW_\alpha(M)_{\dalpha \dbeta} &= - \bar\cD_{(\dalpha} \cW_{\alpha \dbeta)}~,\eol
\cW_\alpha(D) &= \frac{1}{2} \bar\cD^\dgamma \cW_{\alpha \dgamma} 
    - \frac{1}{4} \bar \cD^2 (X_{\um \alpha \dalpha} \bar\cW^{\dalpha \um})
    + \frac{3}{2} Z_\alpha~, \eol
\cW_\alpha(A) &= \frac{3i}{4} \bar\cD^\dgamma \cW_{\alpha \dgamma} 
    - \frac{i}{8} \bar \cD^2 (X_{\um \alpha \dalpha} \bar\cW^{\dalpha \um})
    + \frac{3i}{4} Z_\alpha~, \eol
\cW_\alpha{}_\um{}^\un &= 
    \frac{1}{4} \bar\cD^2 (X_{\um \alpha \dalpha} \bar \cW^{\dalpha \un})
    + \cW_\alpha{}^\up \Big(2 \,\varphi_{[\up, \um]}{}^\un + \Phi_{[\ul{pm}]}{}^\beta \cW_\beta{}^\un\Big)
    \eol
    &\stackrel{*}{=}     \frac{1}{4} \bar\cD^2 (X_{\um \alpha \dalpha} \bar \cW^{\dalpha \un})~.
\end{align}
In the last equality, we have chosen $\varphi_{\um, \un}{}^\up$ to simplify the expression
and build a curvature. This does not determine the symmetric
part of $\varphi_{\um,\un}{}^\up$, but this will drop out of explicit expressions
because lower form indices generally end up antisymmetrized.

The only remaining piece of $\Lambda_\um$ that is undetermined is the complex
linear component $\sigma_\um{}_\alpha$. This contributes to $X_\um(S)_\alpha$,
\begin{align}
X_\um(S)_\alpha &= \pa_\um \cV(S)_\alpha 
    - \frac{1}{2i} (\Lambda_\um(S)_\alpha - \bar \Lambda_\um(S)_\alpha) \eol
    &= \pa_\um \cV(S)_\alpha 
    - \frac{1}{2i} (\sigma_\um{}_\alpha - \tfrac{1}{8} \cD^2 \Lambda_\um{}_\alpha)
\end{align}
This curvature then obeys the Bianchi identity
\begin{align}\label{E:ZalphaBianchi}
    &\frac{1}{4} \bar \cD^2 \Big[
    4 \,X_\um(S)_\alpha + \cD_\alpha X_\um(D) + \frac{2i}{3} \cD_\alpha X_\um(A)
    - \frac{i}{6} \cW_\alpha{}^\un \cD_\beta \Psi_{\un \um}{}^\beta
    - \frac{2}{3} \pa_\um (\bar\cW^{\dalpha \un} X_{\un \alpha \dalpha})
    \Big] \eol & \quad
    =
    - \pa_\um Z_\alpha
    + \frac{1}{3} \cW_\alpha{}^\un \cW_{\beta}{}^\up \Phi_{\ul{pnm}}{}^\beta~.
\end{align}

\subsection{Some lower dimension results for $R^+_{\protect \ul{mn}}$}
The remaining curvature operator we have not directly addressed is $R^+_{\ul{mn}}$,
which is built by taking the curl of the chiral operator $\Lambda_{\um}$. We find
for $T^+_{\um \un}{}^A$ and $F^+_{\um \un}{}^\up$ the results
\begin{subequations}
\begin{align}
T^+_{\um \un}{}^{\dalpha \alpha} &= - \bar\cD^\dalpha \Psi_{\um\un}{}^\alpha~, \\
T^+_{\um \un}{}^{\alpha} &= \frac{i}{8} \bar\cD^2 \Psi_{\um\un}{}^\alpha~, \\
T^+_{\um \un}{}_{\dalpha} &= 2 \,\pa_{[\um} \Lambda_{\un] \dalpha} 
    \stackrel{*}{=} -\frac{i}{8} \cD_\beta \bar\cD_\dalpha \Psi_{\um \un}{}^\beta~, \\
F^+_{\um \un}{}^\up &= 
    2 \Lambda_{[\um, \un]}{}^\up
    \stackrel*=  -\Psi_{\um \un}{}^\alpha \cW_\alpha{}^\up ~.
\end{align}
\end{subequations}
The other linearized curvatures can be computed directly in a similar way, but
their forms will not be terribly enlightening.

\subsection{Comparison to 5D and 6D results and summary}

We have now accounted for all of the prepotentials and field strengths
and uncovered the appropriate curvature constraints to remove all but one 
unconstrained spinor prepotential $\cV(S)_\alpha$ and a complex linear prepotential
$\sigma_\um{}_\alpha$.  These unfixed prepotentials can be eliminated by breaking 
manifest background covariance, but we will find it simpler to just leave them
unfixed in the remainder, keeping in mind that they appear mainly in two curvatures --
a chiral spinor superfield $Z_\alpha$ and the curvature superfield $X_\um(S){}_\alpha$,
which are related in terms of a complicated Bianchi identity \eqref{E:ZalphaBianchi}
involving the field strength $\Phi_{\ul{mnp}}{}_\alpha$. 

At this stage, we can make a few brief comments connecting with the existing
5D and 6D work involving linearized supergravity.
In the explicit linearized 5D construction of Sakamura \cite{Sakamura:2012bj},
the covariant derivative $\hat \pa_y$ can be identified with the linearized 
$\nabla_\um^+$ when acting on chiral superfields, with the rescaling of
$\Psi_\alpha^{\text{5D}} \rightarrow \frac{i}{2}  \Psi_{\um\alpha}$.\footnote{The
precise identification requires going to a (complex) chiral gauge where
$\bar\nabla_\dalpha = \bar D_\dalpha$ and $\nabla_\um^+ = \pa_\um - \Lambda_\um$.}
Similarly, the gravitational superfield $U^\mu$ is identified with  $-H_{a}$ here
(keeping in mind $\sigma^\mu \rightarrow - \sigma^a$) and $U^4$ there is identified with 
$-\cV^\um$ here. Similar comments pertain to the 6D results of 
Abe, Aoki, and Sakamura \cite{Abe:2017pvw}.

The major difference between our linearized results and previous results is that
those papers fully describe 5D and 6D supergravity, whereas we aim only to
describe the minimal extension of $\cN=1$ conformal supergravity necessary to
encode $y$-dependent superfields. We make no effort to identify the internal sector
of the metric, with the understanding that from an $\cN=1$ perspective that sector
must correspond to ``matter'', i.e. some appropriately defined covariant superfields.
Thus, in our formulation, there is no analogue to their gauge parameter
$N$; that parameter is the 5D or 6D analogue of the complex 11D parameter $\Omega^\um$,
and it encodes details of the higher dimensional sector beyond what is purely required
for a covariant $\cN=1$ supergeometry.

At this stage, we have developed enough intuition to address the full  non-linear geometry.
That will be our next task.

\section{Exploring the non-linear Bianchi identities}

Solving the Bianchi identities \eqref{BI.1} -- \eqref{BI.6} in terms of curvature superfields
$W_{\alpha\beta\gamma}$, $X_{\um}{}_{\alpha \dalpha}$, $\Psi_{\um \un}{}_\alpha$,
$\Phi_{\ul{mnp}}{}_\alpha$, and $\cW_\alpha{}^\um$ is a rather involved task, as most of
the Bianchi identities just serve as consistency checks on lower dimension ones. Typically
in superspace, one can invoke some version of Dragon's theorem \cite{Dragon:1978nf},
which states that the curvature superfields are completely determined by the torsion
superfields, so solving the torsion tensor Bianchi identities is the only necessary step.
In its original formulation, Dragon's theorem is limited to dimensions higher than three and for
a tangent space group consisting of the Lorentz and $R$-symmetry groups. For this reason,
it does not directly apply to either conformal superspace (where $S$ and $K$ curvatures
are present) or to its extension here with internal torsion and $\GL(n)$ curvatures.
Moreover, in boiling the Bianchi identities down to \eqref{BI.1} -- \eqref{BI.6}, we have
\emph{already} solved a number of them! It is possible that a modification of Dragon's theorem
is possible, but we found it more direct to analyze the identities \eqref{BI.1} -- \eqref{BI.6}
exhaustively, taking guidance from the linearized case.
In this section, we provide a summary of their solution, with some guideposts for the
enterprising reader to reproduce. The reader interested only in the result may consult
Appendix \ref{App:Curvs} where we summarize the supergeometry.

\subsection{The chiral Bianchi identities \eqref{BI.1} and \eqref{BI.4}}
The easiest Bianchi identities to solve are the ones imposing chirality on $\cW_\alpha$
and $R^+_{\um \un}$, eqs. \eqref{BI.1} and \eqref{BI.4}. In both cases, to
make the chirality analysis simpler, it is convenient to choose a chiral basis
of derivatives -- that is, we will choose to use $\nabla_\um^+$ instead of $\nabla_\um$.
In general, this means $\cW_\alpha$ must possess an expansion of the form
\begin{align}
\cW_{\alpha} &= 
        \cW_\alpha{}^b \nabla_b
        + \cW_\alpha{}^\beta \nabla_\beta
	+ \cW_{\alpha \dbeta} \bar\nabla^\dbeta
	+ \cW_\alpha{}^\um \nabla_\um^+
	+ \Big(\nabla_\un^+ \cW_\alpha{}^\um + \cW_\alpha{}_\un{}^\um\Big) g_{\um}{}^{\un}
	\eol & \quad
	+ \cW_\alpha(D) \mathbb D
	+ \cW_\alpha(A) \mathbb A
	+ \cW_\alpha(M){}^{\beta\gamma} M_{\beta\gamma}
	+ \cW_\alpha(M){}_{\dbeta\dgamma} M^{\dbeta\dgamma}
	\eol & \quad
	+ \cW_\alpha(S)^\beta S_\beta
	+ \cW_\alpha(S)_\dbeta \bar S^\dbeta
	+ \cW_\alpha(K)^b K_b~.
\end{align}
We have chosen to include an explicit $\nabla_\un^+ \cW_\alpha{}^\um$ term
in the $\GL(n)$ piece so that it combines with $\cW_\alpha{}^\um \nabla_\um^+$
to give the covariant internal Lie derivative $\cL^+$ built from $\nabla_\um^+$.

Now we impose the constraint 
\begin{align}
\cW_\alpha{}^b = 0~.
\end{align}
In the linearized theory, recall
this has the effect of fixing the underlying prepotentials $\cV(M)$, $\cV(D)$, and
$\cV(A)$. The Bianchi identity implies several simple conditions:
\begin{gather}
\cW_\alpha{}^\beta = 0~, \qquad
\bar \nabla^\dalpha \cW_\alpha{}^\um = 0~, \qquad 
\bar\nabla^\dalpha \cW_\alpha{}_\um{}^\un = 0~, \qquad
\bar\nabla^\dalpha \cW_\alpha(M)_{\beta\gamma} = 0~.
\end{gather}
No condition is imposed yet on $\cW_\alpha{}_\dalpha$, but higher curvatures are determined
in terms of it:
\begin{alignat}{2}
\cW_\alpha(D) &= \frac{1}{2} \bar\nabla^\dgamma \cW_{\alpha \dgamma} + \phi_\alpha~, &\qquad
\cW_\alpha(A) &= \frac{3i}{4} \bar\nabla^\dgamma \cW_{\alpha \dgamma} + \frac{i}{2} \phi_\alpha~, \eol
\cW_\alpha(M)_{\dalpha \dbeta} &= - \bar\nabla_{(\dalpha} \cW_{\alpha \dbeta)}~, &\qquad
\cW_\alpha(S)_\dalpha &= \frac{1}{8} \bar\nabla^2 \cW_{\alpha \dalpha}~,
\label{eq:WDAMS}
\end{alignat}
where $\phi_\alpha$ is an undetermined chiral superfield. 
The remaining chirality conditions amount to
\begin{gather}\label{eq:WSK}
\bar\nabla^2 \cW_\alpha(S)^\beta = 0~, \qquad
\cW_{\alpha}(K)^{\dbeta \beta} = i \bar\nabla^\dbeta \cW_\alpha(S)^{\beta}~.
\end{gather}

The superfield $\cW_\alpha{}^\um$ corresponds in the flat limit to the Kaluza-Klein
field strength, and we have recovered its chirality condition. As in the linearized
case, we expect the totally
symmetric part of $\cW_\alpha(M)_{\beta \gamma}$ to be the superfield $W_{\alpha\beta\gamma}$,
and this is what happens if we drop the internal derivatives to recover
$\cN=1$ conformal superspace. The other superfields will turn out to be
composite, or correspond to curvatures that can be turned off by redefining
certain connections.

The chirality condition on $R^+_{\um \un}$ is also simple to analyze.
Taking a similar decomposition
\begin{align}\label{eq:RmnExpansion}
R^+_{\um \un}
    &= 
    T^+_{\um \un}{}^B \nabla_B
    + F^+_{\um \un}{}^\up \nabla^+_\up    
    + \Big(\nabla^+_\up F^+_{\um \un}{}^\uq + \cR_{\um \un}{}_\up{}^\uq\Big)g_\uq{}^\up
    \eol & \quad
    + R^+_{\um \un}(D) \mathbb D
    + R^+_{\um \un}(A) \mathbb A
    + R^+_{\um \un}(M)^{\beta \gamma} M_{\beta\gamma}
    + R^+_{\um \un}(M)_{\dbeta \dgamma} M^{\dbeta\dgamma}
    \eol & \quad
    + R^+_{\um \un}(S)^\beta S_\beta
    + R^+_{\um \un}(S)_\dbeta S^\dbeta
    + R^+_{\um \un}(K)^c K_c
\end{align}
one immediately finds that the chirality condition implies
\begin{align}\label{eq:Rmnp.Psi}
T^+_{\um \un}{}^{\dbeta \beta} = - \bar\nabla^\dbeta \Psi_{\um\un}{}^\beta~,\qquad
T^+_{\um \un}{}^{\beta} = \frac{i}{8} \bar\nabla^2 \Psi_{\um\un}{}^\beta~,
\end{align}
for some 2-form spinor superfield $\Psi_{\um \un}{}_\alpha$.
The remaining components of the Bianchi identify impose no condition on
$T^+_{\um \un}{}_\dbeta$.  The other components are
\begin{subequations}
\begin{align}
F^+_{\um \un}{}^\up &= -\Psi_{\um \un}{}^\alpha \cW_\alpha{}^\up + \Phi_{\um \un}{}^\up~, \\
R^+_{\um \un}(M)_{\alpha \beta}
	&= - \Psi_{\um \un}{}^\gamma \cW_\gamma(M)_{\alpha\beta} + \Phi_{\um \un \alpha\beta}~,\\
R^+_{\um \un}(M)_{\dalpha \dbeta}
	&= \bar \nabla_{(\dalpha} \Psi_{\um \un}{}^\gamma W_{\gamma \dbeta)}
	+ \bar \nabla_{(\dalpha} T^+_{\um \un}{}_{\dbeta)}~, \\
R^+_{\um\un}(D) &=
	-\Psi_{\um\un}{}^\alpha \phi_\alpha
	+ \Phi_{\um \un}
	- \frac{1}{2} \bar\nabla^{\dgamma} T^+_{\um\un \dgamma}
	- \frac{1}{2} \bar\nabla^{\dgamma} \Psi_{\um \un}{}^\gamma W_{\gamma\dgamma}~,\\
R^+_{\um \un}(A) &=
	- \frac{i}{2} \Psi_{\um\un}{}^\alpha \phi_\alpha
	+ \frac{i}{2} \Phi_{\um \un}
	- \frac{3i}{4} \bar\nabla^{\dgamma} T^+_{\um\un \dgamma}
	- \frac{3i}{4} \bar\nabla^{\dgamma} \Psi_{\um \un}{}^\gamma W_{\gamma\dgamma}~,\\	
R^+_{\um \un}(S)_{\alpha}
	&= -\Psi_{\um \un}{}^\gamma \cW_\gamma(S)_\alpha
	+ \Sigma_{\um \un \alpha}~,\\
R^+_{\um \un}(S)_\dalpha &=
	\frac{1}{8} \bar\nabla^2 T^+_{\um \un \dalpha}
	+ \frac{1}{8} \bar\nabla^2 \Psi_{\um \un}{}^\gamma W_{\gamma\dgamma}
	- \frac{1}{4} \bar\nabla_\dphi \Psi_{\um \un}{}^\gamma \bar\nabla^\dphi W_{\gamma\dgamma}~,\\
R^+_{\um \un}(K)_{\alpha\dalpha}
	&= -i \bar\nabla_\dalpha R^+_{\um \un}(S)_{\alpha}
	- i \bar\nabla^\dalpha \Psi_{\um \un}{}^\beta \cW_\beta(S)_\alpha~, \\
\cR^+_{\ul{mnp}}{}^\uq
	&=
	- \Psi_{\um \un}{}^\alpha \cW_\alpha{}_\up{}^\uq
	+ \nabla_\up^+ \Psi_{\um \un}{}^\alpha \cW_\alpha{}^\uq
	+ \Phi_{\ul{mnp}}{}^\uq
\end{align}
\end{subequations}
In deriving these results, we used the explicit forms of some of the $\cW_\alpha$
superfields. But there remain certain undetermined pieces.
These are the chiral superfields
$\Phi_{\ul{mn}}{}^\up$, $\Phi_{\ul{mn}}{}_{\alpha \beta}$,
$\Phi_{\um \un}$, and $\Phi_{\ul{mnp}}{}^\uq$,
as well as the complex linear superfield $\Sigma_{\um \un}{}_\alpha$, which obeys
$\bar\nabla^2 \Sigma_{\um \un}{}_\alpha = 0$.
From the linearized analysis, we know that $\Phi_{\um \un}{}^\up$ can be eliminated
by redefining a connection, so we choose it to vanish,
$$\Phi_{\um \un}{}^\up = 0~.$$

\subsection{Interlude: The $X_{\protect\um}$ operator and variant covariant derivatives}

Let us pause to make a few comments that will be useful very soon.
We take the operator $X_\um$ that translates between the chiral internal derivative 
and the antichiral one to have an expansion as
\begin{align}
X_\um = X_\um{}^A \nabla_A + X_\um{}^x g_x + X_{\um \un}{}^\up g_\up{}^\un~.
\end{align}
That is, we explicitly turn off any $X_\um{}^\un \nabla_\un$ term. This is sensible
because $X_\um{}^\un$ has dimension zero and no such superfield seems possible to
construct given our constituents. It is also justified from the linearized analysis.
Recall that $X_\um{}^a$ coincides at the linearized level to \eqref{E:LinSugraCurvs.b}.
Other $X_\um$ fields are of higher dimension and will correspond either to composite
quantities or fields that can be removed by redefinitions.

We observe that the primary condition,
$[S_\alpha, X_\um] = 0$,
implies for the lowest three $X_\um$ fields that
\begin{align}
S_\beta X_\um{}^a &= 0~, \qquad
S_\beta X_\um{}^\alpha = 0~, \qquad
S_\beta X_\um{}_\dalpha = -i X_\um{}_{\beta \dalpha}~,
\end{align}
and similarly for their complex conjugates. So $X_\um{}^a$ is primary, as
expected for a fundamental curvature. The conditions on $X_{\um}{}^\alpha$
suggest that it be written as
\begin{align}
X_\um{}^\alpha = -\frac{i}{8} \bar \nabla_\dalpha X_\um{}^{\dalpha \alpha} 
    + \text{primary superfield}
\end{align}
and it is tempting to set the primary superfield above to zero, just as at
the linearized level.
However, it is going to be more useful to keep the non-primary superfield
$X_\um{}^\alpha$ unfixed and work with it directly.

Assuming the above structure for the $X$ operator, we can already compute some parts of the mixed
curvature $R_{\um \alpha}$. We are interested in the Kaluza-Klein curvature piece,
\begin{align}
F_{\um \alpha}{}^\un = i \, X_{\um \alpha \dalpha} \bar \cW^{\dalpha \un}~.
\end{align}
As anticipated, this is non-vanishing, which means the $\nabla_A$ we are using
do not coincide with the $\znabla$ introduced in section \ref{S:KK.Hconn}
with the simplest $\GL(n)$ connections. Rather, we find that
\begin{align}
\znabla_\alpha = \nabla_\alpha - F_{\um \alpha}{}^\un \,g_\un{}^\um~.
\end{align}
These derivatives do not satisfy the first constraint of \eqref{E:BasicConstraint},
whereas they do lead to vanishing mixed Kaluza-Klein curvatures,
$\mathring F_{\um A}{}^\un = 0$.
The advantage of using $\nabla_\alpha$ is that it anticommutes with itself and
the natural superfields we will be using are chiral or antichiral with respect to it.

Actually, we are going to discover that, at least when working with $\nabla_\um^+$,
there is yet another spinor derivative that makes an appearance. It is defined by
\begin{align}
\hnabla_\alpha := \nabla_\alpha - 2 F_{\um \alpha}{}^\un \, g_\un{}^\um~.
\end{align}
It is not hard to see that $[\hnabla_\alpha, \nabla_\um^+]$ has no Kaluza-Klein curvature.
For this reason, $\{\hnabla_\alpha, \bar \nabla_\dalpha, \nabla_\um^+\}$ turn out to be
a convenient set of derivatives to use when dealing with chiral objects as we have shoved
all of the $\GL(n)$ connection into $\hnabla_\alpha$. We will see this derivative begin
to make appearances very soon.
Similarly, $\{\nabla_\alpha, \hnabla_\dalpha, \nabla_\um^-\}$ turn out to be convenient
to use with anti-chiral objects, where
$\hnabla^\dalpha := \bar\nabla^\dalpha - 2 F_\um{}^{\dalpha \un} \, g_\un{}^\um$.

However, we emphasize that when we discuss the curvature tensors $R^+_{\um \un}$
and $R^+_{\um \alpha}$, they are always here to be understood to be built using $\nabla_\alpha$,
rather than $\hnabla_\alpha$, so as to avoid confusion.

\subsection{The $\cW_\alpha$ reality Bianchi identity $\eqref{BI.2}$}

We introduce the abstract operator
\begin{align}
\cY := -\frac{1}{4} \{\nabla^\alpha, \cW_\alpha\}
\end{align}
The content of the Bianchi identity \eqref{BI.2} is that this is a real operator.
Let's take its lowest engineering dimension components, $\cY^\um$ and $\cY^a$.
The first leads to
\begin{align}
\cY^\um = - \frac{1}{4} \hnabla^\alpha \cW_\alpha{}^\um = - \frac{1}{4} \hnabla_\dalpha \bar\cW^\dalpha{}^\um~,
\end{align}
which reduces in flat space to the Bianchi identity for the Kaluza-Klein field strength.
Note that it is $\hnabla_\alpha$ above, rather than $\nabla_\alpha$.
The other lowest engineering dimension component is
\begin{align}
i \cY_{\alpha \dalpha}
    &= \cW_{\alpha \dalpha}
    - \frac{i}{2} \cW^{\beta \um} T_{\um \beta\, \alpha \dalpha}
    + \frac{1}{4} \nabla \cW^\um X_{\um \alpha \dalpha} \eol
    &= \cW_{\alpha \dalpha}
    - \frac{i}{4} \cW^{\beta \um} T^+_{\um \beta\, \alpha \dalpha}
    + \frac{1}{4} \hnabla \cW^\um X_{\um \alpha \dalpha}
\end{align}
where the mixed torsion tensor is given in Appendix \ref{App:Curvs}. Because $\cY^a$ is real,
this constrains the real part of $\cW_{\alpha \dalpha}$ to be
\begin{align}
\cW_{\alpha \dalpha} - \bar \cW_{\dalpha \alpha}
    &= -2i \,(\cW_\alpha{}^\um X_{\um \dalpha} + \bar\cW_\dalpha{}^\um X_{\um \alpha})
    - \frac{1}{2} \cW^{\beta \um} \hat\nabla_\beta X_{\um \alpha \dalpha}
    - \frac{1}{2} \bar\nabla_\dbeta (\bar \cW^{\dbeta \um} X_{\um \alpha \dalpha})~.
\end{align}
From the linearized analysis, we know that $\cY^a$ can be fixed by a connection redefinition.
One convenient choice is
\begin{align}
\cY_a \stackrel{*}{=} 0 \qquad \implies \quad
\cW_{\alpha \dalpha}
    \stackrel{*}{=} 
    \frac{i}{4} \cW^{\beta \um} T^+_{\um \beta\, \alpha \dalpha}
    - \frac{1}{4} \hnabla \cW^\um X_{\um \alpha \dalpha}~.
\end{align}
Another choice is to take
\begin{align}
\cW_{\alpha \dalpha} \stackrel{*}{=}- \bar \cW_{\dalpha \alpha}
    &\stackrel{*}{=}
    -i \,(\cW_\alpha{}^\um X_{\um \dalpha} + \bar\cW_\dalpha{}^\um X_{\um \alpha})
    - \frac{1}{4} \cW^{\beta \um} \hat\nabla_\beta X_{\um \alpha \dalpha}
    - \frac{1}{4} \bar\nabla_\dbeta (\bar \cW^{\dbeta \um} X_{\um \alpha \dalpha})~.
\end{align}

The Bianchi identity involving $\cY^\alpha$ is a bit more intricate.
It allows one to determine the non-linear version of the combination \eqref{E:XDAM.Lin},
\begin{align}\label{E:XDAM}
X_\um(M)_\alpha{}^\beta + \frac{1}{2} \delta_\alpha{}^\beta (X_\um(D) - 2 i X_\um(A))
    &=
    \hat\nabla_\alpha X_\um{}^\beta
    - \delta_\alpha{}^\beta \bar\nabla^\dgamma X_{\um \dgamma}
    + \frac{i}{4} \bar\nabla^\dgamma \hat\nabla_\alpha X_\um{}^\beta{}_\dgamma
    \eol & \quad
    + X_{\um \alpha \dbeta} \bar \cW^{\dbeta \beta}
    + \frac{i}{2} \cW_\alpha{}^\un \Psi_{\un \um}{}^\beta
\end{align}
where $\hat\nabla_\alpha = \nabla_\alpha - 2 F_{\um \alpha}{}^\un g_\un{}^\um$.
In principle, there is an undetermined chiral superfield on the right-hand side, but it can
be set to zero by a connection redefinition as in the linearized analysis.
Separating $\cW_{\gamma}(M)_{\beta\alpha}$ into spin-1/2 and spin-3/2 pieces as in
the linearized analysis,
\begin{align}
\cW_{\alpha}(M)_{\beta\gamma}
	&= - \eps_{\alpha (\beta} Z_{\gamma)}
	+ 2 i \,W_{\alpha\beta\gamma}
\end{align}
the $\cY^\alpha$ Bianchi then relates $Z_\alpha$ to $\phi_\alpha$ in \eqref{eq:WDAMS} as
\begin{align}
\frac{2}{3} \phi_{\alpha} = Z_\alpha
	-\frac{1}{6} \bar\nabla^2 (\cW^{\dalpha \um} X_{\um \alpha \dalpha})~.
\end{align}

The remaining Bianchi identities in \eqref{BI.2} are more complicated.
The ones at dimension two allow us to determine $\cW_\alpha(S)^\beta$.
Employing the shorthand, 
\begin{align}
\ZZ^x 
    &:= \nabla^\alpha \cW_\alpha{}^x 
    + i \nabla^\alpha \cW_\alpha{}^\um X_\um{}^x
	+ 2 \cW^{\alpha \um} R_{\um \alpha}{}^x \eol
    &= \nabla^\alpha \cW_\alpha{}^x 
    + i \hat \nabla^\alpha \cW_\alpha{}^\um X_\um{}^x
	+ \cW^{\alpha \um} R^+_{\um \alpha}{}^x
\end{align}
the remaining Bianchi identities provide a definition for $\cW_\alpha(S)^\beta$ as
\begin{align}
\cW_\alpha(S){}^\beta
	&:= \frac{1}{8} \delta_\alpha{}^\beta (\ZZ(D) - \bar \ZZ(D))
	+ \frac{i}{12} \delta_\alpha{}^\beta (\ZZ(A) - \bar \ZZ(A))
	- \frac{1}{4} (\ZZ(M)_\alpha{}^\beta - \bar \ZZ(M)_\alpha{}^\beta)~.
\end{align}
In addition, one finds a consistency condition
\begin{align}
\bar\nabla_\dalpha \cW_\alpha(S)^\alpha &= \ZZ(S)_\dalpha - \bar \ZZ(S)_\dalpha
\end{align}
and a Bianchi identity
\begin{align}
\ZZ(K)^a &= \bar \ZZ(K)^a~.
\end{align}
The last corresponds to a complicated modification of the dimension-3 Bianchi identity
that relates derivatives of $W_{\gamma\beta\alpha}$ to its complex conjugate.
This is one of the fundamental Bianchi identities of the geometry, mentioned in
footnote \ref{F:WBianchi}, but it lies at such
high dimension one does not usually need its explicit form.
As we have not worked out
a useful compact way of writing it, we do not give it explicitly here.

These relations are compact, but not necessarily useful. For example, it is not
immediately clear that the expression for $\cW_\alpha(S)^\beta$
satisfies the complex linearity condition \eqref{eq:WSK}.
This can be made more apparent by expanding it out:
\begin{align}\label{E:WS.v2}
\cW_\alpha(S)^\beta &=
    - \frac{1}{4} \Big(
    \nabla^\gamma \cW_\gamma(M)_\alpha{}^\beta
    + 2 F_\um{}^{\gamma \um} \cW_\gamma(M)_\alpha{}^\beta
    + \cW^\gamma{}^\um R^+_{\um \gamma}(M)_\alpha{}^\beta
    \Big)
    \eol & \quad
    + \frac{1}{12} \delta_\alpha{}^\beta
    (\nabla^\gamma \phi_\gamma + 2 F_\um{}^\gamma{}^\um \phi_\gamma)
    + \frac{1}{8} \delta_\alpha{}^\beta \cW^\gamma{}^\um
    (R^+_{\um \gamma}(D) + \frac{2i}{3} R^+_{\um \gamma}(A))
    \eol & \quad
    + \bar\nabla_\dgamma \Big[
    \frac{1}{4} \bar \cW^\dgamma(M)_\alpha{}^\beta
    - \frac{i}{2} \bar\cW^\dgamma{}^\um X_\um(M)_\alpha{}^\beta
    \Big]
    \eol & \quad
    + \delta_\alpha{}^\beta \,\bar\nabla_\dgamma \Big[
    - \frac{1}{8} \bar \cW^\dgamma(D)
    + \frac{i}{4} \bar \cW^\dgamma{}^\um X_\um(D)
    - \frac{i}{12} \bar \cW^\dgamma(A)
    - \frac{1}{6} \bar \cW^\dgamma{}^\um X_\um(A)
    \Big]~.
\end{align}
The last two lines are manifestly complex linear. The first two lines are complex
linear by virtue of the Bianchi identities involving
$\nabla_\um^+ \cW_\gamma(M)_\alpha{}^\beta$ and $\nabla_\um^+ \phi_\alpha$, which we
will encounter below. The expression could be evaluated further but
we will postpone that for now.

\subsection{The $\nabla_{\protect \um}^+ \cW_\alpha$ Bianchi identity \eqref{BI.3}}
The Bianchi identity that directly links $\cW_\alpha$ to $X_\um$ is \eqref{BI.3}, which
can be rewritten as
\begin{align}\label{E:intW}
[\nabla_\um^+, \cW_\alpha] = \frac{i}{8} [\bar \nabla_\dalpha,\{\bar\nabla^\dalpha, R^+_{\um \alpha}\}]~.
\end{align}
We expand $R^+_{\um \alpha}$ in terms of $\nabla_\um^+$, leading to
\begin{align}
R^+_{\um \alpha}
	&= F^+_{\um \alpha}{}^\un \nabla_\un^+
	+ T^+_{\um \alpha}{}^B \nabla_B
	+ R^+_{\um \alpha}{}^x X_x
	+ \nabla^+_\un F^+_{\um \alpha}{}^\up g_\up{}^\un
	+ \cR^+_{\um \alpha \un}{}^\up g_\up{}^\un~.
\end{align}
Expanding out both sides of \eqref{E:intW} leads to a number of identities. Simplifications
occur upon using
\begin{align}
\frac{i}{8} \Big(
    8 i T^+_{\um \alpha}{}^\beta + 2 \bar\nabla_\dbeta T^+_{\um \alpha}{}^{\dbeta \beta}
    \Big)
    = - \cW_\alpha{}^\un \Psi_{\un \um}{}^\beta~,
\end{align}
which holds on account of the explicit expression \eqref{E:XDAM}.
The terms in \eqref{E:intW} involving covariant derivatives become
\begin{subequations}
\begin{align}
0 &= - \cW_\alpha{}_\um{}^\un
    + \cW_\alpha{}^\up \Big(
    F^+_{\ul{pm}}{}^\un + \Psi_{\up \um}{}^\beta \cW_\beta{}^\un \Big)
	-\frac{i}{8} \bar\nabla^2 F^+_{\um \alpha}{}^\un~, \\
0 &= \cW_\alpha{}^\un T^+_{\un \um}{}^a
    - \frac{i}{8} \bar\nabla^2 T^+_{\um \alpha}{}^a
    + \frac{1}{2} \bar\nabla^\dbeta T^+_{\um \alpha}{}^\beta (\sigma^a)_{\beta \dbeta}~, \\
0 &= \cW_\alpha{}^\un T^+_{\un \um}{}^\beta
    - \frac{i}{8} \bar \nabla^2 T^+_{\um \alpha}{}^\beta~, \\
0 &= \nabla_\um^+ \cW_{\alpha \dalpha} 
    + \cW_\alpha{}^\un \Big(T^+_{\un \um}{}_\dalpha 
        + \Psi_{\ul{nm}}{}^\gamma \cW_{\gamma \dalpha} \Big)
    \eol & \quad
    - \frac{i}{8} \Big(
	\bar\nabla^2 T^+_{\um \alpha \dalpha} 
	+ \bar\nabla_\dalpha R^+_{\um \alpha}(D) 
	+ 2 i \bar \nabla_\dalpha R^+_{\um \alpha}(A) 
% 	\eol & \qquad \qquad
	+ 2 \bar\nabla_\dbeta R^+_{\um \alpha}(M)^\dbeta{}_\dalpha
	+ 8 R^+_{\um \alpha}(S)_\dalpha
    \Big)~.\label{E:IntWa}
\end{align}
\end{subequations}
The first identity is solved by
\begin{align}
\cW_\alpha{}_\um{}^\un
    &= -\frac{i}{4} \bar\nabla^2 F_{\um \alpha}{}^\un~.
\end{align}
The second and third identities hold automatically.
The fourth identity leads to a definition of $X_\um(K)_a$ in terms of lower dimension quantities:
\begin{align}
i X_\um(K)_{\alpha \dalpha}
    &= \frac{1}{4} \nabla_\um^+ \cW_{\alpha \dalpha}
    + \frac{1}{4} \cW_\alpha{}^\un (T^+_{\un \um}{}_\dalpha + \Psi_{\un\um}{}^\gamma \cW_\gamma{}_\dalpha)
    \eol & \quad
    + \bar\nabla_\dalpha X_\um(S)_\alpha - F_\um{}_\dalpha{}^\un X_\un(S)_\alpha
    + \frac{1}{16} \bar \nabla^2 \Big[
        \nabla_\alpha X_\um{}_\dalpha
        - 2 F_\um{}_\alpha{}^\un X_{\un}{}_\dalpha
    \Big]
    \eol & \quad
    + \frac{1}{8} \bar\nabla_\dbeta
    \Big(\delta_\um{}^\un \nabla_\alpha - 2 F_\um{}_\alpha{}^\un\Big)
    \Big(
        X_\un(M){}^\dbeta{}_\dalpha 
        + \frac{1}{2} X_\un(D) \,\delta^\dbeta{}_\dalpha
        + i X_\un(A) \,\delta^\dbeta{}_\dalpha \Big)
    \eol & \quad
    - \HC
\end{align}

At dimension two, we find the Bianchi identities
\begin{align}
0 &= \nabla_\um^+ \cW_\alpha(D)
    + \cW_\alpha{}^\un \Big( R^+_{\un \um}(D)
        + \Psi_{\ul{nm}}{}^\gamma \cW_{\gamma}(D)\Big)
%     \eol & \quad
    - \frac{i}{8} \Big[
	\bar\nabla^2 R^+_{\um \alpha}(D)
	+ 4 \eps^{\dbeta \dalpha} \bar\nabla_\dalpha R^+_{\um \alpha}(S)_\dbeta 
        \Big]~, \eol
0 &= \nabla_\um^+ \cW_\alpha(A)
    + \cW_\alpha{}^\un \Big( R^+_{\un \um}(A)
        + \Psi_{\ul{nm}}{}^\gamma \cW_{\gamma}(A) \Big)
%     \eol & \quad
    - \frac{i}{8} \Big[
	\bar\nabla^2 R^+_{\um \alpha}(A)
	+ 6i \eps^{\dbeta \dalpha} \bar\nabla_\dalpha R^+_{\um \alpha}(S)_\dbeta 
        \Big]~, \eol
0 &= \nabla_\um^+ \cW_\alpha(M)_{\dbeta \dgamma}
    + \cW_\alpha{}^\un \Big(R^+_{\un \um}(M){}_{\dbeta \dgamma}
        + \Psi_{\ul{nm}}{}^\gamma \cW_{\gamma}(M)_{\dbeta \dgamma} \Big)
%     \eol & \quad
    - \frac{i}{8} \Big[
	\bar\nabla^2 R^+_{\um \alpha}(M)_{\dbeta\dgamma}
	- 8 \bar\nabla_{(\dbeta} R^+_{\um \alpha}(S)_{\dgamma)}
    \Big]~, \eol
0 &= \nabla_\um^+ \cW_\alpha(M)^{\beta \gamma}
    + \cW_\alpha{}^\un \Big(R^+_{\un \um}(M){}^{\beta \gamma}
        + \Psi_{\ul{nm}}{}^\delta \cW_{\delta}(M)^{\beta\gamma}\Big)
    - \frac{i}{8} \Big[
	\bar\nabla^2 R^+_{\um \alpha}(M)^{\beta\gamma}
    \Big]~.
\end{align}
The first three Bianchi identities hold on account of \eqref{E:IntWa} provided that
\begin{align}\label{E:XmS.v1}
X_\um(S)_\alpha &=
    -\frac{1}{4} \Big[
        \hat \nabla_\alpha X_{\um}(D)
        + X_{\um \alpha \dalpha} \bar \cW^{\dalpha}(D) \Big]
%     \eol & \quad
    - \frac{i}{6} \Big[
        \hat \nabla_\alpha X_{\um}(A)
        + X_{\um \alpha \dalpha} \bar \cW^{\dalpha}(A) \Big]
%     \eol & \quad
    + \Sigma^{(1)}_{\um \alpha}~,
\end{align}
where $\Sigma^{(1)}_{\um \alpha}$ is a non-primary superfield obeying
\begin{align}
- \frac{3}{2} \bar \nabla^2 \Sigma^{(1)}_{\um \alpha}
    = \nabla_\um^+ \phi_\alpha + \cW_\alpha{}^\un \Phi_{\un \um}~.
\end{align}
From the trace part of $\nabla^+_\um \cW_\alpha(M)^{\beta\gamma}$, we find a
similar relation
\begin{align}\label{E:XmS.v2}
X_{\um}(S)_\alpha &=
    \frac{1}{6} \hat \nabla^\beta X_\um(M)_{\beta \alpha}
    - \frac{1}{6} X_{\um \beta \dbeta} \bar \cW^{\dbeta}(M)^{\beta}{}_\alpha
    + \frac{i}{6} \nabla^+_\um F_{\up \alpha}{}^\up
    + \Sigma^{(2)}_{\um \alpha}
\end{align}
where
\begin{align}
- \frac{3}{2} \bar \nabla^2 \Sigma^{(2)}_{\um \alpha}
    = \nabla_\um^+ \phi_\alpha - \cW^\beta{}^\un \Phi_{\un \um}{}_{\beta\alpha}~.
\end{align}
Equating the two competing expressions for $X_{\um}(S)_\alpha$, one can compute
the difference between $\Sigma^{(1)}_{\um \alpha}$ and $\Sigma^{(2)}_{\um \alpha}$. This leads to
\begin{align}
\cW^{\beta \un} \Big(
    \eps_{\alpha \beta} \Phi_{\un \um} + \Phi_{\un \um}{}_{\alpha \beta }
    \Big)
    &= \cW^{\beta \un} \,\bar\nabla^2 \Big[
    - \frac{i}{8} \hat\nabla_\beta \Psi_{\un \um \alpha}
    - \frac{i}{4} \Psi_{\un \um \beta} F_{\up \alpha}{}^\up
    + \frac{1}{16} T^+_{\un (\beta, \gamma) \dgamma} T^+_{\um (\alpha,}{}^{\gamma) \dgamma} 
    \Big]~,
\end{align}
which implies
\begin{align}\label{E:DefPhiDM}
\Phi_{\un \um\, \alpha\beta}
    &= \bar\nabla^2 \Big[
    - \frac{i}{8} \hat\nabla_{(\beta} \Psi_{\un \um \alpha)}
    - \frac{i}{4} \Psi_{\un \um (\beta} F_{\up \alpha)}{}^\up
    - \frac{1}{4} \hat\nabla_{(\beta} X_{[\un \gamma) \dgamma} \hat\nabla_{(\alpha} X_{\um]}{}^{\gamma) \dgamma}
    \Big]
    - \cW_{(\alpha}{}^{\up} \Phi_{\ul{pnm}\, \beta)}~, \eol
\Phi_{\un \um}
    &= \bar\nabla^2 \Big[
    \frac{i}{16} \hat\nabla^\gamma \Psi_{\un \um \gamma}
    + \frac{i}{8} \Psi_{\un \um}{}^\gamma F_{\up \gamma}{}^\up
    \Big]
    + \frac{1}{2} \cW^{\beta \up} \Phi_{\ul{pnm} \, \beta}~,
\end{align}
for some chiral primary 3-form superfield $\Phi_{\ul{pnm}}{}_\alpha$. From the linearized
analysis, we know this should indeed be the curvature $\Phi_{3}{}_\alpha$ whose linearized
form is $\pa \Phi_{2 \alpha}$.
Then one may define a primary superfield $\Sigma_{\um \alpha}$ by the relation
\begin{align}
X_\um(S)_\alpha 
    &=
    -\frac{1}{4} \Big[
        \hat \nabla_\alpha X_{\um}(D)
        + X_{\um \alpha \dalpha} \bar \cW^{\dalpha}(D) \Big]
%     \eol & \quad
    - \frac{i}{6} \Big[
        \hat \nabla_\alpha X_{\um}(A)
        + X_{\um \alpha \dalpha} \bar \cW^{\dalpha}(A) \Big]
    - \frac{1}{4} \bar \nabla^\dalpha \hat\nabla_\alpha X_{\um \dalpha}
    \eol & \quad
    - \frac{2}{3} \cW_\alpha{}^\un \Big(
    \frac{i}{16} \hat\nabla^\beta \Psi_{\un \um \beta}
    + \frac{i}{8} \Psi_{\un \um}{}^\beta F_{\up \beta}{}^\up
    \Big)
%     \eol & \quad
    - \frac{i}{6} \nabla_\um^+ F_{\up \alpha}{}^\up
    + \Sigma_{\um \alpha}~,
\end{align}
where $\Sigma_{\um \alpha}$ obeys
\begin{align}\label{E:SigmaBianchi}
\bar \nabla^2 \Sigma_{\um \alpha}
    &= - \nabla_\um^+ Z_\alpha 
    - \frac{1}{3} \cW_\alpha{}^\un \cW^\beta{}^\up \Phi_{\ul{pnm} \beta} ~.
\end{align}
This is a natural generalization of \eqref{E:XmS.v1}, where we have aimed to
make $\Sigma_{\um \alpha}$ primary and to express it in terms of $Z_\alpha$
rather than $\phi_\alpha$. One could instead have aimed for a generalization of
\eqref{E:XmS.v2} (or some combination of \eqref{E:XmS.v1} and \eqref{E:XmS.v2}).
This would involve shifting $\Sigma_{\um \alpha}$ by some primary complex
linear superfield.

From the totally symmetric part of the $\nabla^+_\um \cW_\alpha(M)^{\beta\gamma}$
Bianchi identity, we find
\begin{align}
\nabla^+_\um W_{\alpha\beta\gamma}
    &= \frac{1}{16} \bar\nabla^2 \Big[
    \hat\nabla_\alpha{}^\dgamma \hat\nabla_\beta X_{\um \gamma \dgamma}
    - \hat\nabla_\alpha (\cW_\beta{}^\un \Psi_{\un \um \gamma})
    + 4 \hat\nabla_\alpha X_{\um \beta \dgamma} \bar \cW^{\dgamma \un} X_{\un \gamma}
    + 2i\,\hat \nabla_\alpha X_{\um \beta \dgamma} \bar\cW^{\dgamma}{}_\gamma
    \Big]_{(\alpha\beta\gamma)}
    \eol & \quad
    + \frac{i}{2} \cW_{(\alpha}{}^\un \Phi_{\un \um \, \beta \gamma)}~.
\end{align}
This is the non-linear generalization of the fundamental Bianchi identity
\eqref{E:LinSugraBI.a}.

The three highest dimension Bianchi identities are
\begin{subequations}
\begin{align}
0 &= \nabla_\um^+ \cW_\alpha(S)^\beta
    + \cW_\alpha{}^\un \Big(R^+_{\un \um}(S){}^{\beta}
    + \Psi_{\ul{nm}}{}^\gamma \cW_{\gamma}(S)^\beta \Big)
    - \frac{i}{8} \Big[
        \bar\nabla^2 R^+_{\um \alpha}(S)^\beta
	+ 2 i \bar\nabla_\dbeta R^+_{\um \alpha}(K)^{\dbeta \beta}
    \Big]~, \eol
0 &= \nabla_\um^+ \cW_\alpha(S)_\dbeta
    + \cW_\alpha{}^\un \Big(R^+_{\un \um}(S){}_{\dbeta} 
    + \Psi_{\ul{nm}}{}^\gamma \cW_{\gamma}(S)_\dbeta \Big)
    - \frac{i}{8} \Big[
        \bar\nabla^2 R^+_{\um \alpha}(S)_\dbeta
    \Big]~, \eol
0 &=  \nabla_\um^+ \cW_\alpha(K)^b
    + \cW_\alpha{}^\un \Big(R^+_{\un \um}(K){}^b  + \Psi_{\ul{nm}}{}^\gamma \cW_{\gamma}(K)^b\Big)
    - \frac{i}{8} \Big[
        \bar\nabla^2 R^+_{\um \alpha}(K)^b
    \Big]~.
\end{align}
\end{subequations}
The first should be a consequence of the explicit form of $\cW_\alpha(S)^\beta$ that we
have derived in \eqref{E:WS.v2}. It is not hard to show that the second and third are consequences 
of lower dimension identities.

\subsection{The $\nabla_{[\protect\um}^+ R^+_{\protect \un \protect\up]}=0$
Bianchi identity \eqref{BI.6}}

Next, we analyze \eqref{BI.6}. The lower dimension ones are
\begin{subequations}
\begin{align}
0 &= - \cR^+_{[\um \un \up]}{}^\uq 
    + T^+_{[\um \un}{}^{B} F^+_{B \up]}{}^\uq
    + F^+_{[\um \un}{}^{\ur} F^+_{\ur \up]}{}^\uq  
    ~, \\
0 &= \nabla_{[\up}^+ T^+_{\um \un]}{}^B 
    + T^+_{[\um \un}{}^{C} T^+_{C \up]}{}^B 
    + F^+_{[\um \un|}{}^{\uq} T^+_{\uq |\up]}{}^B~, \\
0 &= \nabla_{[\up}^+ R^+_{\um \un]}(D) 
    + T^+_{[\um \un}{}^{C} R^+_{C \up]}(D)
    + F^+_{[\um \un|}{}^{\uq} R^+_{\uq |\up]}(D) ~, \\
0 &= \nabla_{[\up}^+ R^+_{\um \un]}(A) 
    + T^+_{[\um \un}{}^{C} R^+_{C \up]}(A)
    + F^+_{[\um \un|}{}^{\uq} R^+_{\uq |\up]}(A) ~, \\
0 &= \nabla_{[\up}^+ R^+_{\um \un]}(M)^{\beta\gamma} 
    + T^+_{[\um \un}{}^{C} R^+_{C \up]}(M)^{\beta\gamma} 
    + F^+_{[\um \un|}{}^{\uq} R^+_{\uq |\up]}(M)^{\beta\gamma} 
    ~, \\
0 &= \nabla_{[\up}^+ R^+_{\um \un]}(M)_{\dbeta\dgamma} 
    + T^+_{[\um \un}{}^{C} R^+_{C \up]}(M)_{\dbeta\dgamma} 
    + F^+_{[\um \un|}{}^{\uq} R^+_{\uq |\up]}(M)_{\dbeta\dgamma} 
    ~,
\end{align}
\end{subequations}
and the higher dimension ones involving $S_\alpha$, $\bar S^\dalpha$ and $K_a$
follow the same pattern.

In analyzing the Bianchi identity on $T^+_{\um \un}{}^a$, one discovers that
\begin{align}
\nabla_{[\up}^+ \Psi_{\um \un] \, \alpha}
    &= 
    - \frac{i}{4} \bar\nabla^\dbeta \Psi_{[\um \un}{}^\beta T^+_{\up] \beta \, \alpha \dbeta}
    + \Psi_{[\um \un|}{}^\beta \cW_\beta{}^\uq \Psi_{\uq| \up]\, \alpha}
    + \frac{1}{3} \Phi_{\ul{pmn}\, \alpha}~,
\end{align}
generalizing the linearized result \eqref{E:LinSugraBI.c}.
This identity is found under an antichiral derivative, so the chiral superfield
$\Phi_{3 \alpha}$ is undetermined. From our linearized analysis, we
know it involves $\pa \Phi_{2\alpha}$.

The Bianchi identity involving $T^+_{\um \un}{}_\dbeta$ is not immediately useful
because we do not yet have an independent expression for it.
The remainder of the Bianchi identities lead to
\begin{align}
\nabla^+_\um \phi_\alpha &=
    -W_\alpha{}^\un \Phi_{\un \um}
    + \frac{i}{8} \bar\nabla^2 \Big[
        \tfrac{3}{2} R^+_{\up\alpha}(D) + i R^+_{\up\alpha}(A)
    \Big]~, \eol
\nabla_{[\up}^+ \Phi_{\um \un]} &=
    \frac{1}{3} \Phi_{\ul{pmn}}{}^\alpha \phi_\alpha
    + \frac{i}{8} \bar\nabla^2 \Big[
        \Psi_{[\um \un}{}^\alpha \Big(\tfrac{3}{2} R^+_{\up]\alpha }(D) + i R^+_{\up]\alpha }(A)
    \Big)\Big]~, \eol
\nabla_{[\up}^+ \Phi_{\um \un] \, \alpha\beta}
    &= 
    \frac{1}{3} \Phi_{\ul{pmn}}{}^\gamma W_\gamma(M)_{\alpha \beta}
    + \frac{i}{8} \bar\nabla^2 (
        \Psi_{[\um \un}{}^\gamma R^+_{\up]\gamma }(M)_{\alpha\beta}
    )~, \eol
\nabla_{[\up}^+ \Sigma_{\um \un] \alpha}
    &= 
    \frac{1}{3} \Phi_{\ul{pmn}}{}^\beta W_\beta(S)_\alpha
    + \frac{i}{8} \bar \nabla^2 (\Psi_{[\um \un}{}^\beta R^+_{\up]\beta }(S)_\alpha)
    - \frac{1}{4} \bar\nabla^\dbeta (
        \Psi_{[\um \un}{}^\beta R^+_{\up] \beta }(K)_{\alpha \dbeta}
    )~, \eol
\Phi_{[\ul{mnp}]}{}^\uq
    &= 
    -\frac{1}{3} \Phi_{\ul{mnp}}{}^\alpha W_\alpha{}^\uq
    - \frac{i}{8} \bar\nabla^2 (\Psi_{[\um \un}{}^\beta F^+_{\up]\beta}{}^\uq)~.
\end{align}
The first corresponds to an identity we have seen already. The remaining ones should hold
on account of the definitions of these various quantities, although we do not here give
explicit forms for $\Sigma_{\um \un}{}_\alpha$ and $\Phi_{\ul{mnp}}{}^\uq$.

As an integrability condition, one can now check $\nabla_1^+ \nabla_1^+ \Psi_{2\alpha}$.
This leads to 
\begin{align}\label{E:dPhi3}
0 &= \nabla^+ \Phi_{3 \alpha}
    + \frac{1}{8} \bar\nabla^2 \Big[
        i \Psi_2{}^\beta \hnabla_\beta \Psi_{2 \alpha}
        + i \Psi_2{}^\beta \Psi_2{}_\beta \, F_{\um \alpha}{}^\um
        - X_{1 \alpha \gamma \dgamma} X_1{}^{\beta \dgamma \gamma} \Psi_{2 \beta}
    \Big]
\end{align}
where $X_{1\,\alpha \beta \dbeta} := \hat\nabla_{(\alpha} X_{1 \beta) \dbeta}$.
This is the non-linear generalization of \eqref{E:LinSugraBI.d}.
It confirms that one \emph{cannot} set the field strength $\Phi_{3\alpha}$ consistently
to zero.

\subsection{The $R^+ - R^-$ Bianchi identity \eqref{BI.5}}
The last batch of Bianchi identities to discuss are those arising from \eqref{BI.5},
\begin{align*}
\nabla_{[\um} X_{\un]} = \frac{i}{4} R_{\um \un}^+  - \frac{i}{4} R_{\um \un}^-~.
\end{align*}
In expanding this expression, we must write both sides in terms of
$\nabla_\um$ rather than $\nabla_\um^+$ or $\nabla_\um^-$.
The lowest dimension pieces read
\begin{subequations}\label{E:intXBianchi}
\begin{align}
X_{[\um \un]}{}^\up
	+ X_{[\um}{}^B F_{\un] B}{}^\up
	&= \frac{i}{4} F^+_{\um \un}{}^\up  - \frac{i}{4} F^-_{\um \un}{}^\up~, 
    \label{E:intXBianchi.a}\\
\nabla_{[\um} X_{\un]}{}^a + X_{[\um}{}^B T_{\un] B}{}^a
	&= \frac{i}{4} (T^+_{\um \un}{}^a - T^-_{\um \un}{}^a) 
	- \frac{1}{4} (T^+_{\um \un}{}^\up + T^-_{\um \un}{}^\up) X_{\up}{}^a~, 
    \label{E:intXBianchi.b} \\
\nabla_{[\um} X_{\un]}{}^\alpha + X_{[\um}{}^B T_{\un] B}{}^\alpha
	&= \frac{i}{4} (T^+_{\um \un}{}^\alpha - T^-_{\um \un}{}^\alpha) 
	- \frac{1}{4} (T^+_{\um \un}{}^\up + T^-_{\um \un}{}^\up) X_{\up}{}^\alpha~,
    \label{E:intXBianchi.c}\\
\nabla_{[\um} X_{\un]}{}_\dalpha + X_{[\um}{}^B T_{\un] B}{}_\dalpha
	&= \frac{i}{4} (T^+_{\um \un}{}_\dalpha - T^-_{\um \un}{}_\dalpha) 
	- \frac{1}{4} (T^+_{\um \un}{}^\up + T^-_{\um \un}{}^\up) X_{\up}{}_\dalpha
    \label{E:intXBianchi.d}~.
\end{align}
\end{subequations}

The first equation defines $X_{\um \un}{}^\up$ up to the symmetric part. There is no
constraint on the symmetric part because lower $\GL(n)$ indices will always be
antisymmetrized in our approach. Writing it as a vector-valued 2-form, we have
several equivalent expressions:
\begin{align}
X_2{}^1 
    &=
    - \frac{i}{4} \Psi_{2}{}^\alpha \cW_\alpha{}^1
    - \frac{1}{8} X_1{}_{\beta \dalpha} X_1{}^{\dalpha \alpha} \hat\nabla^\beta \cW_{\alpha}{}^1
    - \frac{1}{4} X_1{}_{\gamma \dgamma} \hat\nabla^\gamma X_1{}^{\dgamma \alpha} \cW_\alpha{}^1
%     \eol & \quad
    + \frac{i}{2} X_1{}^{\dalpha \alpha} F_1{}_\alpha{}^1 \lrcorner F_1{}_\dalpha{}^1 
    + \HC\eol
    &=
    - \frac{i}{4} \Psi_{2}{}^\alpha \cW_\alpha{}^1
    - \frac{1}{8} X_1{}_{\beta \dalpha} X_1{}^{\dalpha \alpha} \nabla^\beta \cW_{\alpha}{}^1
    - \frac{1}{4} X_1{}_{\gamma \dgamma} \nabla^\gamma X_1{}^{\dgamma \alpha} \cW_\alpha{}^1
    \eol & \quad
    - \frac{i}{4} X_1{}_{\gamma \dgamma} X_1{}^{\gamma \dphi} \bar \cW_\dphi{}^1 \lrcorner X_1{}^{\dgamma \alpha} \cW_\alpha{}^1
    + \HC  
\end{align}
A useful chiral form of this expression is
\begin{align}
X_2{}^1    &= \frac{i}{4} (F^+_{2}{}^1 - F^-_{2}{}^1)
    + \frac{i}{4} X_{1}{}^{\dbeta \beta} \bar\nabla_\dbeta F_{1 \beta}{}^1
    + \frac{1}{4} X_1{}^{\dbeta \beta} \hat\nabla_\beta X_{1\gamma \dbeta} \cW^{\gamma 1}
    + \frac{i}{2} X_1{}^a X_1{}^b F_{b a}{}^1~.
\end{align}

The second equation \eqref{E:intXBianchi.b} gives
\begin{align}
\nabla_{[\um} X_{\un]}{}^a + X_{[\um}{}^B T_{\un] B}{}^a
	&= \frac{i}{4} (T^+_{\um \un}{}^a - T^-_{\um \un}{}^a) 
	- \frac{1}{4} (F^+_{\um \un}{}^\up + F^-_{\um \un}{}^\up) X_{\up}{}^a~.
\end{align}
This gives the generalization of the linearized Bianchi identity \eqref{E:LinSugraBI.b} relating
$X_{\um a}$ to $\Psi_{\um \un \alpha}$.

The remaining two equations \eqref{E:intXBianchi.c} and its complex conjugate
\eqref{E:intXBianchi.d} give
\begin{align}
T^+_{\um \un}{}_\dalpha
	&= T^-_{\um \un}{}_\dalpha
	- i (F^+_{\um \un}{}^\up + F^-_{\um \un}{}^\up) X_\up{}_\dalpha
	- 4 i \Big(\nabla_{[\um} X_{\un]}{}_\dalpha + X_\um{}^B T_{\un B}{}_\dalpha\Big) \eol
	&= - \frac{i}{8} \nabla^2 \Psi_{\um \un}{}_\dalpha
	- i (F^+_{\um \un}{}^\up + F^-_{\um \un}{}^\up) X_\up{}_\dalpha
	- 4 i \Big(\nabla_{[\um} X_{\un]}{}_\dalpha + X_\um{}^B T_{\un B}{}_\dalpha\Big)
\end{align}
as well as it complex conjugate. This defines the expression $T^+_{\um \un}{}_\dalpha$,
which previously had not been determined.

The remaining identities, which we have not explicitly written out, lead to, among other
consistency relations, explicit but complicated expressions for $\Phi_{\um\un}$, $\Phi_{\um \un}{}_{\alpha\beta}$
and $\Sigma_{\um \un}$. For example,
\begin{align}
\frac{i}{2} \Phi_{\um \un}
    &= \frac{i}{2} \Psi_{\um \un}{}^\alpha \phi_\alpha
    + \frac{i}{2} \bar\nabla^\dgamma T^+_{\um \un \dgamma}
    + \frac{i}{4} \nabla_\gamma T^-_{mn}{}^\gamma
    + \frac{i}{2} \bar\nabla^\dgamma \Psi_{\um \un}{}^\gamma W_{\gamma \dgamma}
    + \frac{i}{4} \nabla^\gamma \bar\Psi_{\um \un}{}^\dgamma \bar W_{\dgamma \gamma}
    \eol & \quad
    + \nabla_{[\um} (X_{\un]}(D) - 2i X_{\un]}(A))
    + X_{[\um}{}^B (R_{\un] B}(D) - 2i R_{\un] B}(A))
    \eol & \quad
    + \frac{1}{4} (F^+_{\um \un}{}^\up + F^-_{\um \un}{}^\up) 
        (X_{\up}(D) - 2i X_{\up}(A))~, \\
\frac{i}{4} \Phi_{\um \un}{}_{\alpha\beta}
    &= \nabla_{[\um} X_{\un]}(M)_{\alpha\beta}
    + X_{[\um}{}^D R_{\un] D}(M)_{\alpha\beta}
    + \frac{1}{4} (F^+_{\um \un}{}^\up + F^-_{\um \un}{}^\up) X_\up(M)_{\alpha\beta}
    \eol & \quad
    + \frac{i}{4} \nabla_{(\alpha} \Psi_{\um \un}{}_\dgamma \bar W^\dgamma{}_{\beta)}
    + \frac{i}{4} \nabla_{(\alpha} T^-_{\um \un}{}_{\beta)}
    + \frac{i}{4} \Psi_{\um \un}{}^\gamma \cW_\gamma(M)_{\alpha\beta}~.
\end{align}
It is a complicated exercise to check that the explicit solutions \eqref{E:DefPhiDM}
that we found somewhat indirectly are consistent with these relations. We have confirmed 
this to leading order in curvatures.

\section{Action principles}
Having established the superspace geometry, we now turn to establishing the existence of
superspace actions and the various technical rules for manipulating these actions, both
in superspace and in components. The results in this section will not come as a surprise
to the superspace expert. In short order, we establish:
\begin{itemize}
\item the consistency of both full and chiral superspace integration,
provided one is given a suitable Lagrangian,

\item the formula for converting a full superspace to a chiral superspace integral,

\item the rules for integrating by parts in full and chiral superspace, and

\item the expression for a component action arising from a chiral superspace integral.
\end{itemize}
Because the details are rather technical and only the results are important, we
mainly sketch the computations required.

What we will not be concerned with here is describing how to build the Lagrangians required.
As mentioned elsewhere, this will be the concern of a subsequent paper.
The reader may keep in mind the 11D Chern-Simons action \eqref{E:CSaction} as a prototype.
It will turn out (with some minor modifications) to take the same form in this superspace.

\subsection{Consistency of full and chiral superspace integration}
A full superspace integral can be written
\begin{align}\label{E:FullSuperspaceInt}
S = \int \rd^4x\, \rd^n y\, \rd^4\q\, E\, \mathscr{L} = 
\frac{1}{n!} \int \rd^4x\, \rd^ny\, \rd^4\q\, E\, \eps^{\um_1 \cdots \um_n} \omega_{\um_1 \cdots \um_n}
\end{align}
where $\omega_{\um_1 \cdots \um_n}$ is a real covariant $n$-form on the internal space
and $E = \sdet(E_M{}^A)$ is the full superspace measure, defined as the superdeterminant
(or Berezinian) of the supervielbein. Above we are denoting
$\mathscr{L} \equiv \frac{1}{n!} \,\eps^{\um_1 \cdots \um_n} \,\omega_{\um_1 \cdots \um_n}$
where the antisymmetric tensor density $\eps^{\um_1 \cdots \um_n}$ has constant entries of $\pm1$.
Thus $\omega_{\um_1 \cdots \um_n}$ is a top-form on the internal manifold
and $\mathscr{L}$ is its scalar density.

In order for the action to be gauge invariant, $\omega_{\um_1 \cdots \um_n}$ (equivalently,
$\mathscr{L}$) must be a conformal primary (annihilated by $S$-supersymmetry) of Weyl weight two. 

The vielbein transforms under external diffeomorphisms (with parameter $\xi^M$),
$\cH$-gauge transformations (with parameter $g^x$), and internal diffeomorphisms
(with parameter $\Lambda^\um$) as
\begin{align}\label{eq:deltaE}
\delta E_M{}^A &= \hat D_M \xi^N E_N{}^A + \xi^N \hat D_N E_M{}^A
	+ E_M{}^B g^x f_{x B}{}^A
        + \Lambda^\un \pa_\un E_M{}^A~.
\end{align}
This means that the full superspace measure transforms as
\begin{align}
\delta E = \hat D_M (\xi^M E) + \Lambda^\um \pa_\um E + g^x f_{x A}{}^A (-)^A \, E~.
\end{align}
We require $\mathscr{L}$ to transform as
\begin{align}
\delta \mathscr{L} &= \hat D_M \mathscr{L} + \pa_\um (\Lambda^\um \mathscr{L}) - g^x f_{x A}{}^A\, (-)^A\, \mathscr{L}~.
\end{align}
This is consistent with requiring $\omega_{\um_1 \cdots \um_m}$ to transform as
an $n$-form under internal diffeomorphisms, a scalar field under external
diffeomorphisms, and as a tensor with weight $-f_{x A}{}^A (-)^A$ under $\cH$-gauge
transformations. The action \eqref{E:FullSuperspaceInt} is manifestly invariant
under all but external diffeomorphisms. For these, we find (using internal form notation)
\begin{align}
\delta S &= \int \rd^4x\, \rd^4\q\, \hat D_M \Big( \xi^M E \,\omega \Big) 
	= \int \rd^4x\, \rd^4\q\, \Big\{
	\pa_M \Big( \xi^M E \,\omega \Big)
	- \pa \Big( \imath_{\xi^M A_M} \omega E \,\Big)
	\Big\} = 0
\end{align}
where we have used the property that $\omega$ is a top form on the internal space.

Showing consistency of chiral superspace integration is more involved. The basic integral
looks like
\begin{align}\label{E:ChiralSuperspaceInt}
S_c &= \int \rd^4x\, \rd^ny\, \rd^2\q\, \cE\, \mathscr{L}_c = 
\frac{1}{n!} \int \rd^4x\, \rd^ny\, \rd^2\q\, \cE\, \eps^{\um_1 \cdots \um_n} \omega^c_{\um_1 \cdots \um_n}
\end{align}
where $\omega^c_{\um_1 \cdots \um_n}$ is a covariant chiral $n$-form.
The meaning of chirality here is that $\bar\nabla^\dalpha \omega^c_{\um_1 \cdots \um_n} = 0$.
The measure $\cE$ must be defined. We are
going to take the approach used in Appendix A of \cite{Butter:2015nza}.
Write the full supervielbein and its inverse as
\begin{align}
E_M{}^A =
\begin{pmatrix}
\cE_\cM{}^\cA & E_\cM{}_\dalpha \\
E^\dmu{}^\cA & E^\dmu{}_\dalpha
\end{pmatrix}~, \qquad
E_A{}^M =
\begin{pmatrix}
E_\cA{}^\cM & E_\cA{}_\dmu \\
E_\cA{}^\cM & \bar\cE^\dalpha{}_\dmu
\end{pmatrix}
\end{align}
with $\cM = (m, \mu)$ and $\cA = (a, \alpha)$ describing the coordinates and tangent space 
of chiral superspace.
We have given special names to the blocks $\cE_\cM{}^\cA$ and $\bar\cE^{\dalpha}{}_\dmu$
and assume both of these are invertible with inverses $\cE_\cA{}^\cM$ and
$\bar\cE^\dmu{}_\dalpha$. The chiral measure is $\cE = \sdet \cE_\cM{}^\cA$.

Since $\cE_\cM{}^\cA = E_\cM{}^\cA$, we can use \eqref{eq:deltaE} for its
transformations. Invariance of $S_c$ under internal diffeomorphisms proceeds as before
because $\cE$ is a scalar and $\mathscr{L}_c$ is a scalar density.
Invariance under $\cH$-gauge transformations requires $f_x{}^{\dbeta\, \cA} = 0$.
This can be understood as an integrability condition for the existence of $\cH$-invariant
chiral superfields.
It also means that the chiral part of the vielbein only transforms into itself under 
$\cH$ transformations, leading to
\begin{align}
\delta_\cH \cE = \cE\, g^x f_{x \cA}{}^\cA (-)^\cA \qquad \implies \quad
\delta_\cH \mathscr{L}_c = -\mathscr{L}_c\, g^x f_{x \cA}{}^\cA (-)^\cA~.
\end{align}

To show invariance under external diffeomorphisms, it helps to consider
\emph{covariant} external diffeomorphisms: these are a special combination of 
external diffeomorphisms and $\cH$-gauge transformations with $g^x = \xi^M H_M{}^x$.
For the full supervielbein, these become
\begin{align}
\delta_{cov}(\xi) E_M{}^A &= \hat D_M \xi^N E_N{}^A + \xi^N \hat D_N E_M{}^A
	+ E_M{}^B \xi^N H_N{}^x f_{x B}{}^A \eol
	&= \nabla_M \xi^A + E_M{}^B \xi^C T_{CB}{}^A - E_M{}^B \xi^C F_{CB}{}^\um \chi_\um{}^A
\end{align}
where we have rewritten the last line in terms of $\xi^A = \xi^M E_M{}^A$.
We remind the reader that the field $\chi_\um{}^A$, discussed in detail in
the bosonic case in section \ref{S:KKGeo.1}, can be understood as a component 
of a \emph{super-sehrvielbein} on a larger superspace.

We now consider separately chiral external diffeomorphisms with $\xi^M = (\xi^\cM,0)$
and anti-chiral covariant external diffeomorphisms with 
$\xi^A = (0, \xi_\dalpha)$.\footnote{These span the entire space of external 
diffeomorphisms only when $\bar\cE^\dalpha{}_\dmu$ is invertible.}
Chiral external diffeomorphisms lead to an invariant action just as before.
Under anti-chiral covariant external diffeomorphisms, one finds
\begin{align}
\delta \cE_\cM{}^\cA = E_\cM{}^B \xi_\dgamma \Big(
	T^{\dgamma}{}_B{}^\cA - F^{\dgamma}{}_B{}^\um \chi_\um{}^\cA \Big)
	\quad \implies \quad  \eol
\cE^{-1} \delta \cE = \xi_\dgamma \Big(T^\dgamma{}_\cA{}^\cA - F^\dgamma{}_\cA{}^\um \chi_\um{}^\cA\Big)
	+ \cE_\cA{}^\cM E_\cM{}_\dbeta \xi_\dgamma
	\Big(T^{\dgamma \dbeta}{}^\cA - F^{\dgamma\dbeta}{}^\um \chi_\um{}^\cA\Big)
\end{align}
Provided we satisfy the conditions
\begin{align}
T^{\dgamma \dbeta}{}^\cA = 0~, \qquad F^{\dgamma \dbeta}{}^\um = 0~,
\end{align}
the second batch of terms vanishes. The first batch of terms does not. In our case,
it leads to
\begin{align}
\cE^{-1} \delta \cE
	= \xi_\dgamma W_\gamma{}^\um (\bar\sigma_b)^{\dgamma \gamma} \Big(
	i X_\um{}^b - \chi_\um{}^b
	\Big)
\end{align}
In order for invariance to be maintained $\mathscr{L}_c$ must obey
\begin{align}
\hat \cD^{\dalpha} \mathscr{L}_c = - \mathscr{L}_c\, (\bar\sigma_b)^{\dalpha \beta} W_\beta{}^\um \Big(
	i X_\um{}^b - \chi_\um{}^b
	\Big)~.
\end{align}
This derivative $\hat\cD$ is the original $\hat D$ derivative augmented
with the $\cH$ connection. It does not possess the $\GL(n)$ connection. Recall
the $\GL(n)$ connection involves
\begin{align}
\Gamma_M{}_\un{}^\up \sim - \chi_\un{}^N F_{N M}{}^\up + \Delta\Gamma_M{}_\un{}^\up
\quad \implies \quad
\Gamma^\dalpha{}_\un{}^\up \sim - (\bar\sigma_b)^{\dalpha \alpha} W_\alpha{}^\up
	\Big(i X_\un{}^b - \chi_\un{}^b \Big)
\end{align}
where we have used \eqref{eq:defDeltaGamma} for the shifted part of the $\GL(n)$ connection
$\Delta \Gamma$. (The piece involving $\pa_\un A_M{}^\up$ is already contained in $\hat D$.) It follows that
\begin{align}
\bar\nabla^\dalpha \mathscr{L}_c
	= D^\dalpha \mathscr{L}_c - \Gamma^\dalpha{}_\up{}^\up \,\mathscr{L}_c
	= 0
\end{align}
as the condition for chiral integration to be well-defined.

We emphasize that the redefinition of the $\GL(n)$ connection was key to finding
this simple chirality condition. With the original connection, we would have found
$\mathring \nabla^\dalpha \mathscr{L}_c \neq 0$, which is less convenient to work with.

\subsection{Converting full superspace to chiral superspace}
Now that we know that full superspace and chiral superspace separately exist, we should
establish how to move from one to the other. We claim that
(generalizing the flat superspace result)
\begin{align}\label{E:FullToChiral}
\int \rd^4x\, \rd^ny\, \rd^4\q\, E\, \mathscr{L}
	= -\frac{1}{4} \int \rd^4x\, \rd^ny\, \rd^2\q\, \cE\, \bar\nabla^2 \mathscr{L}~.
\end{align}

The proof goes as follows. Because of the basic condition
$\{\bar\nabla^\dalpha, \bar\nabla^\dbeta\} = 0$, we can
adopt a chiral gauge where
\begin{align}
\bar\nabla^\dalpha = \frac{\pa}{\pa \q_\dalpha} - \Gamma^\dalpha{}_\um{}^\un g_\un{}^\um~.
\end{align}
The $\GL(n)$ connection is
\begin{align}
\Gamma^\dalpha{}_\um{}^\un
	= - \chi_\um{}^b F_b{}^{\dalpha \un}
	+ F^\dalpha{}_\um{}^\un
	= (\chi_\um{}^{\dalpha \alpha} - i X_\um{}^{\dalpha \alpha}) W_\alpha{}^\un~.
\end{align}
The full superspace and chiral superspace measures are equal, $E = \cE$, and furthermore,
\begin{align}
\pa^\dalpha \cE = \cE \Big(
	T^\dalpha{}_b{}^b - F^\dalpha{}_b{}^\up \chi_\up{}^b
\Big) = \cE W_\alpha{}^\um (\bar\sigma_b)^{\dalpha \alpha} \Big(
	i X_\um{}^b - \chi_\um{}^b
	\Big) = - \Gamma^{\dalpha}{}_\um{}^\um
\end{align}
It follows that
\begin{align}
\int \rd^4x\, \rd^ny\, \rd^4\q\, E\, \mathscr{L}
	= -\frac{1}{4} \int \rd^4x\, \rd^ny\, \rd^2\q\, \cE\, 
	(\bar\pa_\dalpha - \Gamma_\dalpha{}_\un{}^\un)
	(\bar\pa^\dalpha - \Gamma^\dalpha{}_\um{}^\um) \mathscr{L}
\end{align}
The operators appearing in parentheses are just $\bar\nabla_\dalpha$ in chiral
gauge, so it follows that the two sides of \eqref{E:FullToChiral}
are equal to each other in chiral gauge. But because they are both gauge invariant
expressions, they must be equal in all gauges.

\subsection{Rules for integrations by parts}

There turn out to be three useful expressions for integrating by parts in superspace.
These are most simply formulated in terms of the vanishing (or near vanishing) of certain total
covariant derivatives.

The first expression is relevant for integrating by parts with external covariant
derivatives in full superspace. Suppose $V^A = (V^a, V^\alpha, V_\dalpha)$ is some
covariant expression, with not necessarily all of these entries nonzero.
(Of course, $V^A$ must be a scalar density under internal diffeomorphisms.)
Then one can show that
\begin{align}
\int \rd^4x\, \rd^ny\, \rd^4\q\, E\, \nabla_A V^A \, (-)^A
	&= -\int \rd^4x\, \rd^ny\, \rd^4\q\, \Big(
	E\,V^B (T_{B A}{}^A (-)^A + F_{B \um}{}^\um)
	+ H_M{}^x g_x (E V^M)
	\Big) \eol
	&= -\int \rd^4x\, \rd^ny\, \rd^4\q\, 
	H_M{}^x g_x (E V^M)~.  \label{eq:IBP1}
\end{align}
The first equality follows rather generally, while the second follows for the particular
constraints on our superspace torsion and curvature tensors we have chosen. The residual
term arises if $E \,V^M = E \,V^A E_A{}^M$ is not a gauge singlet; in practice, this involves
only the $S$ and $K$ connections and such terms cancel out if, after a series of integrations
by parts, the initial and final forms are both primary.

The other expressions involve integrating by parts with internal covariant derivatives.
In full superspace, one can use either $\nabla_\um$ or $\nabla_\um^\pm$ and the results are
structurally similar:
\begin{align}
\int \rd^4x\, \rd^ny\, \rd^4\q\, E\, \nabla_\um V^\um
	&= -\int \rd^4x\, \rd^ny\, \rd^4\q\, 
	E\, V^\um \Big(T^\pm_{\um A}{}^A(-)^A + F^\pm_{\um \un}{}^\un\Big)
	= 0 \label{eq:IBP2} \\
\int \rd^4x\, \rd^ny\, \rd^4\q\, E\, \nabla^\pm_\um V^\um
	&= -\int \rd^4x\, \rd^ny\, \rd^4\q\, 
	E\, V^\um \Big(T^\pm_{\um A}{}^A(-)^A + F^\pm_{\um \un}{}^\un\Big)
        = 0 \label{eq:IBP2b}~.
\end{align}
Here we assume $E V^\um$ is $\cH$-invariant for simplicity (as well as an internal
vector density) so that $\cH$ connections do not appear.
This will always be the case when we need to integrate internal covariant derivatives by parts.
As before, the expressions involving the traces of the torsion and curvature tensors cancel
out for our superspace geometry.
In chiral superspace, we will only need to integrate $\nabla_\um^\pm$ by parts. Its rule is
similar:
\begin{align}
\int \rd^4x\, \rd^ny\, \rd^2\q\, \cE\, \nabla_\um^+ \cV^\um
	&= - \int \rd^4x\, \rd^ny\, \rd^2\q\, 
	\cE\, \cV^\um \Big(T^+_{\um \cA}{}^\cA (-)^\cA + F^+_{\um \un}{}^\un\Big)
        = 0~. \label{eq:IBP3}
\end{align}
To be well-defined, $\cV^\um$ must be chiral, a vector density, and transform so that
$\cE \cV^\um$ is $\cH$-invariant.

The proof of \eqref{eq:IBP1} is completely standard. The proof of \eqref{eq:IBP2}
is only a bit more involved. We give a few steps to guide the reader. Discarding
total derivatives in equalities and suppressing gradings,
\begin{align}
E \nabla_\um V^\um 
    &= - \nabla_\um E V^\um
    - \chi_\um{}^N \nabla_N (E V_\um)
    - 2 \,\Gamma_{[\um\un]}{}^\un (E V^\um) \eol
    &= 
    - \nabla_N \Big(E V^\um \chi_\um{}^N\Big)
    - E V^\um (T_{\um A}{}^A + F_{N \um}{}^\un \chi_\un{}^N)
    - 2 \,\Gamma_{[\um\un]}{}^\un (E V^\um)
    \eol
    &= 
    (F_N{}_\un{}^\un - \chi_\un{}^M F_{M N}{}^\un) \Big(E V^\um \chi_\um{}^N\Big)
    - E V^\um (T_{\um A}{}^A + F_{N \um}{}^\un \chi_\un{}^N)
    - 2 \,\Gamma_{[\um\un]}{}^\un (E V^\um) \eol
    &= -E V^\um (T_{\um A}{}^A + F_{\um \un}{}^\un)~.
\end{align}
The corresponding expression for \eqref{eq:IBP2b} follows just by affixing
$\pm$ superscripts to the internal connections and curvatures, defining them with respect
to $\nabla^\pm_\um$. The rule for \eqref{eq:IBP3} is a bit more involved but
the fact that it vanishes follows from \eqref{eq:IBP2b} by converting to chiral superspace
and identifying $\cV^\um = -\frac{1}{4} \bar \nabla^2 V^\um$.

\subsection{Chiral superspace to components}
The final result we should discuss is how to convert a superspace integral to
components. Since any full superspace integral may be converted to chiral
superspace using \eqref{E:FullToChiral}, it suffices to show how to evaluate the chiral $\theta$
integrations. The result we want to establish is
\begin{align}
S_c = \int \rd^4x\, \rd^ny\, \rd^2\q\, \cE\, \mathscr{L}_c = 
    \int \rd^4x\, \rd^ny\, e\, \cL_c
\end{align}
where
\begin{align}\label{eq:CompLag}
\cL_c
    &= -\frac{1}{4} \nabla^2 \mathscr{L}_c
    + \frac{i}{2} (\bar \psi_m \bar\sigma^m)^\alpha \nabla_\alpha \mathscr{L}_c
    - \bar\psi_m \bar\sigma^{mn} \bar\psi_n\, \mathscr{L}_c
    \eol & \quad
    - 2 i \,\bar \cW_\dalpha{}^\um (\chi_\um{}^\dalpha + i X_\um{}^\dalpha) \,\mathscr{L}_c
    + i \,e_a{}^m \bar\psi_m{}_\dalpha \,\cW^\dalpha{}^\um (\chi_\um{}^a + i X_\um{}^a) \, \mathscr{L}_c~.
\end{align}
Some of the above result may be guessed without much work. The first term is
the flat superspace result, and the rest of the first line is its generalization
to conformal supergravity. Additional terms essentially can only involve 
the terms found in the second line, and some of the relative coefficients can be
determined by $S$-invariance.

A standard way of deriving the above result is to exploit the ectoplasmic approach 
\cite{Gates:1997kr,Gates:1997ag}.
In conventional $N=1$ superspace, this amounts to treating the component Lagrangian as a 4-form in superspace,
writing
\begin{align}
S_c = \int_M \frac{1}{4!} E^A E^B E^C E^D \, J_{DCBA}
\end{align}
where the integral is restricted to the bosonic spacetime $M$ lying at $\theta=0$.
The condition that the action is supersymmetric amounts to $J$ being a closed superform.
Choosing the components of $J$ appropriately then leads to the desired result.
While the ectoplasmic approach does lead to \eqref{eq:CompLag}, it is a bit subtle
because in our case the full superspace is actually extended by the $n$ internal coordinates 
and so $J$ is actually a $(4+n)$-form. Care must be taken to account for this.

A technically simpler approach is to use a brute force normal coordinate method.
Starting from chiral superspace lying at $\bar\q=0$, use the residual $\q$-dependent
coordinate and gauge transformations to set
$\nabla_\alpha = \pa_\alpha - \Gamma_\alpha{}_\um{}^\un g_\un{}^\um$
where $\Gamma_\alpha{}_\um{}^\un = - (\chi_\um{}^a + i X_\um{}^a) (\sigma_a)_{\alpha \dalpha} \cW^{\dalpha \un}$.
In this gauge, $\cE = \det(e_m{}^a) = e$, so evaluating the $\q$ integrals gives
\begin{align}
e \,\cL_c = -\frac{1}{4} \pa^\alpha \pa_\alpha (e \mathscr{L}_c)~.
\end{align}
To evaluate these terms, the following results are useful (in this gauge):
\begin{align}
\pa_\alpha e_m{}^a &= T_{\alpha m}{}^a - F_{\alpha m}{}^\up \chi_\up{}^a
    = i (\sigma^a \bar\psi_m)_\alpha + (\sigma_m)_{\alpha \dalpha} \bar \cW^{\dalpha \up} (\chi_\up{}^a + i X_\up{}^a)~, \eol
e^{-1} \pa_\alpha e &= i (\sigma^m \bar\psi_m)_\alpha - \Gamma_\alpha{}_\um{}^\um~, \eol
\frac{1}{2} \pa_\alpha \psi_m{}_\dbeta &= T_{\alpha m \dbeta} - F_{\alpha m}{}^\up \chi_\up{}_\dbeta
    = (\sigma_m)_{\alpha \dalpha} \bar \cW^{\dalpha \up} (\chi_\up{}_\dbeta + i X_\up{}_\dbeta)~, \eol
\pa_\alpha (e \mathscr{L}_c) &= e \nabla_\alpha \mathscr{L}_c + i (\sigma^m \bar\psi_m)_\alpha e \mathscr{L}_c~.
\end{align}
Putting these results together leads to \eqref{eq:CompLag}.

\section{Conclusion and outlook}

The goal of this paper has been to construct a general framework in 4D $\cN=1$ superspace
that is suitable for describing a higher-dimensional supergravity theory in $4+n$ dimensions.
While this is motivated by previous work on 11D supergravity \cite{Becker:2016xgv, Becker:2016rku, 
Becker:2016edk, Becker:2017zwe, Becker:2018phr} and 5D minimal supergravity, it is expected
to be applicable to other cases.

Let us say a few words on that point.
One potential argument \emph{against} the wider applicability of this framework is that both 11D and 5D 
minimal supergravities correspond to very particular cases where the number of internal dimensions
and the number of hidden supersymmetries coincide (respectively, 7 and 1). This is important
because the superfield $\Psi_\um{}_\alpha$, which here plays only the role of the prepotential
of the lower left block of the higher-dimensional vielbein, should pull double duty as
a prepotential for the additional spin-3/2 gravitino multiplets. The simplest way this can work
is when the number of additional gravitini matches the internal dimension.
Nevertheless, we can learn a lesson from the 6D situation
\cite{Abe:2017pvw}. There, one indeed has two fields $\Psi_\um{}_\alpha$, but in constructing
minimal 6D supergravity, one encounters a constraint that permits one of these fields to be eliminated
(see section 5 of \cite{Abe:2017pvw}) -- this is important as there is only one additional
gravitino (not two) in this framework. This may well persist for other cases where the number
of extra gravitini is smaller than the number of extra dimensions.
For the reverse situation, where the number of extra gravitini is larger than the internal dimension,
we may point to IIA supergravity, which can be constructed by dimensionally reducing 11D
supergravity in this framework. In that case, one of the gravitini, say $\Psi_{7 \alpha}$, is
``ungeometrized'' and becomes a matter superfield, albeit a high superspin one.
It would be interesting to understand both of these cases better.

There are several topics that we did not directly address in this paper. One outstanding issue is 
the application to 11D supergravity itself. This paper only provides the
geometric superspace framework necessary to describe that case. We still must analyze
how the flat results reviewed in section \ref{S:11DSugra.Sketch} are generalized. 
This involves constructing the abelian tensor hierarchy descending from $C_3$ in the
curved supergeometry we have introduced. In principle, this should be fully fixed by the
supergeometry itself so that the intricate structure described in flat space in 
section \ref{S:11DSugra.Sketch} is maintained.
As we have stressed throughout, this will be the subject of a future publication.

A technical issue that we have sidestepped is how to address the differences in the supergeometry
we have encountered relative to the linearized results \cite{Becker:2017zwe, Becker:2018phr}. The
point of mismatch is the three additional prepotentials -- the chiral superfield $\Phi_{\um\un\alpha}$,
the complex linear superfield $\sigma_\um{}_\alpha$, and the unconstrained superfield $\cV(S)_\alpha$.
These appear in the curvatures $\Phi_{\ul{mnp} \alpha}$, $\Sigma_\um{}_\alpha$ (which is a part of
$X_\um(S)_\alpha$), and $Z_\alpha$. These are related by \eqref{E:SigmaBianchi} and it is tempting
to declare them all to vanish. However, we have shown this is not possible due to the integrability
condition \eqref{E:dPhi3}. While $Z_\alpha$ may consistently be turned off, $\Phi_{\ul{mnp} \alpha}$ and thus
$\Sigma_\um{}_\alpha$ appear inescapable.

One potential solution to this is that the additional prepotentials can be eliminated by
field redefinitions even in the presence of additional matter fields of the tensor hierarchy.
This would be similar to the way in which the conventional Wess-Zumino superspace
(see e.g. \cite{Wess:1992cp}), which manifestly describes old minimal Poincar\'e
supergravity, may actually be understood to describe conformal supergravity, by introducing
a super-Weyl transformation that acts as a Weyl rescaling of the metric. Provided one only couples to
matter in a super-Weyl invariant way, only the conformal part of the gravity multiplet survives.
It is plausible that the same sort of mechanism occurs here.

Indeed, we have already seen in 
the linearized case that $Z_\alpha$ and $\Sigma_\um{}_\alpha$ can be shifted around
by redefinitions of underlying prepotentials $\cV(S)_\alpha$ and $\sigma_{\um\alpha}$.
The same can be done by analyzing linearized fluctuations about a generic curved background.
Showing that the same is true for $\Phi_{\um \un}{}_\alpha$ is a bit more involved, as we have
introduced that prepotential by hand in defining the linearized $\Psi_{\um\un}{}_\alpha$.
Seeing this at the non-linear level is a bit involved. The key idea is to introduce a shift
$\delta_\rho \Psi_{\um \un}{}_\alpha = \rho_{\um \un}{}_\alpha$,
where $\rho_{\um\un}{}_\alpha$ is a chiral superfield 2-form.
This corresponds just to a shift $\delta \Phi_{\um\un\,\alpha} = \rho_{\um\un}{}_\alpha$
in the underlying extraneous prepotential. One must then demonstrate that 
$\rho$-transformations can be consistently imposed at the non-linear level on curvatures 
and covariant derivatives.
One finds, for example, that $W_{\alpha\beta\gamma}$ shifts under $\rho$ exactly as
one would expect from its linearized expression \eqref{E:Linearized.Wabc}.
Provided $\rho$ transformations can be extended to the $p$-form superfields of
11D supergravity, one can guarantee that the extraneous prepotential can always
be removed. We will describe this in greater detail elsewhere.

%%%%%%%%%%%%%%%%%%%%%%%%%%%%%%%%%%%%%%%%%%%%%%%%%%%%%%%%%%%%%%%%
%%%%%%%%%%%%%%%%%%%%%%%%%%%%%%%%%%%%%%%%%%%%%%%%%%%%%%%%%%%%%%%%
\section*{Acknowledgements}
We thank William Linch for numerous discussions and Sunny Guha for collaboration
at an early stage.
K.B. also thanks the Institute for Advanced Study for hospitality during part of this work.
This work is partially supported by the NSF under grants NSF-1820921 and PHY-1606531, and the Mitchell Institute for Fundamental Physics and Astronomy at Texas A\&M University.

\appendix

\section{Superspace curvatures}\label{App:Curvs}
Our superspace and spinor conventions follow \cite{Wess:1992cp}.
The curvatures of $\GL(n)$ Kaluza-Klein superspace are abstractly given in terms of three operators
$\cW_\alpha$, $X_\um$, and $R^+_{\um \un}$ in the following manner.
Letting $R_{\hA \hB} := - [\nabla_\hA, \nabla_\hB]$ where $\hA = (a, \alpha, \dalpha, \um)$,
we find that
\begin{gather}
R_{\alpha \beta} = 0~, \qquad
R_{\dalpha \dbeta} = 0~, \qquad
R_{\alpha \dbeta} = 2 i \, \nabla_{\alpha \dbeta}~, \eol
R_{\alpha b} = - (\sigma_{b})_{\alpha \dalpha} \bar \cW^{\dalpha}~, \qquad
R^{\dalpha}{}_{b} = (\bsigma_{b})^{\dalpha \alpha} \cW_{\alpha}~, \eol
R_{a b} = 
    \frac{i}{2} (\sigma_{a b})^{\alpha \beta} \{\nabla_\alpha, \cW_\beta\}
    -\frac{i}{2} (\bsigma_{a b})^{\dalpha \dbeta} \{\bnabla_\dalpha, \bar\cW_\dbeta\}~, \eol
R_{\um \alpha} = i [\nabla_\alpha, X_\um]~, \qquad
R_{\um \dalpha} = -i [\bar\nabla_\dalpha, X_\um]~, \eol
R_{\um a} = 
    -\frac{1}{4} (\bsigma_a)^{\dalpha \alpha} \{\nabla_\alpha,[\bar\nabla_\dalpha, X_\um]\}
    +\frac{1}{4} (\bsigma_a)^{\dalpha \alpha} \{\bar\nabla_\dalpha, [\nabla_\alpha, X_\um]\}~, \eol
R_{\um \un} = \frac{1}{2} (R^+_{\um \un} + R^-_{\um \un}) - [X_\um, X_\un]~.
\end{gather}
When written in terms of $\nabla_\um^+ = \nabla_\um + i X_\um$, the mixed curvatures become
\begin{gather}
R^+_{\um \alpha} = 2 i [\nabla_\alpha, X_\um]~, \qquad
R^+_{\um \dalpha} = 0~, \eol
R^+_{\um a} = \frac{1}{2} (\bsigma_a)^{\dalpha \alpha} \{\bar\nabla_\dalpha, [\nabla_\alpha, X_\um]\}~.
\end{gather}

\subsection{Expressions for $\cW_\alpha$}
The chiral operator $\cW_\alpha$ is defined as:
\begin{align}
\cW_{\alpha} &= 
	\cW_{\alpha \dbeta} \bar\nabla^\dbeta
	+ \cW_\alpha{}^\um \nabla_\um^+
	+ \Big(\nabla_\un^+ \cW_\alpha{}^\um + \cW_\alpha{}_\un{}^\um\Big) g_{\um}{}^{\un}
	\eol & \quad
	+ \cW_\alpha(D) \mathbb D
	+ \cW_\alpha(A) \mathbb A
	+ \cW_\alpha(M){}^{\beta\gamma} M_{\beta\gamma}
	+ \cW_\alpha(M){}_{\dbeta\dgamma} M^{\dbeta\dgamma}
	\eol & \quad
	+ \cW_\alpha(S)^\beta S_\beta
	+ \cW_\alpha(S)_\dbeta \bar S^\dbeta
	+ \cW_\alpha(K)^b K_b~.
\end{align}
The Kaluza-Klein superfield $\cW_\alpha{}^\um$ is chiral and obeys a reduced chirality condition,
\begin{align}
\hat\nabla^\alpha \cW_\alpha{}^\um = \hat{\bar\nabla}_\dalpha \bar \cW^{\dalpha \um}~, \qquad
\hat\nabla_\alpha := \nabla_\alpha - 2 F_{\um \alpha}{}^\un g_\un{}^\um~, \qquad
F_{\um \alpha}{}^\un := i X_{\um \alpha \dalpha} \bar \cW^{\dalpha \un}~.
\end{align}
The superfield $\cW_{\alpha\dalpha}$ is related to its conjugate by
\begin{align}
\cW_{\alpha \dalpha} - \bar \cW_{\dalpha \alpha}
    &= -2i \,(\cW_\alpha{}^\um X_{\um \dalpha} + \bar\cW_\dalpha{}^\um X_{\um \alpha})
    - \frac{1}{2} \cW^{\beta \um} \hat\nabla_\beta X_{\um \alpha \dalpha}
    - \frac{1}{2} \bar\nabla_\dbeta (\bar \cW^{\dbeta \um} X_{\um \alpha \dalpha})~.
\end{align}
The other linear combination $\cW_{\alpha \dalpha} + \bar \cW_{\dalpha \alpha}$ can be fixed
to whatever we wish by a redefinition of the superspace connections.
Other terms are related to those above:
\begin{alignat}{2}
\cW_\alpha(D) &= \frac{1}{2} \bar\nabla^\dgamma \cW_{\alpha \dgamma} + \phi_\alpha~, &\qquad
\cW_\alpha(A) &= \frac{3i}{4} \bar\nabla^\dgamma \cW_{\alpha \dgamma} + \frac{i}{2} \phi_\alpha~, \eol
\cW_{\alpha}(M)_{\beta\gamma}
	&= - \eps_{\alpha (\beta} Z_{\gamma)}
	+ 2 i \,W_{\alpha\beta\gamma}~, & \qquad
\cW_\alpha(M)_{\dalpha \dbeta} &= - \bar\nabla_{(\dalpha} \cW_{\alpha \dbeta)}~, \eol
\cW_\alpha{}_\um{}^\un
    &= -\frac{i}{4} \bar\nabla^2 F_{\um \alpha}{}^\un~, & \qquad
\cW_\alpha(S)_\dalpha &= \frac{1}{8} \bar\nabla^2 \cW_{\alpha \dalpha}~, \eol
\cW_{\alpha}(K)^{\dbeta \beta} &= i \bar\nabla^\dbeta \cW_\alpha(S)^{\beta}~.
\end{alignat}
The chiral superfields $\phi_\alpha$ and $Z_\alpha$ are related by
\begin{align}
\phi_{\alpha} = \frac{3}{2} Z_\alpha - \cW_\alpha{}_\um{}^\um 
\end{align}
$Z_\alpha$ can be fixed to whatever we wish by a connection redefinition.
The superfield $\cW_\alpha(S)^\beta$ is complex linear and can be written
\begin{align}
\cW_\alpha(S)^\beta &=
    - \frac{1}{4} \Big(
    \nabla^\gamma \cW_\gamma(M)_\alpha{}^\beta
    + 2 F_\um{}^{\gamma \um} \cW_\gamma(M)_\alpha{}^\beta
    + \cW^\gamma{}^\um R^+_{\um \gamma}(M)_\alpha{}^\beta
    \Big)
    \eol & \quad
    + \frac{1}{12} \delta_\alpha{}^\beta
    (\nabla^\gamma \phi_\gamma + 2 F_\um{}^\gamma{}^\um \phi_\gamma)
    + \frac{1}{8} \delta_\alpha{}^\beta \cW^\gamma{}^\um
    (R^+_{\um \gamma}(D) + \frac{2i}{3} R^+_{\um \gamma}(A))
    \eol & \quad
    + \bar\nabla_\dgamma \Big[
    \frac{1}{4} \bar \cW^\dgamma(M)_\alpha{}^\beta
    - \frac{i}{2} \bar\cW^\dgamma{}^\um X_\um(M)_\alpha{}^\beta
    \Big]
    \eol & \quad
    + \delta_\alpha{}^\beta \,\bar\nabla_\dgamma \Big[
    - \frac{1}{8} \bar \cW^\dgamma(D)
    + \frac{i}{4} \bar \cW^\dgamma{}^\um X_\um(D)
    - \frac{i}{12} \bar \cW^\dgamma(A)
    - \frac{1}{6} \bar \cW^\dgamma{}^\um X_\um(A)
    \Big]~.
\end{align}

\subsection{Expressions for $X_{\protect\um}$}
The operator $X_\um = -\frac{i}{2} (\nabla_\um^+ - \nabla_\um^-)$ is given by
\begin{align}
X_\um &= X_\um{}^A \nabla_A 
    + X_\um(D) \, \bbD
    + X_\um(A) \, \bbA
    + X_\um(M)^{\alpha \beta} M_{\alpha \beta}
    + X_\um(M)_{\dalpha \dbeta} M^{\dalpha \dbeta}
    \eol & \quad
    + X_{\um \un}{}^\up g_\up{}^\un
    + X_\um(S)^\alpha S_\alpha
    + X_\um(S)_\dalpha \bar S^\dalpha
    + X_\um(K)^a K_a~.
\end{align}
We leave $X_\um{}_\alpha$ and $X_\um{}^\dalpha$ unfixed and give other quantities
in terms of these. $X_\um(M)$, $X_\um(D)$, and $X_\um(A)$ are determined by
\begin{align}
X_\um(M)_\alpha{}^\beta + \frac{1}{2} \delta_\alpha{}^\beta (X_\um(D) - 2 i X_\um(A))
    &=
    \hat\nabla_\alpha X_\um{}^\beta
    - \delta_\alpha{}^\beta \bar\nabla^\dgamma X_{\um \dgamma}
    + \frac{i}{4} \bar\nabla^\dgamma \hat\nabla_\alpha X_\um{}^\beta{}_\dgamma
    \eol & \quad
    + X_{\um \alpha \dbeta} \bar \cW^{\dbeta \beta}
    + \frac{i}{2} \cW_\alpha{}^\un \Psi_{\un \um}{}^\beta~.
\end{align}
The antisymmetric part of the $\GL(n)$ component $X_{\um \un}{}^\up$ is
\begin{align}
X_{[\um \un]}{}^\up
    &=
    - \frac{i}{4} \Psi_{\um \un}{}^\alpha \cW_\alpha{}^\up
    - \frac{1}{4} X_{[\um}{}_{\beta \dalpha} X_{\un]}{}^{\dalpha \alpha} \hat\nabla^\beta \cW_{\alpha}{}^\up
    - \frac{1}{2} X_{[\um}{}_{\gamma \dgamma} \hat\nabla^\gamma X_{\un]}{}^{\dgamma \alpha} \cW_\alpha{}^\up
    \eol & \quad
    + i X_{[\um}{}^{\dalpha \alpha} F_{\un]}{}_\alpha{}^\uq F_\uq{}_\dalpha{}^\up
    + \HC
\end{align}
The $S$-supersymmetry piece is
\begin{align}
X_\um(S)_\alpha 
    &=
    -\frac{1}{4} \Big[
        \hat \nabla_\alpha X_{\um}(D)
        + X_{\um \alpha \dalpha} \bar \cW^{\dalpha}(D) \Big]
%     \eol & \quad
    - \frac{i}{6} \Big[
        \hat \nabla_\alpha X_{\um}(A)
        + X_{\um \alpha \dalpha} \bar \cW^{\dalpha}(A) \Big]
    - \frac{1}{4} \bar \nabla^\dalpha \hat\nabla_\alpha X_{\um \dalpha}
    \eol & \quad
    - \frac{i}{24} \cW_\alpha{}^\un \Big(
    \hat\nabla^\beta \Psi_{\un \um \beta}
    + 2 \Psi_{\un \um}{}^\beta F_{\up \beta}{}^\up
    \Big)
%     \eol & \quad
    - \frac{i}{6} \nabla_\um^+ F_{\up \alpha}{}^\up
    + \Sigma_{\um \alpha}~,
\end{align}
where $\Sigma_{\um\alpha}$ obeys an inhomogeneous complex linearity condition,
\begin{align}\label{E:Sigma.ModLin}
\bar \nabla^2 \Sigma_{\um \alpha}
    &= - \nabla_\um^+ Z_\alpha 
    - \frac{1}{3} \cW_\alpha{}^\un \cW^\beta{}^\up \Phi_{\ul{pnm} \beta} ~.
\end{align}
Because $\Phi_{\ul{pnm}\,\alpha}$ cannot be set to zero, we cannot eliminate
$\Sigma_{\um \alpha}$. However, a connection redefinition shifts it by an arbitrary
complex linear superfield.

Finally, $X_\um(K)^a$ is given by
\begin{align}
i X_\um(K)_{\alpha \dalpha}
    &= \frac{1}{4} \nabla_\um^+ \cW_{\alpha \dalpha}
    + \frac{1}{4} \cW_\alpha{}^\un (T^+_{\un \um}{}_\dalpha + \Psi_{\un\um}{}^\gamma \cW_\gamma{}_\dalpha)
    \eol & \quad
    + \bar\nabla_\dalpha X_\um(S)_\alpha - F_\um{}_\dalpha{}^\un X_\un(S)_\alpha
    + \frac{1}{16} \bar \nabla^2 \Big[
        \nabla_\alpha X_\um{}_\dalpha
        - 2 F_\um{}_\alpha{}^\un X_{\un}{}_\dalpha
    \Big]
    \eol & \quad
    + \frac{1}{8} \bar\nabla_\dbeta
    \Big(\delta_\um{}^\un \nabla_\alpha - 2 F_\um{}_\alpha{}^\un\Big)
    \Big(
        X_\un(M){}^\dbeta{}_\dalpha 
        + \frac{1}{2} X_\un(D) \,\delta^\dbeta{}_\dalpha
        + i X_\un(A) \,\delta^\dbeta{}_\dalpha \Big)
    \eol & \quad
    - \HC
\end{align}

\subsection{Expressions for $R^+_{\protect\um \protect\un}$}
The operator $R^+_{\um \un}$ is given abstractly by
\begin{align}
R^+_{\um \un}
    &= 
    T^+_{\um \un}{}^B \nabla_B
    + F^+_{\um \un}{}^\up \nabla^+_\up    
    + \Big(\nabla^+_\up F^+_{\um \un}{}^\uq + \cR_{\um \un}{}_\up{}^\uq\Big)g_\uq{}^\up
    \eol & \quad
    + R^+_{\um \un}(D) \mathbb D
    + R^+_{\um \un}(A) \mathbb A
    + R^+_{\um \un}(M)^{\beta \gamma} M_{\beta\gamma}
    + R^+_{\um \un}(M)_{\dbeta \dgamma} M^{\dbeta\dgamma}
    \eol & \quad
    + R^+_{\um \un}(S)^\beta S_\beta
    + R^+_{\um \un}(S)_\dbeta S^\dbeta
    + R^+_{\um \un}(K)^c K_c~.
\end{align}
The torsion and KK curvature parts are
\begin{align}
T^+_{\um \un}{}^{\dbeta \beta} &= - \bar\nabla^\dbeta \Psi_{\um\un}{}^\beta~,\qquad
T^+_{\um \un}{}^{\beta} = \frac{i}{8} \bar\nabla^2 \Psi_{\um\un}{}^\beta~, \eol
T^+_{\um \un}{}_\dalpha
	&= - \frac{i}{8} \nabla^2 \Psi_{\um \un}{}_\dalpha
	- i (F^+_{\um \un}{}^\up + F^-_{\um \un}{}^\up) X_\up{}_\dalpha
	- 4 i \Big(\nabla_{[\um} X_{\un]}{}_\dalpha + X_\um{}^B T_{\un B}{}_\dalpha\Big)~, \eol
F^+_{\um \un}{}^\up &= -\Psi_{\um \un}{}^\alpha \cW_\alpha{}^\up~.
\end{align}
The other components are
\begin{align}
R^+_{\um \un}(M)_{\alpha \beta}
	&= - \Psi_{\um \un}{}^\gamma \cW_\gamma(M)_{\alpha\beta} + \Phi_{\um \un \alpha\beta}~,\eol
R^+_{\um \un}(M)_{\dalpha \dbeta}
	&= \bar \nabla_{(\dalpha} \Psi_{\um \un}{}^\gamma W_{\gamma \dbeta)}
	+ \bar \nabla_{(\dalpha} T^+_{\um \un}{}_{\dbeta)}~, \eol
R^+_{\um\un}(D) &=
	-\Psi_{\um\un}{}^\alpha \phi_\alpha
	+ \Phi_{\um \un}
	- \frac{1}{2} \bar\nabla^{\dgamma} T^+_{\um\un \dgamma}
	- \frac{1}{2} \bar\nabla^{\dgamma} \Psi_{\um \un}{}^\gamma \cW_{\gamma\dgamma}~,\eol
R^+_{\um \un}(A) &=
	- \frac{i}{2} \Psi_{\um\un}{}^\alpha \phi_\alpha
	+ \frac{i}{2} \Phi_{\um \un}
	- \frac{3i}{4} \bar\nabla^{\dgamma} T^+_{\um\un \dgamma}
	- \frac{3i}{4} \bar\nabla^{\dgamma} \Psi_{\um \un}{}^\gamma \cW_{\gamma\dgamma}~,\eol	
R^+_{\um \un}(S)_{\alpha}
	&= -\Psi_{\um \un}{}^\gamma \cW_\gamma(S)_\alpha
	+ \Sigma_{\um \un \alpha}~,\eol
R^+_{\um \un}(S)_\dalpha &=
	\frac{1}{8} \bar\nabla^2 T^+_{\um \un \dalpha}
	+ \frac{1}{8} \bar\nabla^2 \Psi_{\um \un}{}^\gamma W_{\gamma\dgamma}
	- \frac{1}{4} \bar\nabla_\dphi \Psi_{\um \un}{}^\gamma \bar\nabla^\dphi \cW_{\gamma\dgamma}~,\eol
R^+_{\um \un}(K)_{\alpha\dalpha}
	&= -i \bar\nabla_\dalpha R^+_{\um \un}(S)_{\alpha}
	- i \bar\nabla^\dalpha \Psi_{\um \un}{}^\beta \cW_\beta(S)_\alpha~, \eol
\cR^+_{\ul{mnp}}{}^\uq
	&=
	- \Psi_{\um \un}{}^\alpha \cW_\alpha{}_\up{}^\uq
	+ \nabla_\up^+ \Psi_{\um \un}{}^\alpha \cW_\alpha{}^\uq
	+ \Phi_{\ul{mnp}}{}^\uq~,
\end{align}
where
\begin{align}
\Phi_{\un \um\, \alpha\beta}
    &= - \frac{i}{8} \bar\nabla^2 \Big[
    \hat\nabla_{(\beta} \Psi_{\un \um \alpha)}
    + 2 \Psi_{\un \um (\beta} F_{\up \alpha)}{}^\up
    - 2i\, \hat\nabla_{(\beta} X_{[\un \gamma) \dgamma} \hat\nabla_{(\alpha} X_{\um]}{}^{\gamma) \dgamma}
    \Big]
    - \cW_{(\alpha}{}^{\up} \Phi_{\ul{pnm}\, \beta)}~, \eol
\Phi_{\un \um}
    &= \frac{i}{16} \bar\nabla^2 \Big[
    \hat\nabla^\gamma \Psi_{\un \um \gamma}
    + 2 \Psi_{\un \um}{}^\gamma F_{\up \gamma}{}^\up
    \Big]
    + \frac{1}{2} \cW^{\beta \up} \Phi_{\ul{pnm} \, \beta}~, \eol
\Phi_{[\ul{mnp}]}{}^\uq
    &= 
    -\frac{1}{3} \Phi_{\ul{mnp}}{}^\alpha \cW_\alpha{}^\uq
    - \frac{i}{4} \bar\nabla^2 (\Psi_{[\um \un}{}^\alpha F_{\up]\alpha}{}^\uq)~.
\end{align}
In practice, only the antisymmetric part of $\Phi_{\ul{mnp}}{}^\uq$ given in the last
line is relevant. We do not give an explicit expression for $\Sigma_{\um \un}{}_\alpha$
but it can be worked out.

\subsection{Fundamental Bianchi identities}
Below we list the fundamental Bianchi identities that generalize \eqref{E:LinSugraBI}:
\begin{subequations}
\begin{align}
\nabla^+_\um W_{\alpha\beta\gamma}
    &= \frac{1}{16} \bar\nabla^2 \Big[
    \hat\nabla_\alpha{}^\dgamma \hat\nabla_\beta X_{\um \gamma \dgamma}
    - \hat\nabla_\alpha (\cW_\beta{}^\un \Psi_{\un \um \gamma})
    + 4 \hat\nabla_\alpha X_{\um \beta \dgamma} \bar \cW^{\dgamma \un} X_{\un \gamma}
    \eol & \quad
    + 2i\,\hat \nabla_\alpha X_{\um \beta \dgamma} \bar\cW^{\dgamma}{}_\gamma
    \Big]_{(\alpha\beta\gamma)}
%     \eol & \quad
    + \frac{i}{2} \cW_{(\alpha}{}^\un \Phi_{\un \um \, \beta \gamma)}~, \\
\nabla_{[\um} X_{\un]}{}^a 
	&= 
        - X_{[\um}{}^B T_{\un] B}{}^a
        + \frac{i}{4} (T^+_{\um \un}{}^a - T^-_{\um \un}{}^a) 
	- \frac{1}{4} (F^+_{\um \un}{}^\up + F^-_{\um \un}{}^\up) X_{\up}{}^a~, \\
\nabla_{[\up}^+ \Psi_{\um \un] \, \alpha}
    &= 
    - \frac{i}{4} \bar\nabla^\dbeta \Psi_{[\um \un}{}^\beta T^+_{\up] \beta \, \alpha \dbeta}
    + \Psi_{[\um \un|}{}^\beta \cW_\beta{}^\uq \Psi_{\uq| \up]\, \alpha}
    + \frac{1}{3} \Phi_{\ul{pmn}\, \alpha}~, \\
\nabla_1^+ \Phi_{3 \alpha}
    &= -\frac{1}{8} \bar\nabla^2 \Big[
        i \Psi_2{}^\beta \hnabla_\beta \Psi_{2 \alpha}
        + i \Psi_2{}^\beta \Psi_2{}_\beta \, F_{\um \alpha}{}^\um
        - X_{1 \alpha \gamma \dgamma} X_1{}^{\beta \dgamma \gamma} \Psi_{2 \beta}
    \Big]
\end{align}
\end{subequations}
The last we have written in form notation.
In addition to these, one must also specify the inhomogeneous complex linearity condition
of $\Sigma_{\um \alpha}$, see \eqref{E:Sigma.ModLin}.

\subsection{Some explicit expressions for torsions and KK curvatures}
For reference, we give some explicit expressions for torsion and Kaluza-Klein curvatures.
Some of the mixed torsion tensors are particularly simple in the $+$ basis:
\begin{subequations}
\begin{align}
T^+_{\um \alpha, \beta \dbeta}
    &= 2 i \hat\nabla_{\alpha} X_{\um \beta \dbeta}
	+ 8\, \eps_{\alpha\beta} X_{\um \dbeta}~, \\
T^+_{\um \alpha}{}^\beta
	&= \frac{i}{4} \bar\nabla_\dphi T^+_{\um \alpha,}{}^{\dphi \beta}
	+ \cW_\alpha{}^\un \Psi_{\un\um}{}^\beta~, \\
T^+_{\um \alpha}{}_{\dbeta}
    &= 2 i \hat\nabla_\alpha X_{\um \dbeta}~, \\
T^+_{\um \, \alpha \dalpha, \,\beta \dbeta}
    &= \frac{i}{2} \bar \nabla_\dbeta T^+_{\um \alpha, \,\beta \dalpha}
    + 2 \,\eps_{\dalpha \dbeta} \cW_\alpha{}^\un \Psi_{\un \um \beta}~, \\
T^+_{\um\, \alpha \dalpha}{\,}^{\beta}
    &= \frac{1}{16} \bar\nabla^2 T^+_{\um \alpha, \dalpha}{}^\beta
    - \frac{i}{2} \cW_\alpha{}^\un \bar\nabla_\dalpha \Psi_{\un \um}{}^\beta
\end{align}
\end{subequations}
The mixed Kaluza-Klein curvatures are
\begin{subequations}
\begin{align}
F^+_{\um \alpha}{}^\un = 2 i \, X_{\um \alpha \dalpha} \bar \cW^{\dalpha \un}~, \qquad
F^+_{\um\, \alpha \dalpha}{}^\un
    = \frac{i}{2} \bar\nabla_\dalpha F^+_{\um\alpha }{}^\un
     - \frac{i}{2} T^+_{\um \alpha, }{}_{\gamma \dalpha} \cW^\gamma{}^\un~.
\end{align}
\end{subequations}
In a real basis written in terms of $\nabla_\um$, one has
\begin{subequations}
\begin{alignat}{2}
F_{\um \alpha}{}^\un &= \frac{1}{2} F^+_{\um \alpha}{}^\un~, &\qquad
T_{\um \alpha}{}^B &= \frac{1}{2} (T^+_{\um \alpha}{}^B + i F^+_{\um \alpha}{}^\un X_\up{}^B)~, \\
F_{\um a}{}^\un &= \textrm{Re\,} F^+_{\um a}{}^\un~, &\qquad
T_{\um a}{}^B &= \textrm{Re\,} \Big(
    T^+_{\um a}{}^B + i F^+_{\um a}{}^\un X_\un{}^B
    \Big)~.
\end{alignat}
\end{subequations}
The external torsion and Kaluza-Klein curvatures found in the $[\nabla_\alpha, \nabla_b]$
commutator are
\begin{subequations}
\begin{alignat}{2}
F_{b \alpha}{}^\un &= (\sigma_b)_{\alpha \dbeta} \bar \cW^{\dbeta \un}~,
&\qquad
F_{b}{}^{\dalpha}{}^\un &= -(\sigma_b)^{\dalpha \beta} \cW_\beta{}^{\un}~, \\
T_{b \alpha}{}^c &= -i (\sigma_b)_{\alpha \dbeta} \bar \cW^{\dbeta \un} X_\un{}^c
&\quad
T_{b}{}^{\dalpha}{}^c &= -i (\sigma_b)^{\dalpha \beta} \cW_\beta{}^{\un} X_\un{}^c \\
T_{b \alpha}{}^\gamma &= 
    (\sigma_b)_{\alpha \dbeta} ( \bar \cW^{\dbeta \gamma}
        - i \bar \cW^{\dbeta \un} X_\un{}^\gamma )
&\quad
T_b{}^{\dalpha}{}_\dgamma &= 
    - (\sigma_b)^{\dalpha \beta} ( \cW_{\beta \dgamma}
        + i \cW_\beta{}^{\un} X_\un{}_\dgamma ) \\
T_{b \alpha}{}_\dgamma &=
    - i (\sigma_b)_{\alpha \dbeta} \bar \cW^{\dbeta \un} X_\un{}_\dgamma
&\quad
T_{b}{}^\dalpha{}^\gamma &=
    - i (\sigma_b)^{\dalpha \beta} \cW_\beta{}^{\un} X_\un{}^\gamma~.
\end{alignat}
\end{subequations}
Those found in the vector-vector commutator are most easily written by decomposing 
the curvature operator into self-dual and anti-self-dual pieces, 
$R_{a b} = - (\sigma_{ab})^{\alpha\beta} R_{\sym{\alpha\beta}} - (\bsigma_{ab})^{\dalpha\dbeta} R_{\sym{\dalpha\dbeta}}$,
\begin{subequations}
\begin{align}
F_{\sym{\alpha \beta}}{}^\um
    &= -\frac{i}{2} \hat \nabla_{(\alpha} \cW_{\beta)}{}^\um~, \\
T_{\sym{\alpha \beta}}{}^c
    &= i F_{\sym{\alpha \beta}}{}^\un X_\un{}^c
    + \frac{i}{2} \cW_{(\alpha}{}^\un T^+_{\un \beta)}{}^c
    - W_{(\alpha \dgamma} (\sigma^c)_{\beta)}{}^\dgamma~,\\
T_{\sym{\alpha \beta}}{}^\gamma
    &= i F_{\sym{\alpha \beta}}{}^\un X_\un{}^\gamma
    + \frac{i}{2} \cW_{(\alpha}{}^\un T^+_{\un \beta)}{}^\gamma
    - \frac{i}{2} \delta_{(\alpha}{}^\gamma 
        \Big(\bar\nabla^\dgamma \cW_{\beta) \dgamma}
        + 2 Z_{\beta)} - \cW_{\beta)}{}_\un{}^\un
        \Big)
    + W_{\alpha \beta}{}^\gamma
    ~, \\
T_{\sym{\alpha \beta}}{}_\dgamma
    &= i F_{\sym{\alpha \beta}}{}^\un X_\un{}_\dgamma
    + \frac{i}{2} \cW_{(\alpha}{}^\un T^+_{\un \beta)}{}_\dgamma
    - \frac{i}{2} \nabla_{(\alpha} \cW_{\beta)}{}_\dgamma
    ~.
\end{align}
\end{subequations}
From the second equation, one can see that $T_{ab}{}^c$ does not vanish.

\bibliography{library.bib}

\bibliographystyle{utphys_mod_v2}

\end{document}

%% file: sugracom_v3.tex
\newcommand {\cA}{{\cal A}}

\newcommand {\cD}{{\cal D}}
\newcommand {\cE}{{\cal E}}
\newcommand {\cF}{{\cal F}}

\newcommand {\cH}{{\cal H}}

\newcommand {\cL}{{\cal L}}
\newcommand {\cM}{{\cal M}}
\newcommand {\cN}{{\cal N}}

\newcommand {\cR}{{\cal R}}

\newcommand {\cV}{{\cal V}}
\newcommand {\cW}{{\cal W}}

\newcommand {\cY}{{\cal Y}}

\def\a{\alpha}

\newcommand{\rd}{{\mathrm d}}

\newcommand{\ad}{{\dot{\alpha}}}

\newcommand{\dalpha}{{\dot{\alpha}}}
\newcommand{\dbeta}{{\dot{\beta}}}
\newcommand{\dgamma}{{\dot{\gamma}}}

\newcommand{\dmu}{{\dot{\mu}}}

\newcommand{\dphi}{{\dot{\phi}}}
\newcommand{\hA}{{\hat{A}}}
\newcommand{\hB}{{\hat{B}}}

\newcommand{\bm}[1]{\mbox{\boldmath$#1$}}
\newcommand{\veps}{\varepsilon}
\newcommand{\eps}{{\epsilon}}
\newcommand{\eol}{\notag \\}

\newcommand{\bsigma}{\bar{\sigma}}
\newcommand{\ul}{\underline}
\newcommand{\pa}{\partial}